\documentclass[twocolumn,showpacs,preprintnumbers,amsmath,amssymb]{revtex4}
\usepackage{graphicx}
\usepackage{dcolumn}
\usepackage{bm}

\newcommand{\Eq}[1]{Eq.~({\protect\ref{#1}})}

\newcommand{\Sect}[1]{Sect.~\protect\ref{#1}}
\newcommand{\Fig}[1]{Fig.~\protect\ref{#1}}
\newcommand{\Figs}{Figs.}
\newcommand{\Table}[1]{Table~\protect\ref{#1}}
\newcommand{\Ndf}{N_{\text{DF}}}
\newcommand{\Nsmear}{N_{\text{smr}}}
\newcommand{\Nconfig}{N_{\text{conf}}}

\newcommand{\Laplacian}{\triangle}

\newcommand{\Tate}{\rule{0cm}{1.1em}}
\newlength{\Tatescale}
\newcommand{\Hs}{\hspace*{1em}}
\newcommand{\Bs}{\hspace*{-0.5em}}

\newlength{\figwidth}
\newcounter{subfigure}
\newcommand{\Cut}[1]{}
\begin{document}
\title{Glueball Properties at Finite Temperature
in SU(3) Anisotropic Lattice QCD}
\author{Noriyoshi Ishii}
\address{
The Institute of Physical and Chemical Research (RIKEN),\\
2-1 Hirosawa, Wako, Saitama 351-0198, Japan}

\author{Hideo Suganuma}
\address{Faculty of Science, Tokyo Institute of Technology,\\
2-12-1 Ohkayama, Meguro, Tokyo 152-8552, Japan}

\author{Hideo Matsufuru}
\address{Yukawa Institute for Theoretical Physics, Kyoto University,\\
Kitashirakawa-Oiwake, Sakyo, Kyoto 606-8502, Japan}


\begin{abstract}
The  thermal  properties  of  the  glueballs  are  studied  using  SU(3)
anisotropic  lattice QCD with  $\beta_{\rm lat}=6.25$,  the renormalized
anisotropy $\xi  \equiv a_s/a_t=4$  over the lattice  of the  size $20^3
\times N_t$ with $N_t = 24, 26, 28,  30, 33, 34, 35, 36, 37, 38, 40, 43,
45, 50, 72$ at the quenched level.  To construct a suitable operator for
the lowest-state glueball on the  lattice, we adopt the smearing method,
providing  an illustration  of  its  physical meaning  in  terms of  the
operator size.  First, we  construct the temporal correlators $G(t)$ for
the lowest $0^{++}$ and $2^{++}$  glueballs, using more than 5,500 gauge
configurations at  each temperature $T$.  We then  perform the pole-mass
measurement  of  the thermal  glueballs  from  $G(t)$.   For the  lowest
$0^{++}$ glueball, we observe a significant pole-mass reduction of about
300 MeV in the vicinity of  $T_c$ or $m_G(T \simeq T_c) \simeq 0.8 m_G(T
\sim 0)$,  while its  size remains almost  unchanged as  $\rho(T) \simeq
0.4$ fm.
Finally, for completeness, as an attempt to take into account the effect
of thermal  width $\Gamma(T)$ at  finite temperature, we perform  a more
general new analysis of $G(t)$ based on its spectral representation.  As
an  ansatz  to  the  spectral  function  $\rho(\omega)$,  we  adopt  the
Breit-Wigner  form, and perform  the best-fit  analysis of  the temporal
correlator  as a  straightforward  extension to  the standard  pole-mass
analysis.
The result indicates a significant  broadening of the peak as $\Gamma(T)
\sim 300$ MeV  as well as rather modest reduction of  the peak center of
about  $100$  MeV near  $T_c$  for  the  lowest $0^{++}$  glueball.
The  temporal correlators  of the  color-singlet modes  corresponding to
these glueballs above $T_c$ are also investigated.
\end{abstract}
\maketitle
\section{Introduction}

QCD at finite temperature is one of the most interesting subjects in the
quark  hadron  physics  \cite{qcd,miyamura,satz}.  At  low  temperature,
quarks and  gluons are confined  in the color-singlet  objects, hadrons,
due to the nonperturbative effect of the strong interactions among them.
With the increasing temperature,  these interactions diminish due to the
asymptotic freedom,  which is expected  to lead to liberation  of quarks
and gluons  above a critical  temperature $T_c$, forming the  new phase,
i.e., the quark-gluon plasma (QGP) phase.

At present,  we are in front  of the experimental  possibility to create
the QGP  phase in  the RHIC project  at Brookhaven  National Laboratory.
Hence, much progress is desired  in the theoretical understanding of QCD
at  finite temperature.   To  study  QCD at  finite  temperature and  to
understand  the  QCD phase  transition,  the  effective  models and  the
lattice  QCD Monte  Carlo  simulation provide  useful and  complementary
approaches.  While  the effective models  offer their own  insights into
the physical phenomena and  often provide analytical methods to estimate
related  physical  quantities, the  lattice  QCD  simulation provides  a
model-independent  method of  calculating  physical quantities  directly
based on QCD.

The SU(3) lattice QCD at  the quenched level indicates the deconfinement
phase transition  of the  weak first order  at the  critical temperature
$T_c  \simeq 260$  MeV  \cite{karsch1},
%
%
and simulations  with dynamical quarks show the  chiral phase transition
at $T_c=173(8)$ MeV for $N_f=2$ and $T_c =154(8)$ MeV for $N_f=3$ in the
chiral limit \cite{full.qcd}.

Above  $T_c$,   most  of   nonperturbative  properties  such   as  color
confinement   and   spontaneous   chiral-symmetry  breaking   disappear.
Consequently,  quarks  and  gluons  are liberated,  and  the  tremendous
changes are expected in the mass spectrum.

Even  below  $T_c$,  one  is  interested  in  the  hadronic  mass  shift
\cite{miyamura,hatsuda-kunihiro,ichie},  because it  can be  one  of the
most  important pre-critical phenomena  of the  QCD phase  transition at
finite temperature (and also at  finite density).  In this respect, from
the experimental side for instance, CERES Collaboration \cite{ceres} has
proposed the  high-energy heavy-ion  collision data, which  may indicate
the mass  shift of $\rho$-meson, and extensive  theoretical efforts have
been   made   aiming  at   understanding   of   its  true   implications
\cite{kapusta}.

The theoretical background  to believe in the hadronic  mass shift is as
follows.   As   the  temperature  increases  up  close   to  $T_c$,  the
inter-quark    potential    is    known    to    change    significantly
\cite{matsufuru,karsch2}.  As far as  the heavy quarkonia are concerned,
for instance  $J/\psi$, the change  of the inter-quark potential  may be
followed by the changes in the  structures of hadrons, which may lead to
significant  mass shifts  as  a consequence  \cite{miyamura}.  One  also
expects  significant mass  shifts of  the light  hadrons,  for instance,
$\sigma$-meson  \cite{hatsuda-kunihiro},  which is  due  to the  partial
chiral restoration.

At finite temperature, due to the absence of the Lorentz invariance, the
concept of  the ``mass'' becomes  less definitive, and there  emerge two
distinct concepts of the ``screening  mass'' and the ``pole-mass''.  The
screening  mass governs  the correlations  along the  spatial direction,
while  the  pole-mass  governs  the  correlations  along  the  temporal
direction.
The relation  between them  is analogous to  the relation  between the
Debye  screening mass  and the  plasma frequency  in  electro dynamics
\cite{detar-kogut}.   Both   of  these  two   masses  are  interesting
quantities, which  can reflect the  important nonperturbative features
of QCD.
However,  we are more  interested here  in the  pole-mass than  in the
screening mass,  because the pole-mass  is closely related  to the QGP
creation experiments and calculations mentioned above.
%
In  fact, the  pole-mass  is  expected to  be directly  observed as  the
physical mass of the thermal hadrons.

In  contrast  to  the   effective  model  approaches,  which  have  been
extensively   used  to   study   the  thermal   properties  of   hadrons
\cite{miyamura,hatsuda-kunihiro},  only a few  lattice QCD  studies have
been  performed   for  the   pole-mass  of   thermal  hadrons   so  far
\cite{taro,umeda,australia,ishii}.
Instead,  the  screening mass  has  been  extensively  studied with  the
lattice  QCD \cite{detar-kogut,gupta,gupta2,laermann}  by  analyzing the
spatial correlations of the hadronic correlators.
The reason behind this is the technical difficulty of measuring hadronic
two-point correlators in the temporal direction at finite temperature on
the lattice.
In fact, since the temporal extension of the lattice shrinks as $1/T$ at
high temperature, the pole-mass measurements have to be performed in the
limited distance  shorter than $1/(2T)$,  which corresponds to  $N_t/2 =
2\mbox{--}4$  near   $T_c$  in   the  ordinary  isotropic   lattice  QCD
\cite{detar-kogut}.

The severe  limitation on the measurement of  the temporal correlation
can be avoided with an  anisotropic lattice where the temporal lattice
spacing $a_t$  is smaller than  the spatial one  $a_s$ \cite{klassen}.
On the anisotropic lattice, by taking a small $a_t$, it is possible to
use efficiently a large number  of the temporal lattice points even in
the vicinity of $T_c$,  while the physical temporal extension $1/T=N_t
a_t$ is  kept fixed.   In this way,  the number of  available temporal
data are  largely increased on  the anisotropic lattice,  and accurate
pole-mass measurements from the temporal correlation become possible.

In this paper, we study  the glueball properties at finite temperature
from the  temporal correlation in SU(3) anisotropic  lattice QCD.  The
glueballs  are special  and  interesting hadrons.   They  belong to  a
different class  of hadrons.  Unlike mesons and  baryons, which mainly
consist  of  valence  quarks  and anti-quarks,  the  glueballs  mainly
consist of gluons in a color-singlet combination.
Their existences  are theoretically predicted in QCD  as a consequence
of  the self-interactions  among gluons,  and numbers  of  lattice QCD
calculations   have  been  performed   aiming  at   investigating  the
properties of glueballs \cite{morningstar,weingarten,teper}.
In  the  real  world, the  glueballs  are  expected  to mix  with  the
$q\bar{q}$ mesons due  to the presence of the  light dynamical quarks.
This mixing  problem raises a  difficulty in distinguishing  them from
ordinary mesons in full lattice QCD as well as in experiments.
Still, enormous experimental  efforts have been and are  being made in
the  pursuit of  the  glueballs  in the  real  world \cite{seth}.   At
present, $f_0(1500)$  and $f_0(1710)$ are taken  as serious candidates
for the $0^{++}$ glueball,  and $\xi(2230)$ for the $2^{++}$ glueball,
and these  glueball candidates  are considered to  contain substantial
fractions of glueball component.
In this paper, to avoid this difficulty of identifying the glueball, we
adopt quenched lattice QCD as a necessary first step before attempting
to include the effect of dynamical quarks in future.
%
It  is  worth  mentioning  here  that, even  without  dynamical  quarks,
quenched lattice  QCD reproduces well various masses  of hadrons, mesons
and baryons, as well as the important nonperturbative quantities such as
the string tension and the chiral condensate.

In quenched  QCD, the elementary  excitations are only glueballs  in the
confinement phase  below the  critical temperature $T_c  \simeq260$ MeV.
From     the     lattice    QCD     studies     at    quenched     level
\cite{morningstar,weingarten,teper}, the lightest physical excitation at
zero temperature is known to be the scalar glueball with $J^{PC}=0^{++}$
and the  mass $m_{\rm G}(0^{++}) \simeq 1500\mbox{--}1700$  MeV, and the
next  lightest  one is  the  tensor  glueball  with $J^{PC}=2^{++}$  and
$m_{\rm G}(2^{++}) \simeq 2000\mbox{--}2400$ MeV.  These light glueballs
are expected to play the  important role in the thermodynamic properties
of quenched QCD below $T_c$.
Our aim  is to understand  the thermodynamic properties of  quenched QCD
below  $T_c$  from the  view  point of  the  thermal  properties of  the
lowest-lying $0^{++}$ and the $2^{++}$ glueballs.

In  our previous  paper, only  the numerical  result of  the pole-mass
reduction of the $0^{++}$ glueball at finite temperature was presented
\cite{ishii}.   In the  present paper,  we  go into  more detail.   We
reformulate the pole-mass measurement itself from the viewpoint of the
spectral representation  of the  temporal correlator.  In  addition to
the  $0^{++}$  glueball,  we   also  consider  the  $2^{++}$  glueball
including the deconfinement phase above the critical temperature $T_c$
as well.
Furthermore, based on the spectral representation, we attempt to take
into account the effects of the possible appearance of the thermal
width of the bound-state peak at finite temperature by proposing the
best-fit analysis of the Breit-Wigner type.

The contents are organized  as follows. In \Sect{section.smearing}, we
begin  with a  brief  review  of the  spectral  representation of  the
temporal glueball correlator. We then introduce the smearing method as
a  method to  enhance the  ground-state contribution  in  the glueball
correlator,  and  consider  the   physical  meaning  of  the  smearing
method. We  also give a prescription  how to make a  rough estimate of
the glueball size based on the informations from the temporal glueball
correlators with the smearing method.
\Sect{section.lattice-setup} is  devoted to the  brief descriptions of
the  lattice QCD  action,  its parameters,  the  determination of  the
scale, and the critical temperature $T_c$ in our anisotropic lattice.
In  \Sect{section.pole-mass-measurement},  we  construct the  temporal
glueball  correlators based  on  SU(3) lattice  QCD,  and perform  the
pole-mass measurements assuming that the  thermal width of the peak in
the spectral function is  sufficiently narrow.  We study the low-lying
$0^{++}$  and $2^{++}$  glueballs both  below and  above  $T_c$.  This
procedure  itself   has  been  widely   used  in  the   standard  mass
measurements  of  various  hadrons  in  the lattice  QCD  Monte  Carlo
calculations   at   zero  temperature.   It   was   also  adopted   by
\cite{taro,umeda,ishii}  in  the  pole-mass  measurements  of  various
mesons at finite temperature.
The  section is closed  with the  comparison of  our results  with the
related  lattice QCD  results on  the  screening masses  and the  pole
masses at finite temperature.
In \Sect{section.spectral-function}, for completeness, we perform a more
general  new analysis  of the  temporal correlators  of glueballs  as an
attempt to take  into account the effects of  the possible appearance of
the  non-zero   thermal  width  of   the  bound-state  peak   at  finite
temperature.  We adopt Breit-Wigner ansatz  for the form of the spectral
function to extract the center and the thermal width of the ground-state
peak through  the spectral representation.   This procedure itself  is a
straightforward  extension to  the  procedure adopted  in the  pole-mass
measurements, which  seems to be  widely applicable to  various temporal
correlators of thermal hadrons.
\Sect{section.summary} is devoted to the summary and concluding remarks.

\section{The Smearing Method in Quenched SU(3) Lattice QCD}
\label{section.smearing}
\subsection{The spectral representation}
Using the glueball operator $\phi_0(t,\vec x)$, we consider the temporal
correlator of the glueball as
\begin{equation}
	G(t)
\equiv
	\left\langle\Tate \phi(t) \phi(0) \right\rangle,
\label{glueball.correlator}
\end{equation}
where $\phi(t)$ is the zero-momentum projected operator defined as
\begin{equation}
	\phi(t)
\equiv
	\phi_0(t)  -  \langle   \phi_0  \rangle,
\Hs
	\phi_0(t)
\equiv
	{1\over N_s} \sum_{\vec x} \phi_0(t,\vec x),
\label{glueball.field}
\end{equation}
with $N_s \equiv N_x N_y N_z$ being the number of sites in the spatial
submanifold.  The angular bracket  $\langle \cdot \rangle$ denotes the
statistical average.  The subtraction of the  vacuum expectation value
is necessary for  the $0^{++}$ glueball.  The summation  over $\vec x$
is  performed  for  the  zero-momentum  projection,  i.e.,  the  total
momentum is set to be  zero.  The glueball operator $\phi_0(t,\vec x)$
has  to be  properly constructed  so as  to have  a definite  spin and
parity  in  the  continuum  limit.   For instance,  the  $0^{++}$  and
$2^{++}$       glueball       operator       are      defined       as
\cite{morningstar,weingarten,rothe,montvey}
\begin{eqnarray}
	\phi_0(t,\vec x)
&\equiv&
	\text{Re}
	\text{Tr}
	\left(\Tate
		P_{12}(t,\vec x) + P_{23}(t,\vec x)  + P_{31}(t,\vec x)
	\right),
\label{glueball.operator}
\\\nonumber
	\phi_0(t,\vec x)
&\equiv&
	\text{Re}
	\text{Tr}
	\left(\Tate
		2P_{12}(t,\vec x) - P_{23}(t,\vec x)  - P_{31}(t,\vec x)
	\right),
\end{eqnarray}
respectively, where $P_{ij}(t,\vec x)$ denotes the plaquette operator in
the $i$-$j$ plane.

To  consider  the  spectral  representation,  we  express  the  temporal
correlator $G(\tau)$ in the canonical operator representation as
\begin{equation}
	G(\tau)
=
	Z(\beta)^{-1}
	\text{tr}\left(\Tate
		e^{-\beta H} \phi(\tau)\phi(0)
	\right),
\end{equation}
where $\beta  = 1/T$ denotes  the inverse temperature,  $Z(\beta) \equiv
\text{tr}(  e^{-\beta  H}  )$   the  partition  function,  and  $H$  the
Hamiltonian of  QCD.  The field operator $\phi(\tau)$  is represented in
the  imaginary-time  Heisenberg  picture  as $\phi(\tau)  =  e^{\tau  H}
\phi(0) e^{-\tau H}$.  By using the identity $G(\tau) = G(\beta - \tau)$
for the manifest periodicity, we obtain the spectral representation as
\begin{eqnarray}
	G(\tau)
&=&
\label{spectral.rep}
\label{spectral.representation}
\renewcommand{\arraystretch}{1.5}
\Bs
\begin{array}[t]{l}	\displaystyle
	\sum_{n,m}
	{
		\left|\left\langle n \left|
			\phi
		\right| m \right\rangle\right|^2
	\over
		2Z(\beta)
	}
	\exp\left( - \beta {E_m + E_n \over 2} \right)
\\\displaystyle
\times
	\cosh\left[\Tate
		(\tau - \beta/2)(E_n - E_m)
	\right]
\end{array}
\\\nonumber
&=&
	\int_{-\infty}^{\infty}
	{d\omega \over 2\pi}
	{\rho(\omega) \over 2\sinh(\beta \omega/2)}
	\cosh\left(\Tate
		\omega(\beta/2 - \tau)
	\right),
\end{eqnarray}
where $E_n$ denotes the energy  of the $n$-th excited state $|n\rangle$.
In this  paper, $|0\rangle$ denotes the vacuum,  and $|1\rangle$ denotes
the lowest  glueball state.   Here, $\rho(\omega)$ denotes  the spectral
function defined as
\begin{eqnarray}
	\rho(\omega)
&\equiv&
	\sum_{n,m}
	{
		\left|\left\langle n \left| \phi \right|m\right\rangle\right|^2
	\over
		Z(\beta)
	}
	e^{-\beta E_m}
\label{spectral.function}
\\\nonumber
&&
\times
	2\pi\left(\Tate
		\delta \left( \omega - E_n + E_m \right)
	-	\delta \left( \omega - E_m + E_n \right)
	\right).
\end{eqnarray}
Note  that  $\rho(\omega)$  is  positive for  positive  $\omega$,  and
negative  for negative  $\omega$.  Due  to the  bosonic nature  of the
glueballs, $\rho(\omega)$ is odd in $\omega$.

The  spectral function  $\rho(\omega)$ is  the residue  of  the glueball
correlator in the Fourier representation as
\begin{equation}
	G(\omega_l)
=
	\int_0^\beta
	d\tau
	e^{-i\omega_l \tau}
	G(\tau)
=
	\int_{-\infty}^{\infty}
	{d\omega' \over 2\pi}
	{\rho(\omega') \over i\omega_l - \omega'},
\end{equation}
where $\displaystyle \omega_l \equiv  {2\pi l\over \beta}$ denotes the
Matsubara  frequency  for bosons.   It  is  also  the residue  of  the
retarded and advanced Green  functions $G_{\rm R}(\omega)$ and $G_{\rm
A}(\omega)$ as \cite{nakamura}
\begin{equation}
	G_{{\rm R}/{\rm A}}(\omega)
=
	\int_{-\infty}^{\infty} {d\omega'\over 2\pi}
	{\rho(\omega') \over \omega - \omega' \pm i\epsilon}.
\label{real.time}
\end{equation}
($+$  for the  retarded  and  $-$ for  the  advanced Green  functions.)
Hence, the  information of the  physical observables such as  the mass
and the width can be  obtained by parameterizing the spectral function
$\rho(\omega)$ properly and by performing the best-fit analysis to the
lattice QCD Monte Carlo data of the temporal correlator $G(t)$ through
the spectral representation \Eq{spectral.rep}.

\subsection{The smearing method}

For the study of the low-lying glueballs, it is important to extract the
contributions  of  the  low-lying  states  in lattice  QCD  Monte  Carlo
simulations. The  smearing method is  one of the most  popular numerical
techniques to enhance the contributions from the low-lying states. Here,
we give  a brief review of the  smearing method in SU(3)  lattice QCD in
the zero-temperature limit.
In  this  limit $\beta\equiv  1/T  \to  \infty$,  the spectral  function
$\rho(\omega)$ and the correlator $G(t)$ are simply expressed as
\begin{eqnarray}
	\rho(\omega)
&=&
	\sum_n 2\pi A_n
	\left(\Tate
		\delta(\omega - E_n) - \delta(\omega + E_n)
	\right),
\\\nonumber
	\frac{G(t)}{G(0)}
&=&
	\sum_n C_n \exp( -E_n t ),
\end{eqnarray}
with
$A_n
\equiv
	\left|\left\langle n \left|
		\phi
	\right|0 \right\rangle\right|^2
	/
	Z
$
and
\begin{equation}
	C_n
\equiv
	\frac{ A_n }{\displaystyle \sum_k A_k }
=
	\frac{ |\langle n| \phi |0\rangle|^2 }
	{\displaystyle \sum_k |\langle k| \phi |0\rangle|^2 }.
\end{equation}
Note that  $C_n$ is a  non-negative real number  with $0 \le C_n  \le 1$
with  the normalization  as $\displaystyle  \sum C_n  = 1$.  It  will be
referred to  as the overlap between  the state $\phi  |0\rangle$ and the
$n$-th asymptotic state $|n\rangle$.

In general, to measure the  ground-state mass from $G(t)$, one seeks for
the  region where  contributions  from excited  states  almost die  out,
leaving  only  the  ground-state  component  as  $G(t)/G(0)  \simeq  C_1
\exp(-E_1 t)$.   In this region,  one may perform the  best-fit analysis
with
\begin{equation}
	C \exp(-m_{\rm G} t)
\label{single.exponential}
\end{equation}
to obtain the  ground-state mass $m_{\rm G} =  E_1$ and the ground-state
overlap $C= C_1$.   (Here, as was mentioned in  the previous subsection,
$|1 \rangle$ denotes the state  of the lowest-lying glueball, while $| 0
\rangle$   denotes  the   vacuum  state.    Because  $\phi(t)$   is  the
zero-momentum projected operator,  defined in \Eq{glueball.field}, $E_n$
denotes  the mass of  the $n$-th  state.)  In  principle, such  a region
always   exists   for  enough   large   $t$   in  the   zero-temperature
case. However,  in practice, it  is difficult to  use such large  $t$ in
lattice  QCD  Monte  Carlo  calculations, since  the  correlator  $G(t)$
decreases exponentially with $t$ and  becomes so small for the large $t$
that it is comparable to its statistical error.
Hence, to measure the mass of the ground-state, it is essential that the
ground-state overlap $C_1$ should be sufficiently large.
In  the  case  of the  glueballs  in  quenched  SU(3) lattice  QCD,  the
ground-state    overlaps    $C_1$   of    the    operators   given    in
\Eq{glueball.operator} are quite small,  as long as they are constructed
from the simple plaquette operators $P_{ij}(t,\vec x)$.
In this case, due to the considerable contributions from excited states,
the measured  mass with \Eq{single.exponential} always behaves  as if it
were much heavier.
The difficulty of  this small overlap originates from  the fact that the
plaquette operator  has a smaller  ``size'' of $O(a)$ than  the physical
size of the glueball \cite{ape}.   This problem becomes severer near the
continuum  limit.   Hence, for  the  accurate  mass  measurement, it  is
necessary  to enhance  the  ground-state contribution  by improving  the
glueball  operator.   This  is  achieved  by making  the  operator  have
approximately the same  size as the physical size  of the glueball.  The
smearing  method provides  an iterative  procedure to  generate  such an
extended  operators,  which is  referred  to  as  the smeared  operators
\cite{morningstar,rothe,ape,takahashi}.

\begin{figure}[h]
\[
	\parbox{0.05\figwidth}{\includegraphics[height=0.3\figwidth]{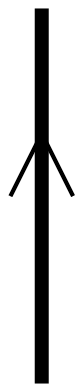}}
=
	\alpha \times
	\parbox{0.05\figwidth}{\includegraphics[height=0.3\figwidth]{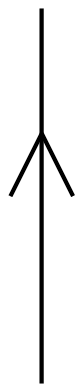}}
+
	\parbox{0.15\figwidth}{\includegraphics[height=0.3\figwidth]{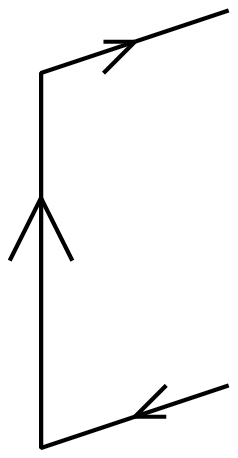}}
+
	\parbox{0.15\figwidth}{\includegraphics[height=0.3\figwidth]{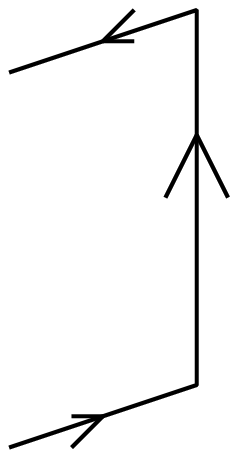}}
+
	\parbox{0.15\figwidth}{\includegraphics[height=0.3\figwidth]{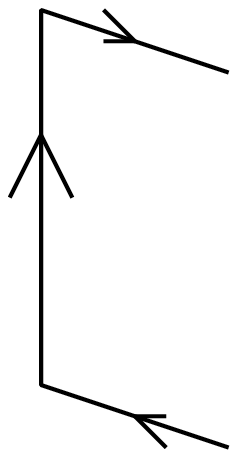}}
+
	\parbox{0.15\figwidth}{\includegraphics[height=0.3\figwidth]{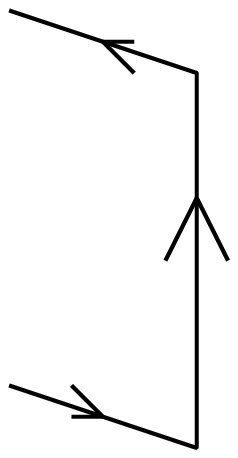}}
\]
\caption{ The schematic illustration  of $V^{(n)}_i(s)$.  The thick line
represents the  variable $V_i^{(n)}(s)$.  Thin lines  represent the link
variables $U_i^{(n)}(s)$. }
\label{smearing.figure}
\end{figure}

The smeared operator is obtained  by replacing the original spatial link
variables $U_i(s)$ in the plaquette operators by the associated fat link
variables $U_i^{(n)}(s)$.   Starting from $U_i^{(0)}(s)  \equiv U_i(s)$,
the $(n+1)$-th  fat link variable $U^{(n+1)}_{i}(s) \in  {\rm SU(3)}$ is
defined iteratively from $U_i^{(n)}(s)$ so as to maximize
\begin{equation}
	\text{Re}
	\text{Tr}
	\left(\Tate
		U_i^{(n+1)}(s)
		V_i^{(n)\dagger}(s)
	\right),
\label{fat.link}
\end{equation}
with
\begin{eqnarray}
	V_i^{(n)}(s)
&\equiv&
	\alpha U_i^{(n)}(s)
\\\nonumber
&+&
	\sum_{\pm,j\neq i}
	U_{\pm j}^{(n)}(s)
	U_i^{(n)}(s\pm\hat j)
	U_{\pm j}^{(n)\dagger}(s+\hat i),
\end{eqnarray}
where  $U_{-\mu}^{(n)}(s)  \equiv  U^{(n)\dagger}_{\mu}(s  -  \hat\mu)$.
Here, $\alpha \in \mathbb{R}$ is  referred to as the smearing parameter,
which controls the speed of  the smearing.  The summation index $j$ runs
over only the spatial directions  to avoid the artificial nonlocality in
time.   A   schematic  illustration   of  $V_i^{(n)}(s)$  is   shown  in
\Fig{smearing.figure}. $U_i^{(n+1)}(s)$ is  the closest SU(3) element to
$V_i^{(n)}(s)$.
It is easy to see that $V_i^{(n)}(s)$ and $U_i^{(n+1)}(s)$ hold the same
gauge transformation properties with $U_i^{(n)}(s)$. Hence, the smearing
method  respects  the  gauge  covariance.  As  the  physically  extended
glueball  operator, we  adopt  the $n$-th  smeared  operator, i.e.,  the
plaquette  operator  which  is  constructed  with the  $n$-th  fat  link
variables $U_i^{(n)}(s)$.

\subsection{The physical meaning of the smearing method}
Here, we estimate  the size of the $n$-th  smeared operator and consider
the  physical meaning  of the  smearing  method.  To  obtain the  direct
correspondence with  the continuum theory, we adopt  the lattice Coulomb
gauge, which is defined so as to maximize
\begin{equation}
	R
\equiv
	\sum_{s}\sum_{k=1}^3{\rm  Re  Tr}U_k(s),
\label{lattice.coulomb.gauge}
\end{equation}
with respect to SU(3) gauge transformations.  In the continuum limit, we
recover the continuum  Coulomb gauge condition as
\begin{equation}
	\sum_{k=1}^3 \partial_k  A_k(x) = 0.
\label{continuum.coulomb.gauge}
\end{equation}
In  this gauge,  the  spatial link-variable  $U_k(s)$ becomes  maximally
continuous around 1, and can be safely expanded in $a$ as $U_k(s)\equiv
e^{iaA_k(s)}= 1+iaA_k(s)+O(a^2)$ near  the continuum limit.  We consider
the smearing with the  smearing parameter $\alpha$.  For convenience, we
define
\begin{equation}
	p    \equiv   {\alpha \over \alpha+4},
\Hs
	q \equiv {1 \over \alpha+4},
\end{equation}
which  satisfy   $p+4q=1$.   As  shown   in  \Fig{smearing.figure},  the
\mbox{$(n+1)$-th}   smeared   spatial   gluon  field   $A_k(s;n+1)$   is
iteratively described with the $n$-th smeared gluon field $A_k(s;n)$ as
\begin{widetext}
\begin{eqnarray}
\lefteqn{
	A_x(s;n+1)
}
\\\nonumber
&=&
\renewcommand{\arraystretch}{1.5}
	pA_x(s;n)
+
	q\left(\Tate\right.
	\Bs\begin{array}[t]{l}
		A_y(s;n)
		+ A_x(s+\hat y;n)
		- A_y(s+\hat x;n)
		-A_y(s-\hat y;n)
		+ A_x(s-\hat y;n)
		+ A_y(s+\hat x-\hat y;n)
	\\
		+A_z(s;n)
		+ A_x(s+\hat z;n)
		- A_z(s+\hat x;n)
		-A_z(s-\hat z;n)
		+ A_x(s-\hat z;n)
		+ A_z(s+\hat x-\hat z;n)
		\left.\Tate\right)
	\end{array}
\end{eqnarray}
\end{widetext}
in  the leading  order in  $a$.   The expressions  for $A_y(s;n+1)$  and
$A_z(s;n+1)$ are  similar.  After some rearrangements, we  obtain in the
leading order in $a$
\begin{eqnarray}
\lefteqn{ \partial_n A_x(s;n) }
\label{discrete.diffusion.eq}
\\\nonumber
&\equiv&
	A_x(s;n+1) - A_x(s;n)
\\\nonumber
&=&
\renewcommand{\arraystretch}{1.5}
	q\left\{\Tate\right.
	\Bs
	\begin{array}[t]{l}
		\partial_y A_x(s;n)-\partial_y A_x(s-\hat y;n)
	\\
		+\partial_z A_x(s;n)-\partial_z A_x(s-\hat z;n)
	\\
		-\partial_x A_y(s;n)+\partial_x A_y(s-\hat y;n)
	\\
		-\partial_x A_z(s;n)+\partial_x A_z(s-\hat z;n)
		\left.\Tate\right\}
	\end{array}
\\\nonumber
&=&
\renewcommand{\arraystretch}{1.5}
	q\left\{\Tate\right.
	\Bs
	\begin{array}[t]{l}\displaystyle
		\left(
			\partial_x^B \partial_x
			+ \partial_y^B \partial_y
			+ \partial_z^B \partial_z
		\right)
		A_x(s;n)
	\\\displaystyle
	-	\partial_x \sum_{k=1}^3 \partial^B_k A_k(s;n)
		\left.\Tate\right\}.
	\end{array}
\end{eqnarray}
Here, $\partial_k$  and $\partial_k^B$  denote the forward  and backward
difference operators along $k$-axis ($k=x,y,z$), respectively, which are
defined as
\begin{equation}
	\partial_k f(s)
\equiv
	f(s + \hat k) - f(s),
\Hs
	\partial_k^B f(x)
\equiv
	f(s) - f(s - \hat k),
\end{equation}
respectively. Here, $\hat k$ denotes the unit vector along the $k$-axis.
By applying  the continuum  approximation both with  respect to  $s$ and
$n$,   the   difference   operators   $\partial_n$,   $\partial_k$   and
$\partial_k^B$   are    then   replaced   by    the   derivatives   $a_n
\partial/\partial    \tilde   n$,    $a_s\partial/\partial    x_k$   and
$a_s\partial/\partial x_k$, respectively. Here,  $\tilde n \equiv n a_n$
is  the   continuum  parameter  corresponding  to  $n$,   and  $a_n  \in
\mathbb{R}$ is formally  introduced as a small constant  interval in the
$n$-direction, although  the final  result does not  depend on  $a_n$ at
all.  In the continuum Coulomb gauge using \Eq{continuum.coulomb.gauge},
\Eq{discrete.diffusion.eq} reduces to the diffusion equation
\begin{equation}
	a_n{\partial \over \partial \tilde n}
	A_i(\vec x;n)
=
	D\Laplacian A_i(\vec x;n)
\Hs
	{\rm for}\Hs i=x,y,z,
\label{kakusan.houtei.siki}
\end{equation}
where  $\displaystyle D\equiv  {a_s^2\over  \alpha +  4}$  works as  the
diffusion parameter.  By solving this equation, the $n$-th smeared gauge
field $A_i(\vec  x,n)$ can be expressed  by a linear  combination of the
original gluon field $A_i(\vec x)$ as
\begin{eqnarray}
	A_i(\vec x; n)
&=&
	\int d^3 y
	K(\vec x - \vec y;n)
	A_i(\vec y)
\label{gaussdistribution}
\\\nonumber
	K(\vec x; n)
&\equiv&
	{1\over \pi^{3/2} \rho^3}
	\exp\left( -{\vec x^2 \over \rho^2}\right),
\end{eqnarray}
where the Gaussian extension $\rho$ is defined as 
\begin{equation}
	\rho
\equiv
	2\sqrt{Dn }
=
	2a_s \sqrt{ n \over \alpha + 4 }.
\label{size}
\end{equation}
In this  way, the $n$-th  smeared gluon field physically  corresponds to
the Gaussian-distributed  operator of  the original gluon  field.  After
the linearization  of the gluon  field, the $n$-th smeared  plaquette is
also expressed as the Gaussian-distributed operator.  Here, the Gaussian
extension $\rho$ can be regarded  as a characteristic size of the $n$-th
smeared operator.

The expression of $\rho$ in \Eq{size} explains the physical roles of the
two parameters, $\alpha$ and $n$.  The smearing parameter $\alpha$ plays
the role of controlling the speed of the smearing, while, for each fixed
$\alpha$,  $n$ plays  the  role of  extending  the size  of the  smeared
operator. Note that,  for the larger $\alpha$, the  speed of smearing is
the slower.

The smearing method was originally  introduced to carry out the accurate
mass  measurement  by maximizing  the  ground-state overlap  \cite{ape}.
However,  we will  use  it to  give  a rough  estimate  of the  physical
glueball  size.   In  fact,  once  we  find the  pair  of  the  smearing
parameters  $n$ and  $\alpha$, which  achieves the  maximum ground-state
overlap, the  physical size  of the glueball  is roughly  estimated with
\Eq{size}.

With  \Eq{gaussdistribution},  we  try  to express  the  $n$-th  smeared
$0^{++}$  glueball  operator  in  terms  of  the  original  gluon  field
$A_{\mu}^a(x)$.  Near  the continuum limit,  the zero-momentum projected
$0^{++}$ glueball  operator $\Phi_{\rm GB}(0^{++};n)$  consisting of the
smeared link-variable $U^{(n)}_i(s)$ is expressed as
\begin{equation}
	\Phi_{\rm GB}(0^{++};n) =
	\frac1{V}
	\int d^3 x
	G_{ij}^a(\vec x,t;n) G_{ij}^a(\vec x,t;n),
\end{equation}
where $G_{ij}^a(\vec  x,t;n)$ is the field strength  associated with the
smeared gluon field  $A_i^{a}(\vec x,t; n)$.  Here, $t$  can be regarded
as a  fixed parameter in this  argument, and will  be omitted hereafter.
By   inserting   \Eq{gaussdistribution},   $\Phi_{\rm  GB}(0^{++})$   is
expressed with the original gauge field $A_i^a(\vec x)$ as
\begin{widetext}
\begin{eqnarray}
	\Phi_{\rm GB}(0^{++};n)
&=&
	\frac1{V}
	\int d^3 x
	\int d^3 y d^3 z
	\renewcommand{\arraystretch}{1.5}
	\begin{array}[t]{l}\displaystyle
	\left(
		{\partial\over \partial x_i}
		K(\vec x - \vec y; n) A_j(\vec y)
	-
		{\partial \over \partial x_j}
		K(\vec x - \vec y; n) A_i(\vec y)
	\right)
	\\\displaystyle
	\times
	\left(
		{\partial\over \partial x_i}
		K(\vec x - \vec z; n) A_j(\vec z)
	-
		{\partial \over \partial x_j}
		K(\vec x - \vec z; n) A_i(\vec z)
	\right)
+	\cdots
	\end{array}
\label{interpolating.field}
\\\nonumber
&=&
	{2 \over V}
	\int {d^3 y d^3 z\over \pi^{3/2} \rho^3}
	\exp\left(
		-{(\vec y - \vec z)^2 \over 2\rho^2}
	\right)
	{\partial \over \partial y_j}
	A_i(\vec y)
	{\partial \over \partial z_j}
	A_i(\vec z)
+	\cdots
\\\nonumber
&=&
	{2 \over V}
	\int {d^3 y d^3 z\over \pi^{3/2} \rho^5}
	\left(
		3-\frac{(\vec y-\vec z)^2}{\rho^2}
	\right)
        \exp\left(
		-{(\vec y - \vec z)^2 \over 2\rho^2}
	\right)
	A_i(\vec y)
	A_i(\vec z)
+	\cdots.
\end{eqnarray}
\end{widetext}
Here, ``$\cdots$'' represents the terms which are more than quadratic in
the  gluon field,  and are  dropped  off.  In  this derivation,  partial
integrations and the Coulomb gauge  condition have been used.  We expect
that $\Phi_{\rm GB}(0^{++}, n)$ is  associated with the best pair of the
smearing  parameters  $n$  and  $\alpha$  and works  as  an  approximate
``creation''  operator  of  the  $0^{++}$  glueball,  which  creates  an
approximate single glueball state from the vacuum.


\subsection{The relation to Bethe-Salpeter amplitude}
In the  previous subsection, we proposed  one possible method  to give a
rough estimate of the glueball size. In this subsection, we consider the
relation  between  our estimate  of  the  glueball  size and  the  other
estimate from  the Bethe-Salpeter (BS)  amplitude, which was  adopted in
Refs.\cite{forcrand}.

The size of the glueball is actually a nontrivial quantity. Although the
charge  radius of  the glueball  can be  formally defined,  its electric
charges are  carried by the quarks  and anti-quarks, which  play only of
secondary roles in describing the glueball state in the idealized limit,
since the glueball  by its nature does not  contain any valence contents
of (anti-)quarks.
%
%

In  order to  investigate the  size of  the glueball,  the  BS amplitude
provides a convenient tool. The BS amplitude $\Psi(\vec  x, \vec y)$ is
expressed as
\begin{equation}
	\Psi_{\rm BS}(\vec y, \vec z)
\equiv
	\langle 0 |
		A_i(\vec y) A_i(\vec z)
	| {\rm G}(\vec P =\vec 0) \rangle,
\end{equation}
which  can  be obtained  as  a  solution of  the  BS  equation.  In  the
non-relativistic limit,  it is expected  to reduce to the  glueball wave
function  in  the  first-quantized  picture.   This  property  has  been
exploited    to    estimate    the    size   of    the    glueball    in
Ref.\cite{forcrand}.

The   connection  between   these   two  estimates   is  provided   with
\Eq{interpolating.field}  in the  following  way.
We consider $\Phi_{\rm{GB}}(0^{++};n)$ associated with the best smearing
with the  Gaussian extension $\rho$.  Since the  ground-state overlap is
maximized, the $0^{++}$-glueball state can be approximated as
\begin{widetext}
\begin{eqnarray}
|{\rm G}(\vec P=\vec 0)\rangle 
&\simeq&
	\Phi_{\rm{GB}}(0^{++};n)|0\rangle
\label{overlap}
\\\nonumber
& = &
	{2\over V}
	\int {d^3 y d^3 z\over \pi^{3/2}\rho^5}
         \left(3-\frac{(\vec y-\vec z)^2}{\rho^2}\right)
	\exp\left( - {(\vec y - \vec z)^2 \over 2\rho^2}\right)
	| A_i(\vec y) A_i(\vec z) \rangle
+	\cdots
\\\nonumber
&=&
	{2 \over \pi^{3/2}\rho^5}
	\int d^3 r
         \left(3-\frac{\vec r^2}{\rho^2}\right)
	\exp\left( -{\vec r^2 \over 2\rho^2} \right) 
	|A_i(\vec r) A_i(\vec 0) \rangle
+	\cdots, 
\end{eqnarray}
\end{widetext}
where we exploit the  translational invariance and partial integrations.
Similar to \Eq{interpolating.field}, the  states in \Eq{overlap} such as
$|{\rm G}(\vec P=\vec 0)\rangle$ and $| A_i(\vec y) A_i(\vec z) \rangle$
are  obtained with  the gauge  configuration generated  in  lattice QCD.
Hence,   this  relation   would  provide   a  kind   of  nonperturbative
construction of the glueball state in terms of the gluon field.

In  the  constituent  gluon   picture  \cite{hou},  the  BS  amplitude
associated with \Eq{overlap} reduces to the following expression as
\begin{equation}
	\Psi_{\rm BS}(\vec r, \vec 0)
\simeq
	{2 \over \pi^{3/2}\rho^5}
         \left(3-\frac{\vec r^2}{\rho^2}\right)
	\exp\left( -{\vec r^2 \over 2\rho^2} \right),
\end{equation}
in the heavy effective gluon mass limit.

\section{Anisotropic Lattice QCD}
\label{section.lattice-setup}
\subsection{Lattice Parameter Set}
We   use  the  SU(3)   plaquette  action   on  an   anisotropic  lattice
\begin{equation}
\renewcommand{\arraystretch}{1.5}
	S_{\rm G}
=
\Bs
\begin{array}[t]{l}\displaystyle
	{\beta_{\rm lat}\over N_c}
	{1\over \gamma_{\rm G}}
	\sum_{s, i<j \le 3}
	\mbox{Re}\mbox{Tr}
	\left(\Tate 1 - P_{ij}(s) \right)
\\\displaystyle
	+ 
	{\beta_{\rm lat}\over N_c}
	\gamma_{\rm G}
	\sum_{s, i \le 3}
	\mbox{Re}\mbox{Tr}
	\left(\Tate 1 - P_{i4}(s) \right),
\end{array}
\end{equation}
where $P_{\mu\nu}(s)\in  {\rm SU(3)}$ denotes the  plaquette operator in
the $\mu$-$\nu$-plane.   The lattice  parameter and the  bare anisotropy
parameter are  fixed as  $\beta_{\rm lat} \equiv  {2N_c}/{g^2}=6.25$ and
$\gamma_{\rm  G}  =  3.2552$,  respectively,  so  as  to  reproduce  the
renormalized  anisotropy $\xi  \equiv a_s/a_t=4$  \cite{klassen}.
We adopt  the pseudo-heat-bath algorithm  for the update of  the gauge
field configurations in the Monte Carlo calculation.

\begin{figure}
\includegraphics[width=\figwidth]{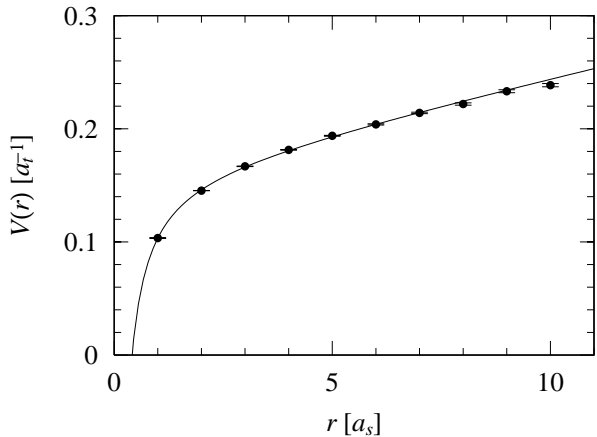}
\caption{  The inter-quark potential  plotted against  the inter-quark
separation  $r$. The  solid circles  denote  the on-axis  data of  the
inter-quark potential.   The solid line  represents the result  of the
best-fit analysis with \Eq{potential.v.r}.}
\label{potential}
\end{figure}

To  measure  the  glueball   correlators,  we  perform  the  numerical
calculations  on the  lattice  of  the sizes  $20^3  \times N_t$  with
various $N_t$ as  $N_t = 24, 26, 28,  30, 33, 34, 35, 36,  37, 38, 40,
43, 45, 50,  72$, which correspond to the  various temperatures listed
in \Table{table-S}.  
For all  temperatures, we pick  up gauge field configurations  every 100
sweeps for measurements, after skipping  more than 20,000 sweeps for the
thermalization.  The numbers $N_{\rm conf}$ of gauge configurations used
in our  calculations are summarized  in \Table{table-S}.

We divide the data into bins  of the size $100$ to reduce the possibly
strong  auto-correlations.   However,  the auto-correlations  are  not
observed so strong  in this simulation with the  100 sweep separation,
since the separation by 100 sweeps seems already rather large.  
\begin{em}
Unless otherwise  stated, the statistical errors are  understood to be
estimated  with the  jackknife analysis  by regarding  each bin  as an
independent data point throughout in this paper.
\end{em}

As  for  the  smearing  parameters,  which play  the  important  role  in
extracting the  ground-state contribution, we  have checked that  one of
the best sets is provided as
\begin{equation}
	\alpha=2.1, \Hs \Nsmear=40,
\label{suitable.parameter.set}
\end{equation}
in the present lattice calculation  for the lowest $0^{++}$ and $2^{++}$
glueballs.
As we shall see later,  the temporal correlator $G(t)/G(0)$ and the pole
mass $m_{\rm  G}$ are almost insensitive  to a particular  choice of the
smearing   number  $\Nsmear$,  as   far  as   $\Nsmear  \sim   40$  with
$\alpha=2.1$.
In  fact,  in the  rather  wide  range as  $30  \alt  \Nsmear \alt  50$,
$G(t)/G(0)$ and  $m_{\rm G}$ are almost unchanged,  and the ground-state
overlap $C$ is kept almost maximized.  (See \Fig{against-nsmear}.)
%
\begin{em}
Unless    otherwise    stated,   we    adopt    the   smearing    with
\Eq{suitable.parameter.set} in this paper.
\end{em}
Only  for the  accurate measurement,  we seek  for the  best  value of
$\Nsmear$ with $\alpha=2.1$ fixed, although the results are almost the
same as those with $\Nsmear=40$.

In  this paper,  we only  consider  the case  with $\alpha=2.1$.   The
essential point  of the  smearing method is  to construct  an operator
which has  the same size with  the physical glueball size  in order to
improve the ground-state overlap  \cite{ape}.  Hence, a different pair
of the smearing parameters  $(\alpha',\Nsmear')$ would work as good as
the original $(\alpha,\Nsmear)$ does, provided that these two pairs of
the smearing parameters correspond to the same operator size.
However,  we have  to  take  into account  the  following points  when
picking up the value of $\alpha$.
On  the one hand,  the speed  of the  smearing is  the faster  for the
smaller $\alpha$, which, however, results in the coarser discretization
in $n$ in \Eq{kakusan.houtei.siki}.   Hence, $\alpha$ has to be enough
large so as  to construct the ground-state glueball  operator. We have
checked that $\alpha>2.0$ seems to work well.
On the other hand, if we adopt the larger value of $\alpha$, the speed
of the smearing  becomes the slower. Hence, we have  to scan the wider
range of  $\Nsmear$ to maximize  the overlap. A practical  solution is
provided by  $\alpha=2.1$, since  it is one  of the  smallest possible
values of  $\alpha$, which,  in the same  time, corresponds  to enough
fine discretization in $n$ in \Eq{kakusan.houtei.siki}.

\subsection{Determination of lattice spacings}
In order to evaluate the  spatial and temporal lattice spacings, $a_s$
and $a_t$, we perform a separate Monte Carlo simulation on the lattice
of the  size $20^3 \times  80$.  Here, we  skip 10,000 sweeps  for the
thermalization, and pick up  gauge configurations every 500 sweeps for
measurement of the static  inter-quark potential. The statistical data
are divided into bins  of the size 2, where each bin  is thought of as
an  independent  data  point.   In  \Fig{potential},  the  inter-quark
potential is shown.  We  adopt the parameterization of the inter-quark
potential as
\begin{equation}
	V(r)
=
	C - {A \over r} + \sigma r,
\label{potential.v.r}
\end{equation}
where $C$,  $A$, $\sigma$ denote the offset,  the Coulomb coefficient,
and  the string tension,  respectively.  By  fitting the  on-axis data
with   $V(r)$,   we   obtain  $C=0.16217(53)$,   $A=0.06758(42)$   and
$\sigma=0.00883(12)$.  Hence, by assuming $\sqrt{\sigma}=440$ MeV, the
lattice spacings are obtained as
\begin{equation}
	a_s^{-1} = 2.341(16) \mbox{GeV},
\Hs
	a_t^{-1} = 9.365(66) \mbox{GeV},
\end{equation}
namely, $a_s \simeq  0.084$ fm and $a_t \simeq  0.021$ fm.  Hence, the
spatial lattice size of $20$  corresponds to $1.68$ fm in the physical
unit.

\subsection{The critical temperature}
Here, for completeness, we give  an estimate of the critical temperature
$T_c$ on our anisotropic lattice, although the value of $T_c \simeq 260$
MeV  with $\sqrt{\sigma}\simeq  420$ MeV  is rather  established  on the
isotropic lattice \cite{karsch1}.

\begin{figure}
\includegraphics[width=\figwidth]{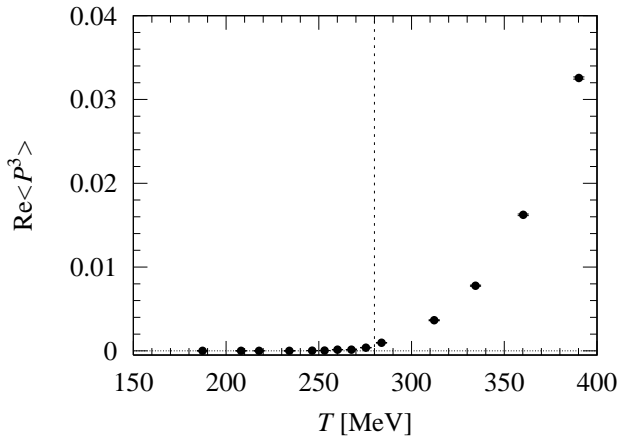}
\caption{The cubed Polyakov loop $\mbox{Re} \langle P^3 \rangle$ plotted
against  temperature.  The  vertical dotted  line denotes  the estimated
critical temperature, $T_c  \simeq$ 280 MeV.  The error  bars are hidden
inside the symbols.}
\label{fig.polyakov.loop}
\end{figure}
To  estimate the critical  temperature $T_c$,  we consider  the Polyakov
loop, which is defined as
\begin{eqnarray}
	P
&\equiv&
\label{polyakov.loop}
	\frac1{V}
	\int d^3 x
	{\rm Tr}\left(
		T \exp i\int_0^{\beta} dt A_{0}(\vec x,t)
	\right)
\\\nonumber
&=&
	\frac1{V}
	\int d^3 x
	{\rm  Tr}
	\left( \Tate
		U_4(\vec  x,0)  \cdots  U_4(\vec x,N_t-1)
	\right),
\end{eqnarray}
where  $V$ denotes the  volume of  the space  and $T$  the time-ordering
operation.    As   is   well-known,   its  thermal   expectation   value
$\displaystyle \langle  P\rangle = \frac1{\Nconfig}\sum_{k=1}^{\Nconfig}
P_k$, averaged over lots of  gauge configurations, is an order parameter
of quark confinement associated  with the center $\mathbb{Z}_3$ symmetry
\cite{montvey,rothe}, where $P_k$ denotes the value of the Polyakov loop
$P$ with the $k$-th configuration.
However, to  be strict, it is  only in the thermodynamic  limit with the
infinite volume that \Eq{polyakov.loop}  can work as the order parameter
to estimate $T_c$. Note that, since  the size of the lattice is limited,
tunnelings  from  one  vacuum  to  another  are  unavoidable  for  large
simulation time even in the $\mathbb{Z}_3$ broken phase.
As  a consequence,  due to  the  associated cancellation  of the  center
$\mathbb{Z}_3$-phase factor, $\langle  P\rangle$ finally vanishes on the
finite lattice, which raises  a technical difficulty in estimating $T_c$
from $\displaystyle \langle P\rangle$.   To avoid this, instead of using
the total  expectation value itself,  one often analyzes  the scattering
plot, a plot of $P_k$ for each configuration labeled by $k$.

Here, for convenience,  we analyze the thermal expectation value
of the cubed Polyakov loop as
\begin{equation}
	\left\langle P^3\right\rangle
=
	\frac1{\Nconfig}
	\sum_{k=1}^{\Nconfig}
	(P_k)^3,
\end{equation}
which provides  the equivalent result  for the estimate of  the critical
temperature  $T_c$ with  the  scattering plot.   We  note that  $\langle
P^3\rangle$ is  not the  order parameter of  the confinement phase  in a
strict  sense, since  it is  invariant under  the  center $\mathbb{Z}_3$
transformation.   In general,  in the  symmetric phase,  the expectation
value of  the order  parameter vanishes due  to the cancellation  of the
phase factor  associated with the symmetry  transformations, which never
happens in the broken symmetry phase.
However, in  the specific case of  $\langle P\rangle$, it  is known from
the analysis  of the  scattering plot  that the norm  of $P_k$  for each
configuration  is already  quite small  in the  symmetric phase,  due to
which  its expectation  value  $\langle P\rangle$  almost vanishes  even
without involving  the cancellation  of the center  $\mathbb{Z}_3$ phase
factor.  In  contrast, $P_k$ takes a manifestly  non-vanishing value for
each configuration  in the broken symmetry  phase, which can  be used to
identify the critical temperature $T_c$.

Such a tendency is quantitatively  described with the cubed Polyakov loop
$\langle P^3  \rangle$.  In fact, in  the plot of  $\langle P^3 \rangle$
versus temperature $T$, the  critical temperature $T_c$ is characterized
as the point where it begins to have a non-vanishing value.

In  \Fig{fig.polyakov.loop}, $\langle  P^3 \rangle$  is  plotted against
temperature.   By taking  into  account the  finite  size effects  which
smoothen the  phase transition, we estimate the  critical temperature as
$T_c \simeq 260\mbox{--}280$ MeV,  which is consistent with the previous
studies in Ref.~\cite{karsch1}, i.e., $T_c/\sqrt{\sigma} = 0.629(3)$.

\begin{figure}
\includegraphics[width=\figwidth]{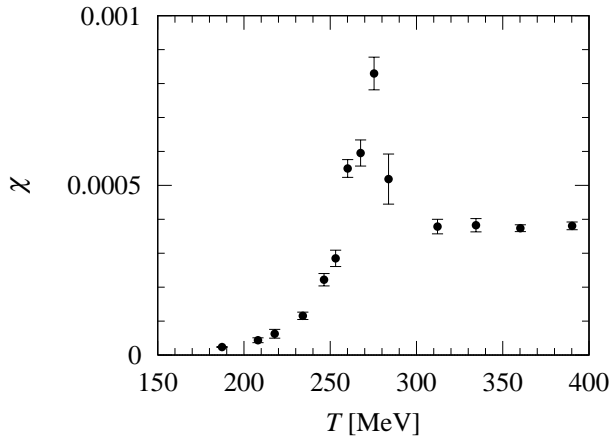}
\caption{  The  susceptibility  $\chi$  of the  Polyakov-loop  plotted
against the  temperature.  There is  a sharp peak around  $T=280$ MeV,
which indicates the critical temperature $T_c$ of the phase transition
in quenched SU(3) QCD.}
\label{susceptibility}
\end{figure}

For more  accurate determination of  the critical temperature  $T_c$, we
investigate the Polyakov-loop susceptibility  $\chi$, which is defined as
\cite{iwasaki}
\begin{eqnarray}
	\chi
&\equiv&
	\langle \Omega^2\rangle - \langle \Omega \rangle^2
\\\nonumber
	\Omega
&\equiv&
\renewcommand{\arraystretch}{1.5}
	\left\{
	\begin{array}{ll}\displaystyle
		\mbox{Re} P \exp\left(-{2 \pi i/ 3}\right),
		& \mbox{arg} P \in [\pi/3,\pi),
	\\\displaystyle
		\mbox{Re} P,
		& \mbox{arg} P \in [-\pi/3,\pi/3),
	\\\displaystyle
		\mbox{Re} P \exp\left( {2\pi i/ 3} \right),
		& \mbox{arg} P \in [-\pi,-\pi/3).
	\end{array}
	\right.
\end{eqnarray}
In \Fig{susceptibility},  the Polyakov loop  susceptibility is plotted
against   temperature.    To   obtain   $\chi$,  about   3,000   gauge
configurations  are used  at each  temperature near  $T_c$,  which are
divided into bins of the size 50. The statistical errors are estimated
with the jackknife  analysis. We observe that a  sharp peak is located
between $T=275$ MeV and $T=284$  MeV.  Hence, we identify the critical
temperature of our lattice as $T_c \simeq 280$ MeV.

\section{Numerical Results ---Temporal Correlations and Pole-Masses}
\label{section.pole-mass-measurement}
In this section, we construct the temporal glueball correlator $G(t)$ at
finite temperature, and perform the pole-mass measurement.
We  adopt the  procedure  used in  \cite{taro}  as the  standard one  to
extract the  pole-mass $m_{\rm G}(T)$  of the thermal  glueball from the
temporal  correlator $G(t)$  at temperature  $T$ by  using  the best-fit
analysis with the fit function of single-cosh type considered later.
Here, we  regard each glueball as  a quasi-particle and  assume that the
thermal width $\Gamma$ of the  ground-state peak is enough narrow in the
spectral function $\rho(\omega)$.
From  the viewpoint  of the  spectral  representation \Eq{spectral.rep},
such a narrowness of the peak  is necessary to identify the pole-mass as
a  definite object  and also  to justify  this standard  procedure  in a
strict sense.  (In \Sect{section.spectral-function}, we will analyze our
lattice  QCD data  without  assuming the  narrowness  of the  peak as  a
straightforward extension to the current analysis.)

Suppose  that the thermal  width is  sufficiently narrow,  i.e., $\Gamma
\simeq  0$.  Then, the  spectral  function  $\rho(\omega)$ receives  the
contribution from the ground-state in the following manner:
\begin{equation}
	\rho(\omega)
=
	2\pi A
	\left(\Tate
		\delta(\omega - m_{\rm G}) - \delta(\omega + m_{\rm G})
	\right)
	+ \cdots,
\label{narrow.width}
\end{equation}
where ``$\cdots$'' represents the contributions from excited-states, and
$A\equiv A(T)$ and  $m_{\rm G}\equiv m_{\rm G}(T)$ are  the strength and
the  pole-mass of the  glueball peak  at temperature  $T$, respectively.
Note that the appearance of the second delta function is due to the fact
that  the spectral  function $\rho(\omega)$  is odd  in  $\omega$, which
reflects the bosonic nature of the glueball.
Corresponding to  \Eq{narrow.width}, by  inserting it into  the spectral
representation \Eq{spectral.rep},  the ground state  contribution to the
temporal correlator $G(t)$ is expressed as
\begin{equation}
	G(t)
=
	{A \over 1 - e^{-\beta m_{\rm G}}}
	\left(\Tate
		e^{-tm_{\rm G}} + e^{-(\beta - t)m_{\rm G}}
	\right)
	+ \cdots.
\label{g-t.narrow.width}
\end{equation}
Hence, after the ground state contribution is sufficiently enhanced, the
appropriate  fit  function  for  $G(t)/G(0)$  is  the  following  single
hyperbolic cosine as
\begin{equation}
	g(t)
\equiv
	C\left(\Tate
		e^{-t m_{\rm G}}
	+	e^{-(\beta - t) m_{\rm G}}
	\right),
\label{single.cosh}
\end{equation}
where $C$ and $m_{\rm G}$ are  understood as the fit parameters. We will
refer to  $C$ as the  ground-state overlap, and  to the fit  function of
$g(t)$ as the fit function of single-cosh type for simplicity.

For the accurate measurement, it is required that $G(t)$ is dominated by
the  ground-state  contribution  in  the fit-range.   (We  consider  the
determination of  the fit range in  the next subsection  with an explicit
example.)  In this case,  $g(t)$ can properly represent the ground-state
contribution in  \Eq{g-t.narrow.width}, and $C$  and $m_{\rm G}$  can be
extracted  accurately.    In  the  actual   calculations,  the  complete
elimination of all the remaining  contributions of the excited states is
impossible. Since each hyperbolic cosine component contributes to $G(t)$
positively as is seen  in the spectral representation \Eq{spectral.rep},
we always have the following inequalities as
\begin{equation}
	g(0)
\le
	1,
\mbox{  and  }
	C
=
	{g(0) \over 1 + e^{-\beta m_{\rm G}}}
\le
	{1\over 1 + e^{-\beta m_{\rm G}}}.
\end{equation}
Here,  $g(0)$ and  $C$  take their  maxima,  if and  only  if $G(t)$  is
completely  dominated  by the  ground-state  contribution  in the  whole
region  $0 \le  t  \le \beta$.   In  this case,  we  have the  following
equalities instead:
\begin{equation}
	g(0)
=
	1,
\mbox{  and  }
	C
=
	\frac1{1 + e^{-\beta m_{\rm G}}}.
\end{equation}
We thus see  that $g(0) \simeq 1$ and $C \simeq  1/(1 + e^{-\beta m_{\rm
G}}) \simeq  1$ can both  work as a  criterion to determine  whether the
ground-state enhancement is sufficient or  not.  We will refer to $g(0)$
as the normalized ground-state overlap.
Note  that $C$  differs  from $g(0)$  only  through the  factor $1/(1  +
e^{-\beta  m_{\rm  G}})$, which  is  independent  of  both the  smearing
parameters $\alpha$ and $\Nsmear$.
In  this  section, we  examine  $C$  to  determine if  the  ground-state
enhancement  is sufficient  or not.   We prefer  $C$ rather  than $g(0)$
because of the  following reasons.  (i) The ground-state  overlap $C$ is
directly obtained  from the  best-fit analysis with  \Eq{single.cosh} as
one of the best-fit parameters.  (ii) In our previous work \cite{ishii},
$C$ has been  used for the criterion.  However, $g(0)  \simeq 1$ has the
advantage  over $C\simeq  1$ in  \Sect{section.spectral-function}, where
the counterpart of  $C$ cannot be properly defined,  whereas the counter
part of $g(0)$ can.  At any rate, as is seen in Tables~\ref{table-S} and
\ref{table-T}, they  are not so  different from each other  in numerical
value, especially below $T_c$.

\subsection{Thermal \boldmath{$0^{++}$} glueball}
\begin{figure}
\leftline{(a)}
\includegraphics[width=\figwidth]{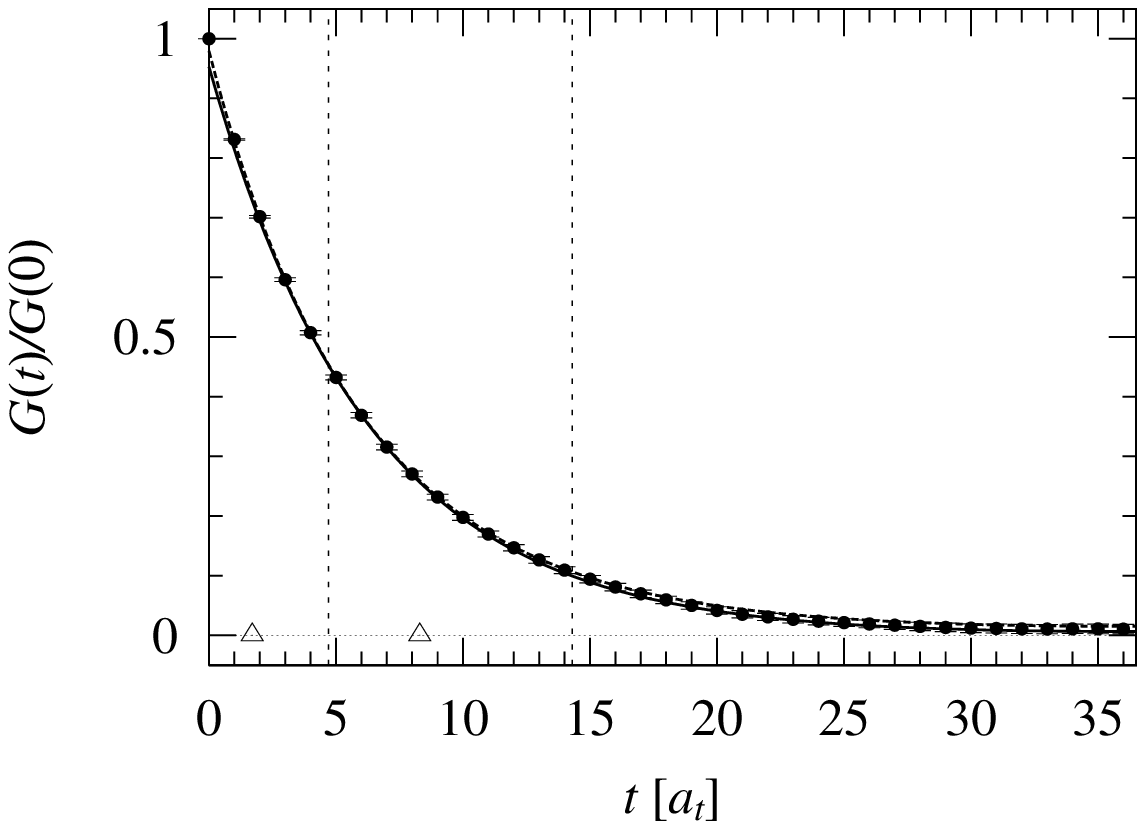}
\leftline{(b)}
\includegraphics[width=\figwidth]{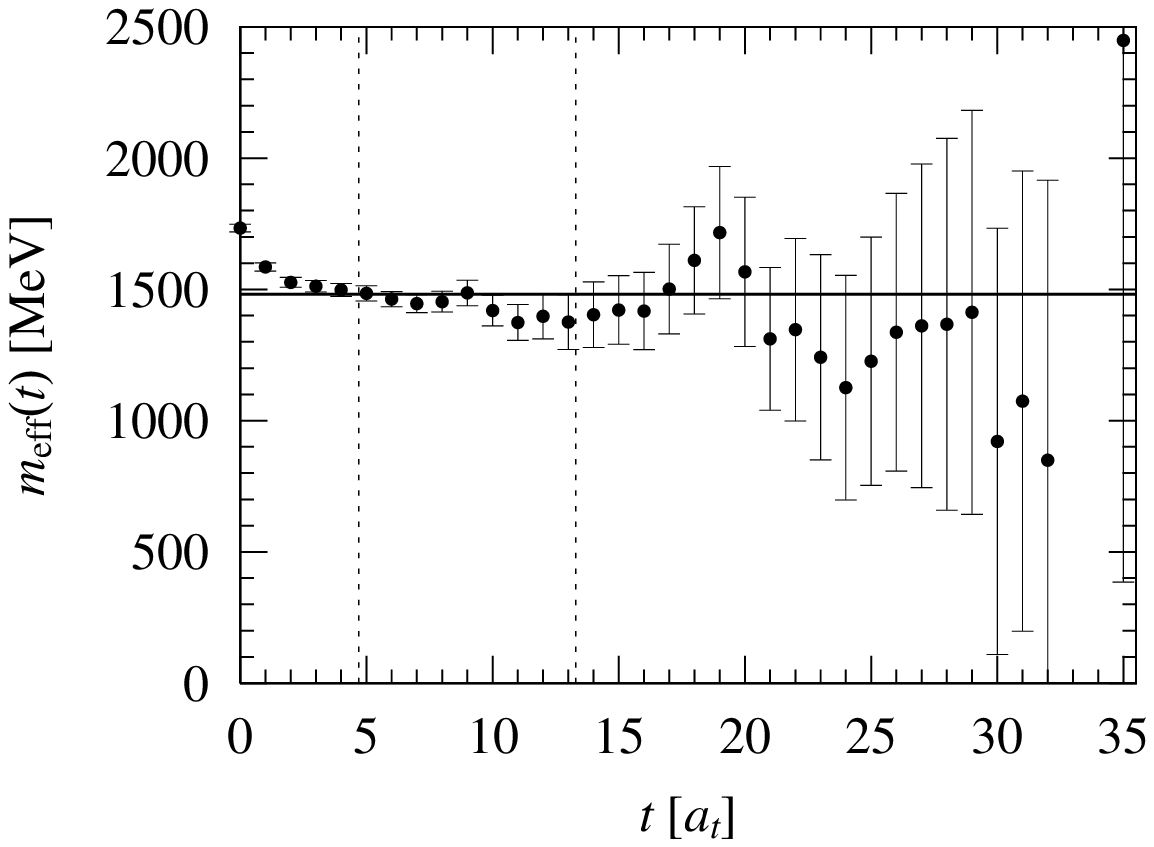}
\caption{  (a) The  temporal  correlator $G(t)/G(0)$  of the  $0^{++}$
glueball at the low  temperature ($T=130$ MeV).  (b) The corresponding
cosh-type  effective-mass  plot.  In  both  figures,  the solid  lines
represent the  best single  hyperbolic-cosine fitting in  the plateau,
which is indicated  by the vertical dotted lines.  The dashed curve in
(a) denotes the  result of the best-fit analysis  of Breit-Wigner type
performed in the  interval indicated by the two  open triangles.  (See
\Sect{section.spectral-function}   and   \Eq{single.lorentz.fit}   for
detail.)  }
\label{green.72}
\setcounter{subfigure}{0}
\refstepcounter{subfigure}
\label{correlator.72}
\refstepcounter{subfigure}
\label{effmass.72}
\end{figure}
\begin{figure}
\leftline{(a)}
\includegraphics[width=\figwidth]{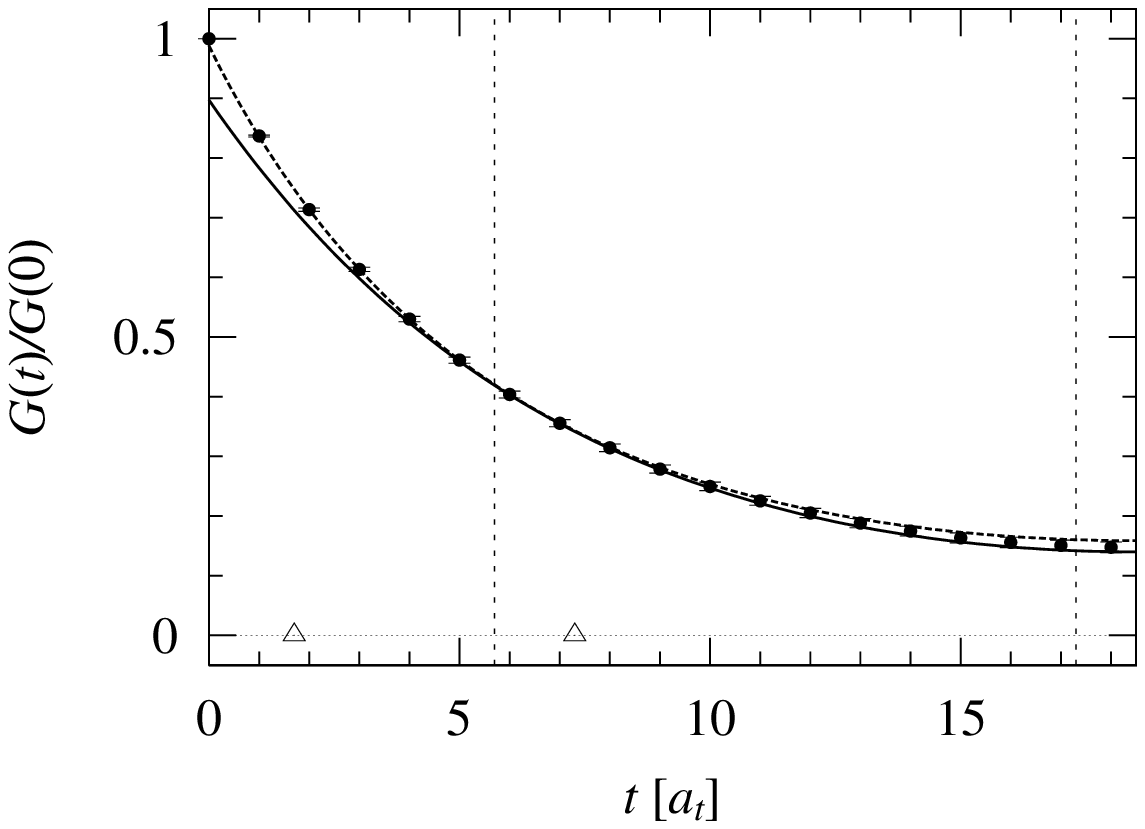}
\leftline{(b)}
\includegraphics[width=\figwidth]{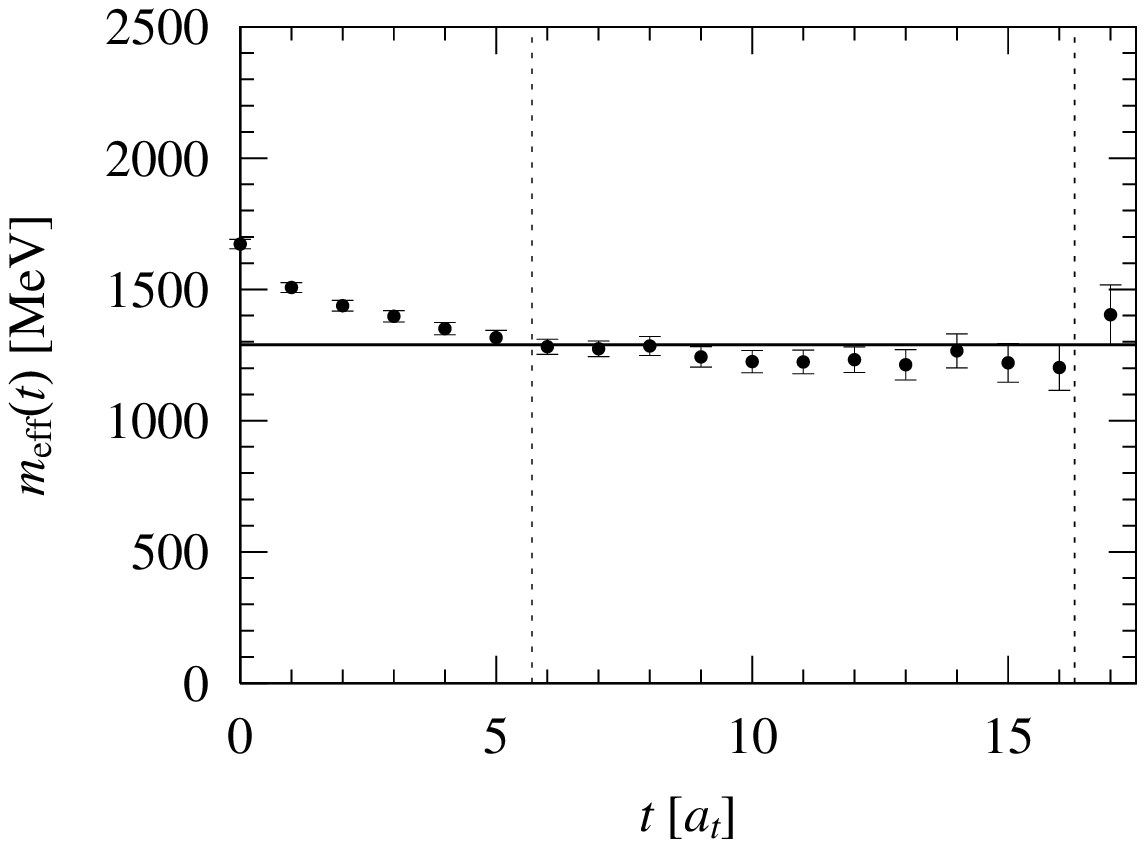}
\caption{  (a) The  temporal  correlator $G(t)/G(0)$  of the  $0^{++}$
glueball at the high temperature ($T=253$ MeV).  (b) The corresponding
cosh-type effective-mass plot.  The meaning of the solid lines in both
figures   and  the  dashed   curve  in   (a)  are   the  same   as  in
\Fig{green.72}.}
\label{green.37}
\setcounter{subfigure}{0}
\refstepcounter{subfigure}
\label{correlator.37}
\refstepcounter{subfigure}
\label{effmass.37}
\end{figure}
In \Fig{correlator.72}, we show the $0^{++}$ glueball correlator for the
low  temperature  case  $T=130$  MeV  after  the  suitable  smearing  as
\Eq{suitable.parameter.set}.
We   perform   the   best-fit   analysis   of   the   single-cosh   type
\Eq{single.cosh} with  the two fit parameters,  the ground-state overlap
$C$ and the  pole-mass $m_{\rm G}$ of the  lowest thermal glueball.  The
solid line denotes  the result of this best-fit  analysis. The fit-range
is indicated by the vertical dotted lines.

\subsubsection{The fit range}
We  consider  the  determination  of  the  fit-range.   Even  after  the
ground-state enhancement  is performed, the complete  elimination of all
the contributions of the excited states is impossible, especially in the
neighborhood of $t\sim 0$ (mod $\beta$).
It  follows that  the  equality \Eq{g-t.narrow.width}  holds  only in  a
limited interval, which does not  include the neighborhood of $t\sim 0$.
Hence, for  the accurate  measurement of the  pole-mass $m_{\rm  G}$, we
need to find the appropriate fit-range, where contributions from excited
states almost die out.
To this  end, we examine  the effective-mass plot.  The  effective mass
$m_{\rm eff}(t)$ is defined as the solution to the following equation as
\begin{equation}
	G(t+1)/G(t)
=
	\frac{
		\cosh\left(\Tate (t+1 - N_t/2) a_t m_{\rm eff}(t) \right)
	}{
		\cosh\left(\Tate (t - N_t/2) a_t m_{\rm eff}(t) \right)
	},
\end{equation}
for a  given $G(t+1)/G(t)$ at  each fixed $t$.  In  \Fig{effmass.72}, we
plot  the effective mass  $m_{\rm eff}(t)$  against $t$  associated with
\Fig{correlator.72}.
We see  that there  appears a plateau,  a region where  $m_{\rm eff}(t)$
takes  almost a constant  value.  In  this region,  it is  expected that
$G(t)$  consists of  nearly  a single  spectral  component, and,  hence,
$g(t)$   can  properly  represent   the  ground-state   contribution  in
\Eq{g-t.narrow.width}.
Note that, if  more than a single state  contribute with their different
masses, a nontrivial $t$-dependence will appear in $m_{\rm eff}(t)$.
In the zero-temperature case, the  plateau in the effective-mass plot is
widely used as a candidate for the fit range in the best-fit analysis of
single-cosh type \Eq{single.cosh}.
Here, the role  of the smearing method is  to suppress the excited-state
contributions as much  as possible, so that the  plateau can be obtained
with the smallest possible $t$.

In \Fig{green.72}~(b), the plateau is indicated by the vertical dotted
lines,  which  is  used  to  determine the  fit-range.   In  this  low
temperature  case  $T=130$  MeV,  the  best-fit analysis  shows  $C  =
0.953(7)$ and  $m_{\rm G}=1482(25)$  MeV, which seems  consistent with
$m_{\rm G} \simeq 1500\mbox{--}1700$ MeV at $T \simeq 0$ MeV.  We note
that the  ground-state overlap $C=0.95$ becomes enough  large owing to
the suitable smearing.

In \Fig{correlator.37}, we show the $0^{++}$ glueball correlator for the
high  temperature  case $T  =  253$ MeV  ($<  T_c$)  after the  suitable
smearing  as \Eq{suitable.parameter.set}.   The solid  line  denotes the
result of the  best-fit analysis of the single-cosh  type. The fit-range
is indicated by the vertical  dotted lines, which is determined from the
plateau in  the corresponding effective-mass  plot \Fig{effmass.37}.  We
obtain  $C =  0.89(1)$ and  $m_{\rm G}  = 1289(26)$  MeV.  Owing  to the
suitable  smearing,  contributions  from  the  excited  state  are  seen
suppressed ($C\simeq 1$).

\begin{figure}
\leftline{(a)}
\includegraphics[width=\figwidth]{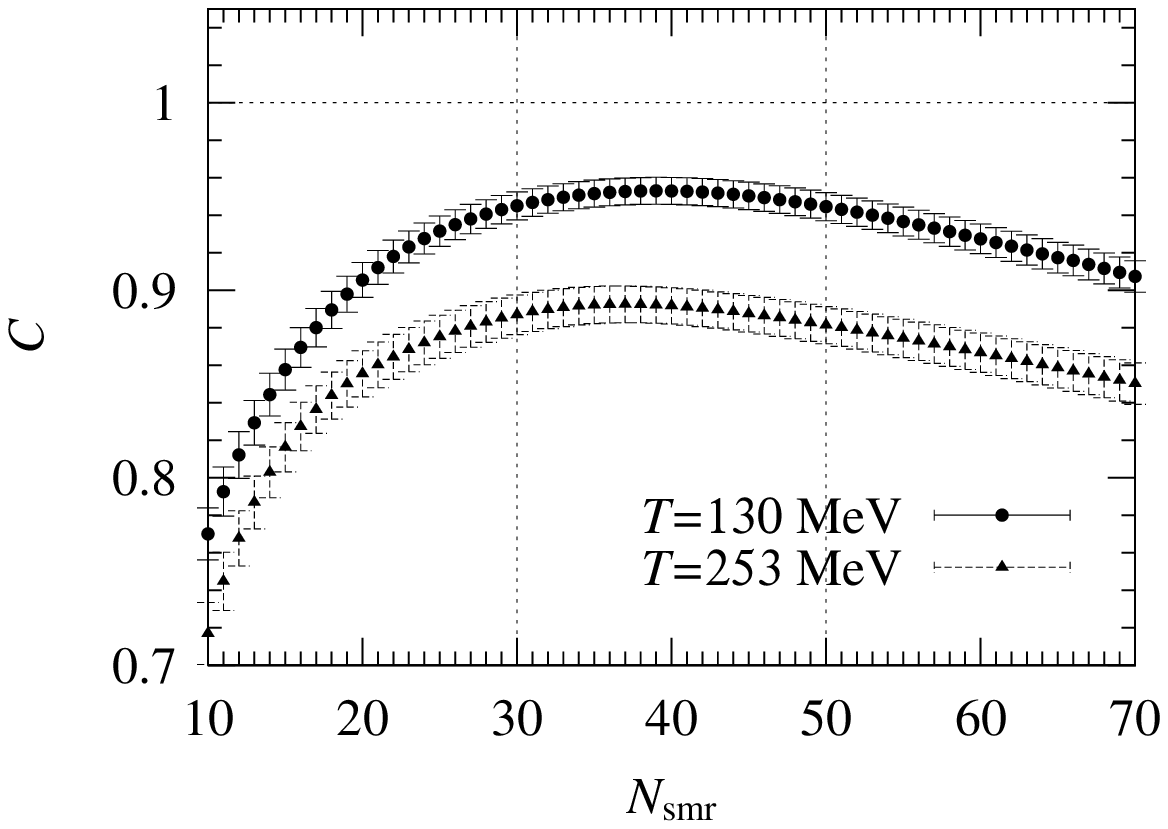}
\leftline{(b)}
\includegraphics[width=\figwidth]{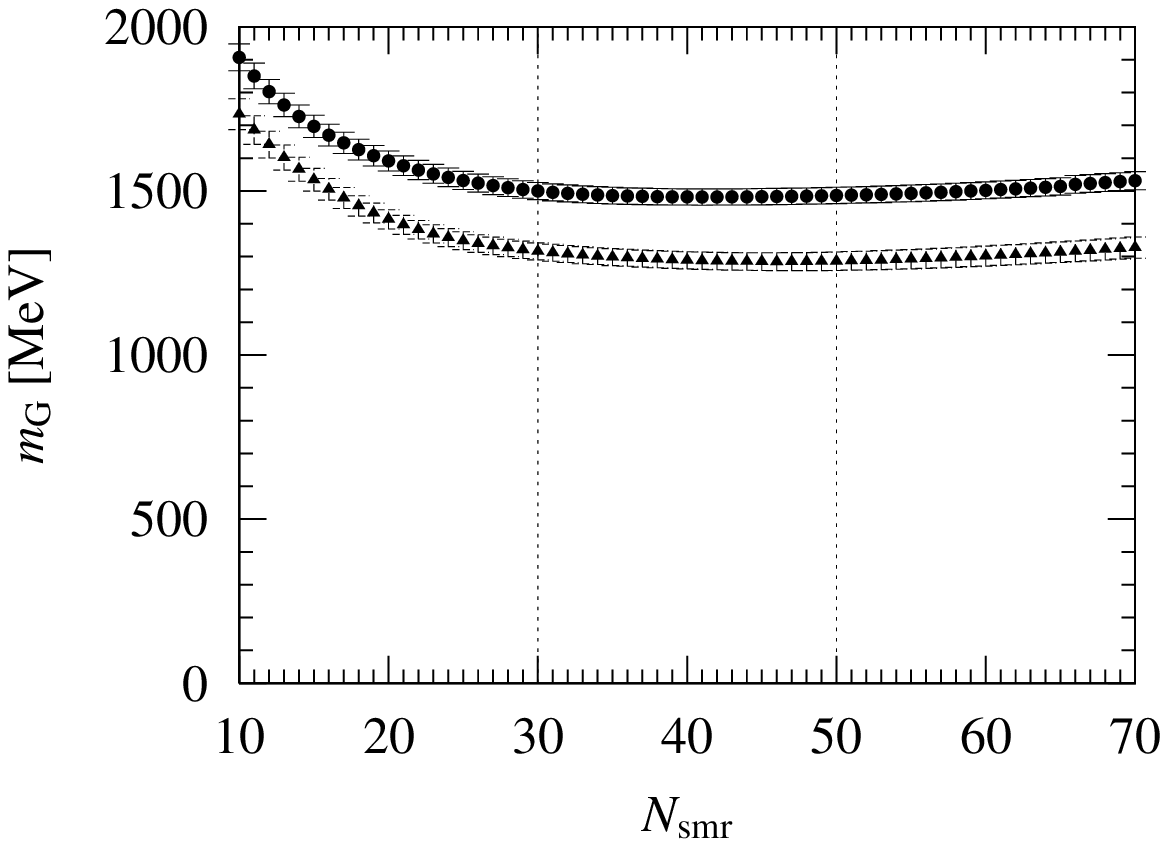}
\caption{(a) The  ground-state overlap $C$  and (b) the  glueball mass
$m_{\rm  G}$ plotted  against the  smearing number  $\Nsmear$  for the
temperatures $T=130,253$ MeV.  The solid circle corresponds to $T=130$
MeV, and the solid triangle  to $T=253$ MeV.  The suitable smearing is
found to  be in  a range as  $30 <  \Nsmear < 50$,  where $C$  is kept
maximized and $m_{\rm G}$ is kept minimized simultaneously. }
\label{against-nsmear}
\setcounter{subfigure}{0}
\refstepcounter{subfigure}
\label{ovl-nsmear}
\refstepcounter{subfigure}
\label{mass-nsmear}
\end{figure}
\subsubsection{The ground-state overlap}
In \Fig{against-nsmear}, the ground-state  overlap $C$ and the pole-mass
$m_{\rm G}$ for  the $0^{++}$ glueball are plotted  against the smearing
number $\Nsmear$ for the two  cases (i) the low temperature case $T=130$
MeV  (circle), (ii)  the high  temperature case  $T=253$  MeV (triangle)
below  $T_c$.   Here,   we  fix  one  of  the   smearing  parameters  as
$\alpha=2.1$.   We  see  that,  in  the  region  $\Nsmear  <  30$,  with
increasing $\Nsmear$,  the ground state  overlap $C$ is growing  and the
pole-mass $m_{\rm G}$ is reducing.  Then,  in the region $30 < \Nsmear <
50$,  $C$ is kept  maximized and  $m_{\rm G}$  is kept  minimized, which
implies  that  the  ground-state  contribution in  $G(t)$  is  maximally
enhanced.  Beyond this region, i.e.,  for $\Nsmear > 50$, since the size
of the smeared  operator exceeds the physical size  of the glueball, $C$
begins to decline and $m_{\rm G}$ begins to grow gradually.

The maximum  overlap and the minimum  mass should be  achieved at almost
the same $N_{\rm  smear}$. In fact, both of  these two conditions should
work   as  an   indication  of   the  maximally   enhanced  ground-state
contribution.   In  practical  calculations,  these two  conditions  are
achieved at slightly different  $N_{\rm smear}$.  However, the numerical
results  on the  pole-mass  are  almost the  same.   For instance,  from
\Fig{against-nsmear},   the   former    condition   leads   to   $m_{\rm
G}(T=130{\rm{MeV}}) =  1482(25)$ MeV  at $N_{\rm smear}=39$  and $m_{\rm
G}(T=253{\rm{MeV}})  = 1294(26)$  MeV at  $N_{\rm smear}=37$,  while the
latter condition leads to $m_{\rm G}(T=130{\rm{MeV}}) = 1481(25)$ MeV at
$N_{\rm smear}=41$  and $m_{\rm  G}(T=253{\rm{MeV}}) = 1284(27)$  MeV at
$N_{\rm smear}=46$.

As we mentioned before, we divide  the statistical data into the bins of
the size  $100$ to reduce the possible  auto-correlations.  However, the
bin-size dependence  is not seen so  strong even in the  vicinity of the
critical temperature $T_c$.  Recall that the measurement interval of the
gauge field configuration is 100  sweeps. Since this interval is already
rather large, the bin-size of 100 seems to be sufficient to overcome the
auto-correlations.  This is also the case in the $2^{++}$ glueball.

\begin{figure}
\includegraphics[width=\figwidth]{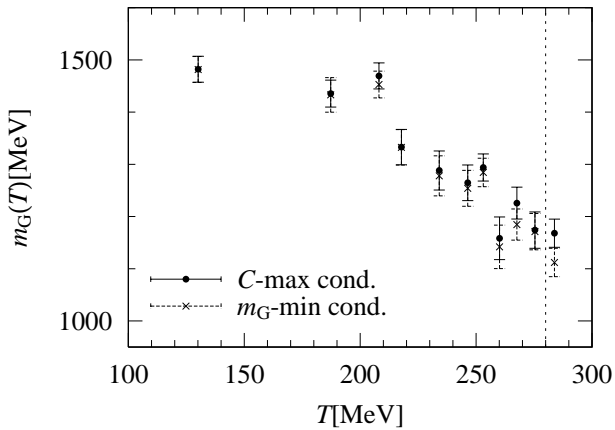}
\caption{The  $0^{++}$  glueball  mass  $m_{\rm G}(T)$  plotted  against
temperature  $T$.  The  solid circle  and the  cross denote  the thermal
$0^{++}$ glueball  mass obtained with the maximum  overlap condition and
the  mass minimum  condition,  respectively.  The  vertical dotted  line
indicates the critical temperature $T_c \simeq 280$ MeV in quenched QCD.
\Cut{The data above $T_c$ will be eliminated at the final stage.}}
\label{mass-temperature}
\end{figure}

\subsubsection{The main result on the pole-mass}
From the analysis at various temperatures as
$T$ = 130, 187, 208, 218, 234, 246, 253, 260, 268, 275, 284 MeV,
we plot the SU(3) lattice QCD result for the pole-mass $m_{\rm G}(T)$ of
the    lowest   $0^{++}$   glueball    against   temperature    $T$   in
\Fig{mass-temperature}.   The  solid circle  and  the  cross denote  the
thermal glueball  mass $m_{\rm G}(T)$ obtained with  the maximum overlap
condition  and the mass  minimum condition,  respectively.  We  see that
they  are almost  the  same.  In  \Fig{mass-temperature},  we observe  a
significant pole-mass  reduction for  the lowest $0^{++}$  glueball near
$T_c$ as
\begin{equation}
	m_G(T) = 1200 \pm 100 {\rm MeV}
	\Hs{\rm for}\Hs
	0.9 T_c <T < T_c,
\end{equation}
in comparison with  $m_G(T \simeq 0.5T_c) \simeq 1480  {\rm MeV}$ in our
data or $m_G(T \sim 0) \simeq 1500\mbox{--}1700$ MeV in the lattice data
in \cite{morningstar,weingarten,teper}.  In  this way, we observe nearly
300  MeV reduction  or about  20 \%  reduction of  the pole-mass  of the
lowest $0^{++}$ glueball in the vicinity of $T_c$ as
\begin{equation}
\renewcommand{\arraystretch}{1.5}
\begin{array}[t]{l}
	m_{\rm G}(T\sim 0) - m_{\rm G}(T \simeq T_c)
\simeq
	300 {\rm MeV},
\\
	m_{\rm G}(T \simeq T_c)
\simeq
	0.8 m_{\rm G} (T \sim 0).
\end{array}
\end{equation}
It is remarkable  that the pole-mass shift obtained  here is much larger
than  any other pole-mass  shifts which  has been  ever observed  in the
meson sector in lattice QCD \cite{taro,umeda}.

We next consider the glueball size.  To this end, we seek for $\Nsmear$,
which realizes the maximum  overlap condition.  Such $\Nsmear$ is listed
in \Table{table-S}  for each temperature.   By using \Eq{size},  we give
rough  estimates of  the  thermal glueball  sizes  $\rho(T)$ at  various
temperatures, which are shown in \Table{table-S}.  We find
\begin{equation}
	\rho \simeq 0.4 {\rm fm},
\end{equation}
both at low and high temperature below $T_c$.

In  \Table{table-S},  we summarize  the  lowest  $0^{++}$ glueball  mass
$m_{\rm  G}(T)$,  the  ground-state  overlap $C^{\rm  max}$,  correlated
$\chi^2/\Ndf$, the fit-range $(t_1,t_2)$, the normalized overlap $g(0)$,
the corresponding smearing number  $\Nsmear$, the $0^{++}$ glueball size
$\rho(T)$, and the number of gauge field configurations $\Nconfig$.

\begin{table*}
\caption{ The pole-mass $m_{\rm G}(T)$ of the lowest $0^{++}$ glueball
at finite temperature $T$ in  SU(3) lattice QCD.  The temperature $T$,
the  temporal  lattice  size  $N_t$, the  thermal  glueball  pole-mass
$m_{\rm G}(T)$,  the maximum  ground-state overlap $C^{\rm  max}$, the
correlated  $\chi^2/\Ndf$, the fit  range $(t_1,t_2)$,  the normalized
ground-state  overlap  $g(0)$,  the  smearing  number  $\Nsmear$,  the
thermal  glueball size $\rho(T)$  with \Eq{size},  and the  number the
gauge  configurations $\Nconfig$  are  listed.  The  best smearing  on
$\Nsmear$ is  determined with  maximum overlap condition.   Above $T_c
\simeq 280$ MeV,  the data such as $m_{\rm G}$ is  to be understood as
the best-fit parameters.}
\label{table-S}
\begin{ruledtabular}
\begin{tabular}{cccccccccc}
$T$[MeV] &
$N_t$ &
$m_G$[MeV] &
$C^{\rm{max}}$ &
$\chi^2/\Ndf$ &
$(t_1,t_2)$ &
$g(0)$ &
$\Nsmear$ &
$\rho$[fm] &
$\Nconfig$ \\
\hline
130 & 72 & 1482(25) & 0.95(1) & 1.60 & ( 5,14) & 0.95(1) & 39 & 0.43 & 5500 \\
187 & 50 & 1436(26) & 0.94(1) & 0.60 & ( 5,25) & 0.94(1) & 41 & 0.44 & 5700 \\
208 & 45 & 1469(25) & 0.96(1) & 0.73 & ( 5,22) & 0.96(1) & 36 & 0.41 & 6400 \\
218 & 43 & 1333(34) & 0.89(1) & 1.22 & ( 7,21) & 0.89(1) & 41 & 0.44 & 9200 \\
234 & 40 & 1288(37) & 0.87(2) & 1.01 & ( 8,20) & 0.87(2) & 36 & 0.41 & 8600 \\
246 & 38 & 1265(34) & 0.88(1) & 1.21 & ( 7,13) & 0.88(1) & 39 & 0.43 & 8900 \\
253 & 37 & 1294(26) & 0.89(1) & 1.45 & ( 6,17) & 0.90(1) & 37 & 0.41 & 8900 \\
260 & 36 & 1158(41) & 0.80(2) & 1.54 & (10,17) & 0.81(2) & 42 & 0.44 & 9900 \\
268 & 35 & 1226(30) & 0.85(1) & 2.74 & ( 7,11) & 0.86(1) & 34 & 0.40 & 9900 \\
275 & 34 & 1174(35) & 0.81(2) & 0.12 & ( 9,12) & 0.82(2) & 44 & 0.45 & 9900 \\
\hline
284 & 33 & 1168(27) & 0.82(1) & 1.01 & ( 8,16) & 0.83(1) & 41 & 0.44 & 9900 \\
312 & 30 & 1158(28) & 0.79(2) & 1.22 & ( 8,15) & 0.80(2) & 45 & 0.46 & 9900 \\
334 & 28 & 1091(38) & 0.75(2) & 1.37 & ( 8,14) & 0.78(2) & 44 & 0.45 & 6200 \\
360 & 26 & 1256(26) & 0.84(1) & 1.72 & ( 6,13) & 0.87(1) & 36 & 0.41 & 6800 \\
390 & 24 & 1199(33) & 0.79(2) & 0.88 & ( 8,12) & 0.82(2) & 33 & 0.39 & 7700 \\
\end{tabular}
\end{ruledtabular}
\end{table*}
\subsection{Thermal \boldmath{$2^{++}$} glueball}
\begin{figure}
\leftline{(a)}
\includegraphics[width=\figwidth]{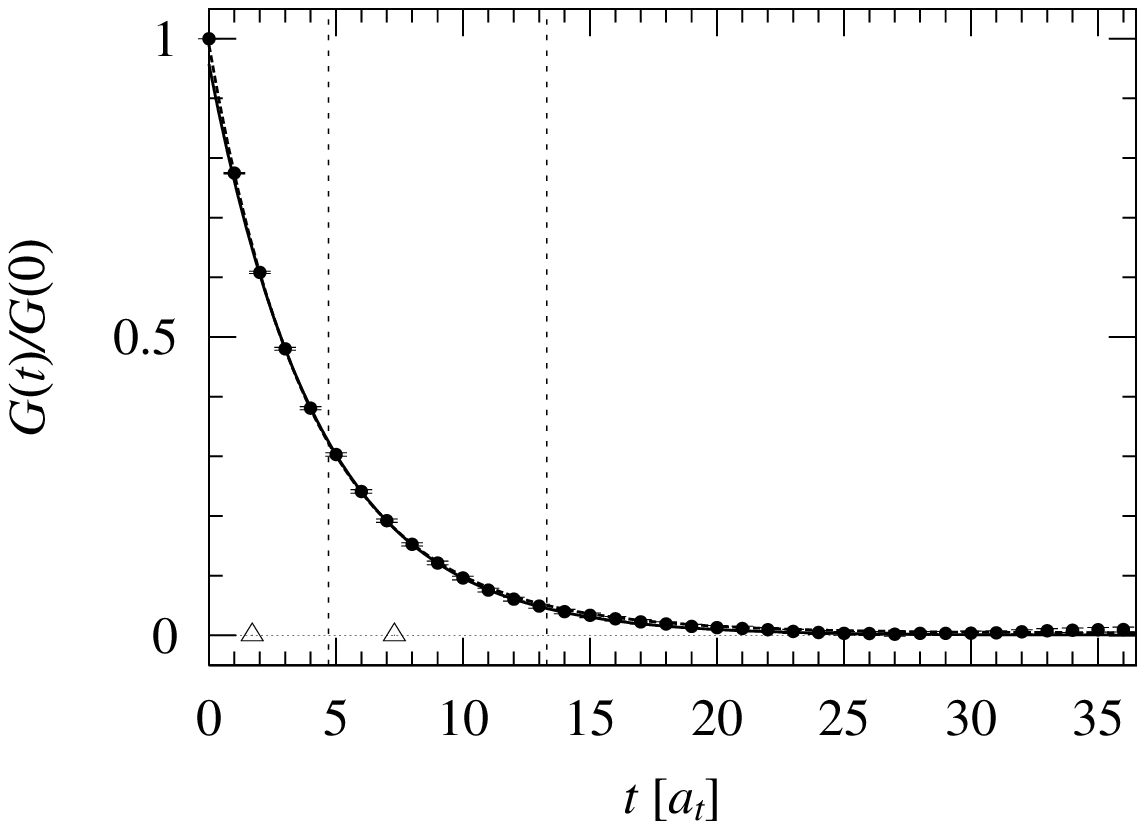}
\leftline{(b)}
\includegraphics[width=\figwidth]{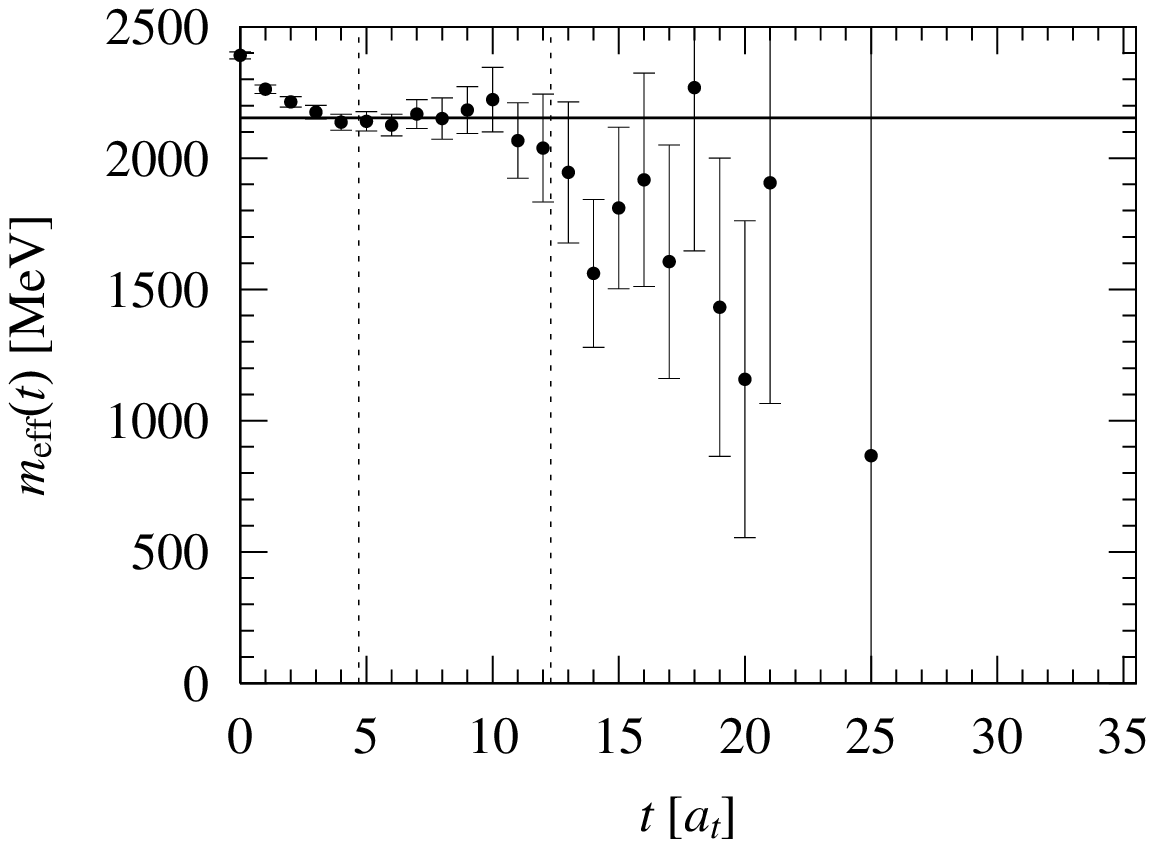}
\caption{ (a)  The $2^{++}$ glueball correlator $G(t)/G(0)$  of the at
the low  temperature ($T=130$  MeV).  (b) The  corresponding cosh-type
effective-mass plot.  The  meaning of the solid lines  in both figures
and the  dashed curve in  (a) are the  same as in  \Fig{green.72}. In
(b),  the  points beyond  $t=25$  are  suppressed  due to  their  huge
statistical errors. }
\label{green.72-T}
\setcounter{subfigure}{0}
\refstepcounter{subfigure}
\label{correlator.72-T}
\refstepcounter{subfigure}
\label{effmass.72-T}
\end{figure}
\begin{figure}
\leftline{(a)}
\includegraphics[width=\figwidth]{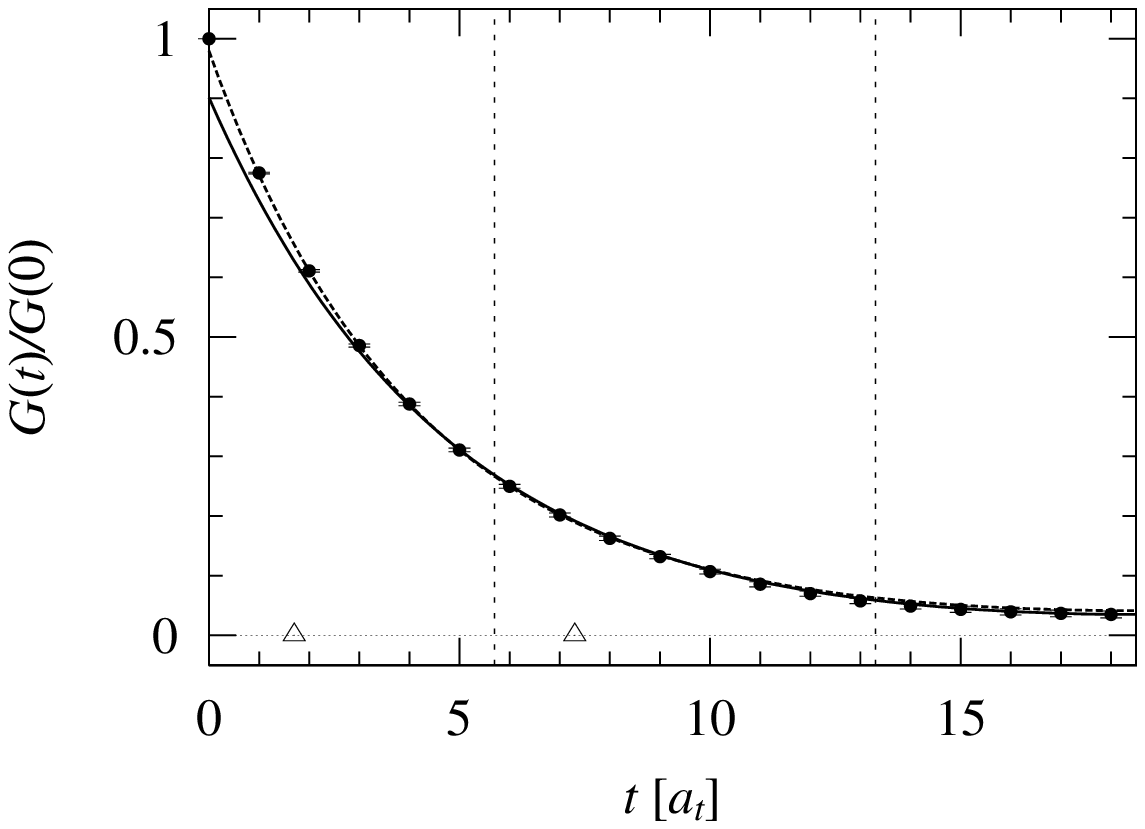}
\leftline{(b)}
\includegraphics[width=\figwidth]{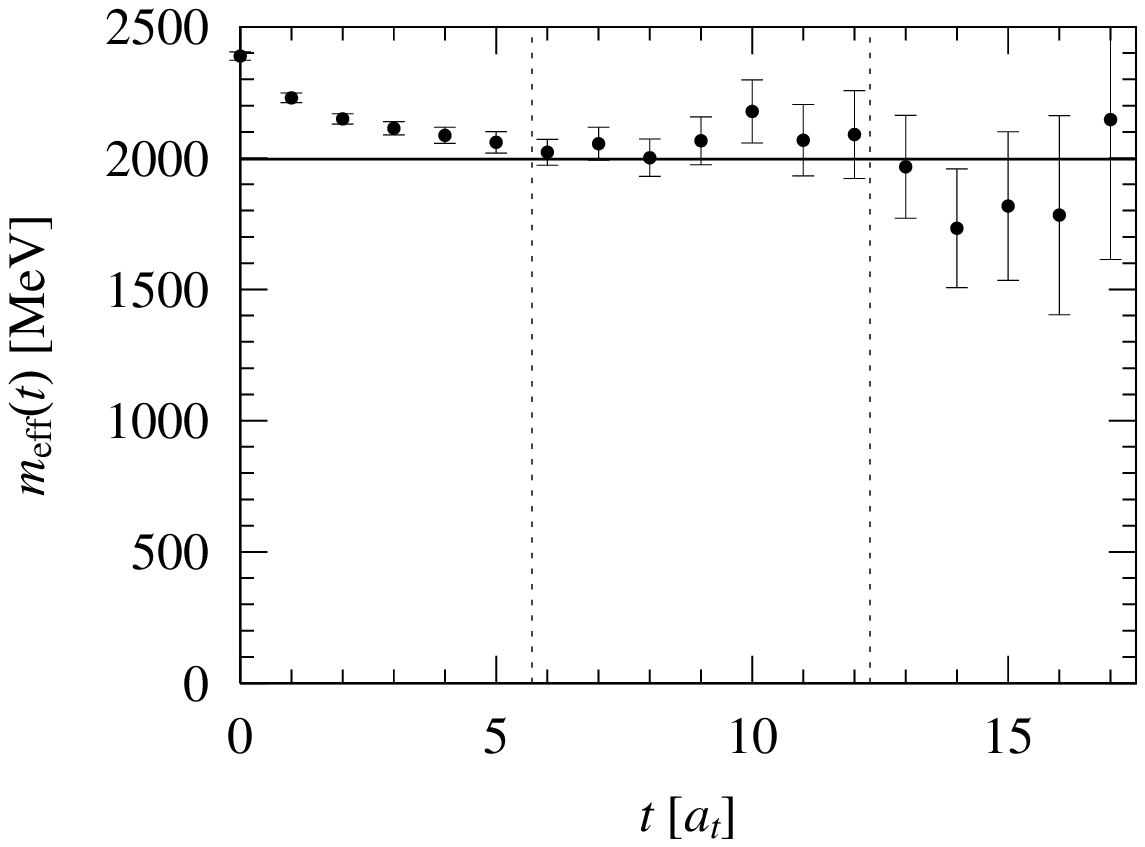}
\caption{ (a) The $2^{++}$ glueball correlator $G(t)/G(0)$ at the high
temperature   ($T=253$   MeV).    (b)  The   corresponding   cosh-type
effective-mass plot.  The  meaning of the solid lines  in both figures
and the dashed curve in (a) are the same as in \Fig{green.72}.  }
\label{green.37-T}
\setcounter{subfigure}{0}
\refstepcounter{subfigure}
\label{correlator.37-T}
\refstepcounter{subfigure}
\label{effmass.37-T}
\end{figure}
\begin{figure}
\includegraphics[width=\figwidth]{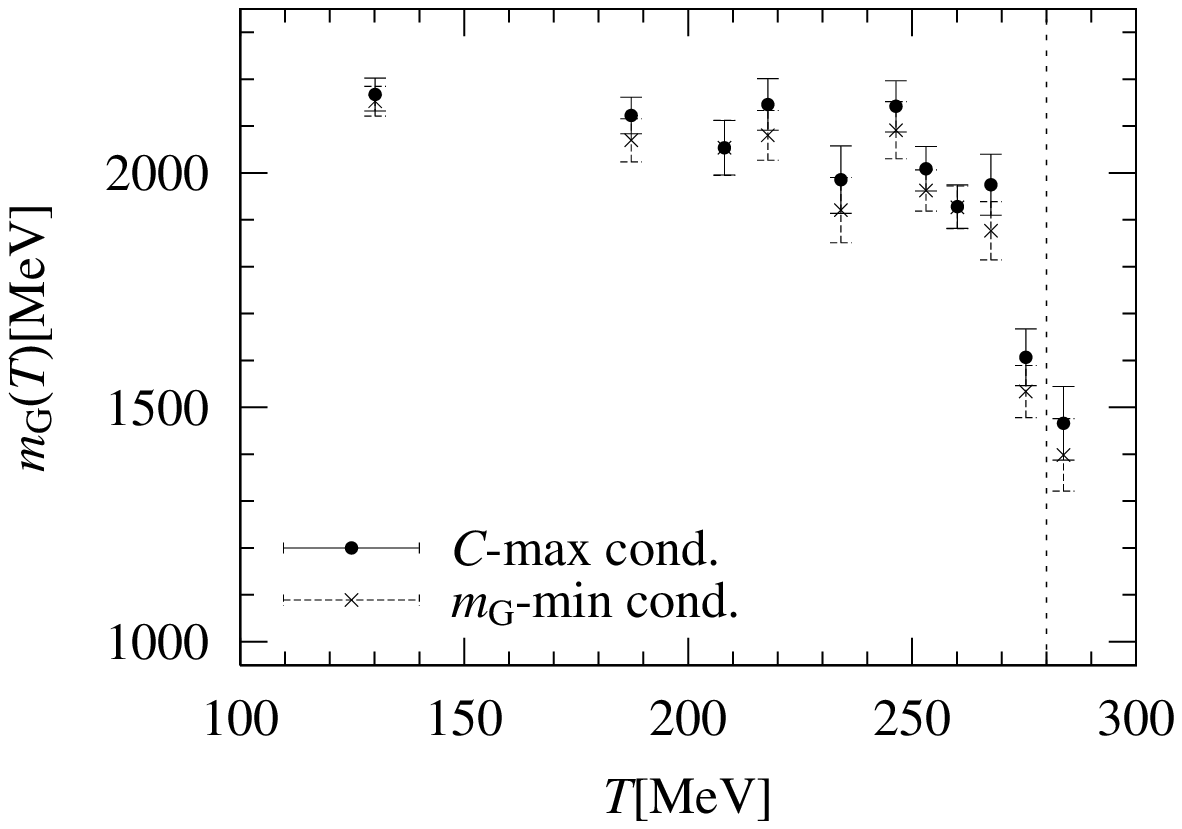}
\caption{The  $2^{++}$ glueball  mass plotted  against  temperature $T$.
The solid circle and the cross denote the thermal $2^{++}$ glueball mass
obtained  with  the  maximum  overlap  condition and  the  mass  minimum
condition,  respectively.   The   vertical  dotted  line  indicates  the
critical  temperature $T_c \simeq  280$ MeV  in quenched  QCD.  \Cut{The
data above $T_c$ will be eliminated at the final stage.}}
\label{mass-temperature-T}
\end{figure}

We show,  in \Figs~\ref{correlator.72-T} and  \ref{correlator.37-T}, the
$2^{++}$  glueball correlators  $G(t)/G(0)$ for  $T=130$ and  $253$ MeV,
respectively,       after      the      suitable       smearings      as
\Eq{suitable.parameter.set}.  The solid lines  denote the results of the
best-fit analysis of single-cosh  type.  The fit-ranges are indicated by
the vertical  dotted lines, which  are determined from the  plateaus in
the corresponding effective mass plots shown in \Figs~\ref{effmass.72-T}
and \ref{effmass.37-T}.

From \Fig{correlator.72-T}, we find  $C\simeq 0.96$ and $m_{\rm G}\simeq
2154$ MeV  for the $2^{++}$ glueball  in low temperature case  $T = 130$
MeV, which seems consistent with $m_{\rm G} = 2000\mbox{--}2400$ MeV for
the  $2^{++}$  glueball at  zero  temperature \cite{morningstar}.   From
\Fig{correlator.37-T}, we find $C\simeq 0.90$ and $m_{\rm G}\simeq 1996$
MeV for the $2^{++}$ glueball in the high temperature case $T = 253$ MeV
($< T_c$).

By combining the results at  various temperatures, we plot the pole-mass
of    the   thermal   $2^{++}$    glueball   against    temperature   in
\Fig{mass-temperature-T}.   The solid  circle and  the cross  denote the
pole-masses of  the thermal $2^{++}$ glueball obtained  with the maximum
overlap condition  and the mass minimum condition,  respectively. We see
again that they are qualitatively almost the same.  The thermal $2^{++}$
glueball shows  a tendency to  decline by about  $100$ MeV for $T  < 0.9
T_c$,  which is  rather modest  in  contrast to  the $0^{++}$  glueball.
However, in the  very vicinity of $T_c$, it shows  a sudden reduction of
about $500$ MeV.

In  \Table{table-T},  we summarize  the  lowest  $2^{++}$ glueball  mass
$m_{\rm  G}(T)$,  the  ground-state  overlap $C^{\rm  max}$,  correlated
$\chi^2/\Ndf$, the fit-range $(t_1,t_2)$, the normalized overlap $g(0)$,
the corresponding smearing number  $\Nsmear$, the $2^{++}$ glueball size
$\rho(T)$, and the number of gauge configurations $\Nconfig$.

\begin{table*}
\caption{ The pole-mass $m_{\rm G}(T)$ of the  lowest $2^{++}$ glueball
at finite  temperature $T$  in SU(3) lattice  QCD.  The meaning  of $T$,
$N_t$,  $m_{\rm  G}(T)$,   $C^{\rm  max}$,  $\chi^2/\Ndf$,  $(t_1,t_2)$,
$g(0)$,  $\Nsmear$,  $R$,  and  $\Nconfig$  are the  same  as  those  in
\Table{table-S}.   The best  smearing  on $\Nsmear$  is determined  with
maximum overlap condition.  }
\label{table-T}
\begin{ruledtabular}
\begin{tabular}{cccccccccc}
$T$[MeV] &
$N_t$ &
$m_G$[MeV] &
$C^{\rm{max}}$ &
$\chi^2/\Ndf$ &
$(t_1,t_2)$ &
$g(0)$ &
$\Nsmear$ &
$\rho$[fm] &
$\Nconfig$ \\
\hline
130 & 72 & 2167(35) & 0.97(1) & 1.01 & ( 5,13) & 0.97(1) & 54 & 0.50 & 5500 \\
187 & 50 & 2123(39) & 0.95(2) & 0.57 & ( 5,13) & 0.95(2) & 39 & 0.43 & 5700 \\
208 & 45 & 2054(58) & 0.90(3) & 1.07 & ( 7,19) & 0.90(3) & 47 & 0.47 & 6400 \\
218 & 43 & 2146(55) & 0.91(3) & 1.49 & ( 7,12) & 0.91(3) & 68 & 0.56 & 9200 \\
234 & 40 & 1986(72) & 0.85(4) & 1.03 & ( 8,14) & 0.85(4) & 63 & 0.54 & 8600 \\
246 & 38 & 2142(55) & 0.95(3) & 0.33 & ( 6,14) & 0.95(3) & 41 & 0.44 & 8900 \\
253 & 37 & 2009(47) & 0.90(2) & 1.32 & ( 6,13) & 0.90(2) & 38 & 0.42 & 8900 \\
260 & 36 & 1928(46) & 0.88(2) & 3.12 & ( 6,12) & 0.88(2) & 52 & 0.49 & 9900 \\
268 & 35 & 1975(65) & 0.91(4) & 2.12 & ( 8,16) & 0.91(4) & 44 & 0.45 & 9900 \\
275 & 34 & 1607(61) & 0.69(4) & 0.80 & ( 9,14) & 0.69(4) & 48 & 0.47 & 9900 \\
\hline
284 & 33 & 1466(78) & 0.60(5) & 1.51 & (11,16) & 0.60(5) & 51 & 0.49 & 9900 \\
312 & 30 & 1319(58) & 0.61(4) & 0.33 & (10,15) & 0.62(4) & 53 & 0.50 & 9900 \\
334 & 28 & 1408(52) & 0.68(3) & 1.93 & ( 8,14) & 0.69(3) & 47 & 0.47 & 6200 \\
360 & 26 & 1440(38) & 0.75(2) & 0.59 & ( 7,13) & 0.76(2) & 61 & 0.53 & 6800 \\
390 & 24 & 1466(55) & 0.68(4) & 1.47 & ( 8,12) & 0.70(3) & 32 & 0.39 & 7700 \\
\end{tabular}
\end{ruledtabular}
\end{table*}

\subsection{Glueball correlations above \boldmath{$T_c$}}

In the  deconfinement phase above  $T_c$, neither hadrons  nor glueballs
are elementary excitations any  more.  Instead, quarks and gluons appear
as  elementary excitations  in the  quark gluon  plasma.   Therefore, we
might  wonder if  there  is anything  interesting  in investigating  the
color-singlet modes such as hadrons and glueballs above $T_c$.
Nevertheless,  based  on the  lattice-QCD  Monte  Carlo  studies on  the
screening mass, it was pointed  out that some of the strong correlations
survive even  above $T_c$ in  the scalar and  pseudoscalar color-singlet
modes such as $\sigma$ and pions\cite{detar-kogut}.
Hence, at  least in the neighborhood  of $T_c$, it may  be possible that
some of the nonperturbative  effects survive in the deconfinement phase,
which provides interesting information of the high-temperature QCD.

In this subsection,  we attempt to examine the  temporal correlator of
the glueball-like  color-singlet modes  above $T_c$.  We  simply apply
the pole-mass analysis of  the temporal correlators on the anisotropic
lattice with $N_t  = 24, 26, 28, 30, 33$,  which correspond to $T=390,
360, 334, 312, 284$ MeV, respectively.  (A more sophisticated analysis
will be performed in \Sect{section.spectral-function}.)

\begin{figure}
\leftline{(a)}
\includegraphics[width=\figwidth]{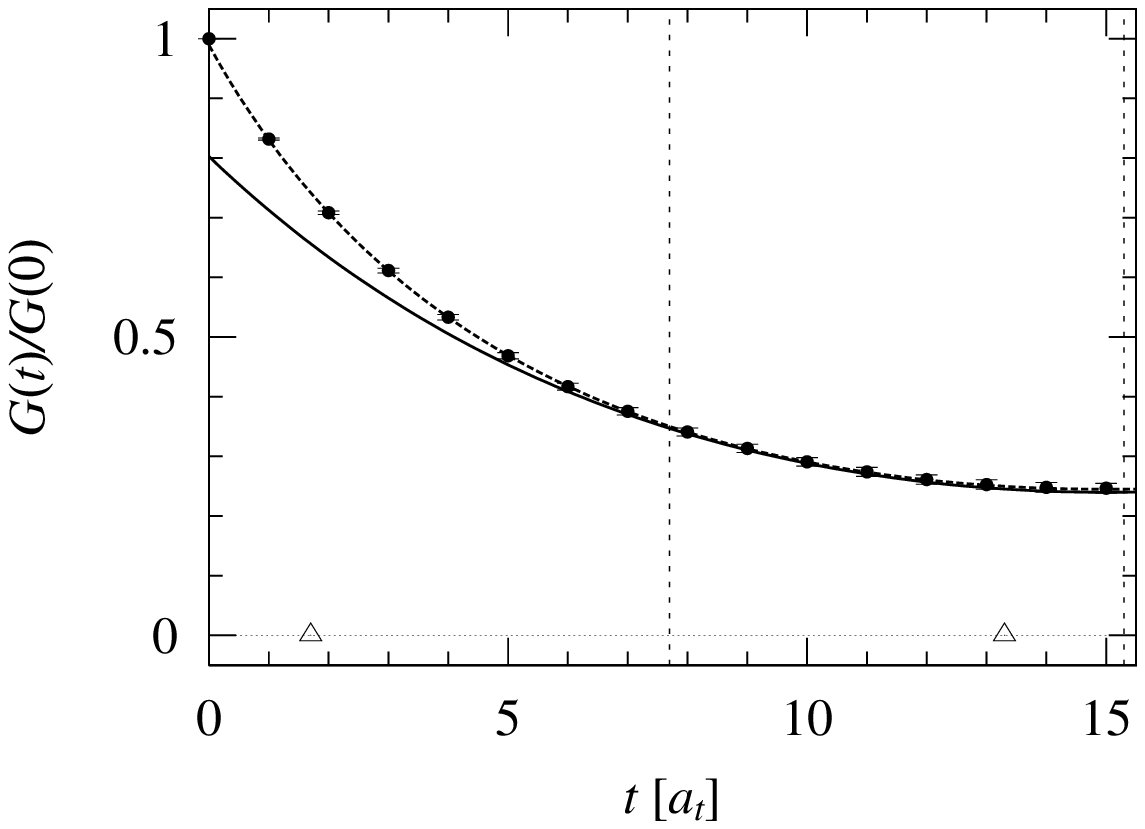}
\leftline{(b)}
\includegraphics[width=\figwidth]{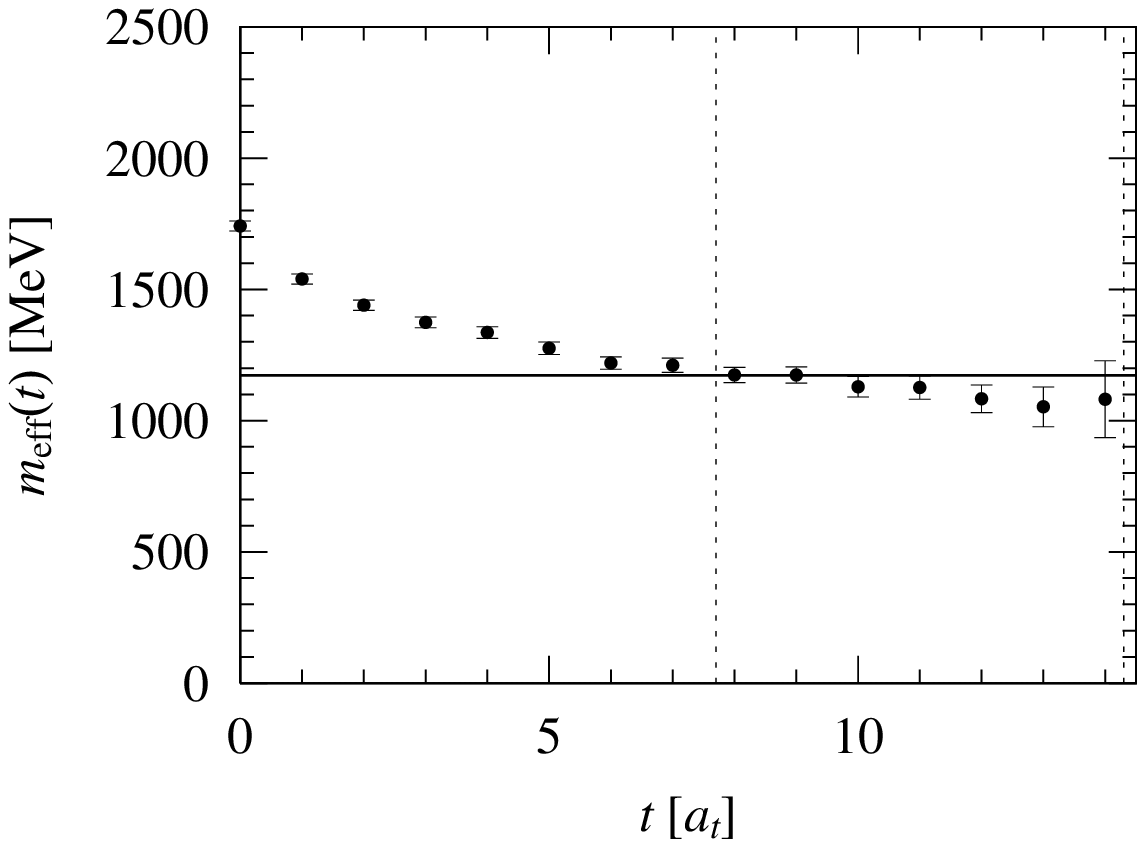}
\caption{ (a) The $0^{++}$  glueball correlator $G(t)/G(0)$ at $T=312$
MeV above $T_c$.  (b) The corresponding cosh-type effective-mass plot.
The meaning of the solid lines in both figures and the dashed curve in
(a) are the same as in \Fig{green.72}.  }
\label{green.30}
\setcounter{subfigure}{0}
\refstepcounter{subfigure}
\label{correlator.30}
\refstepcounter{subfigure}
\label{effmass.30}
\end{figure}
\begin{figure}
\leftline{(a)}
\includegraphics[width=\figwidth]{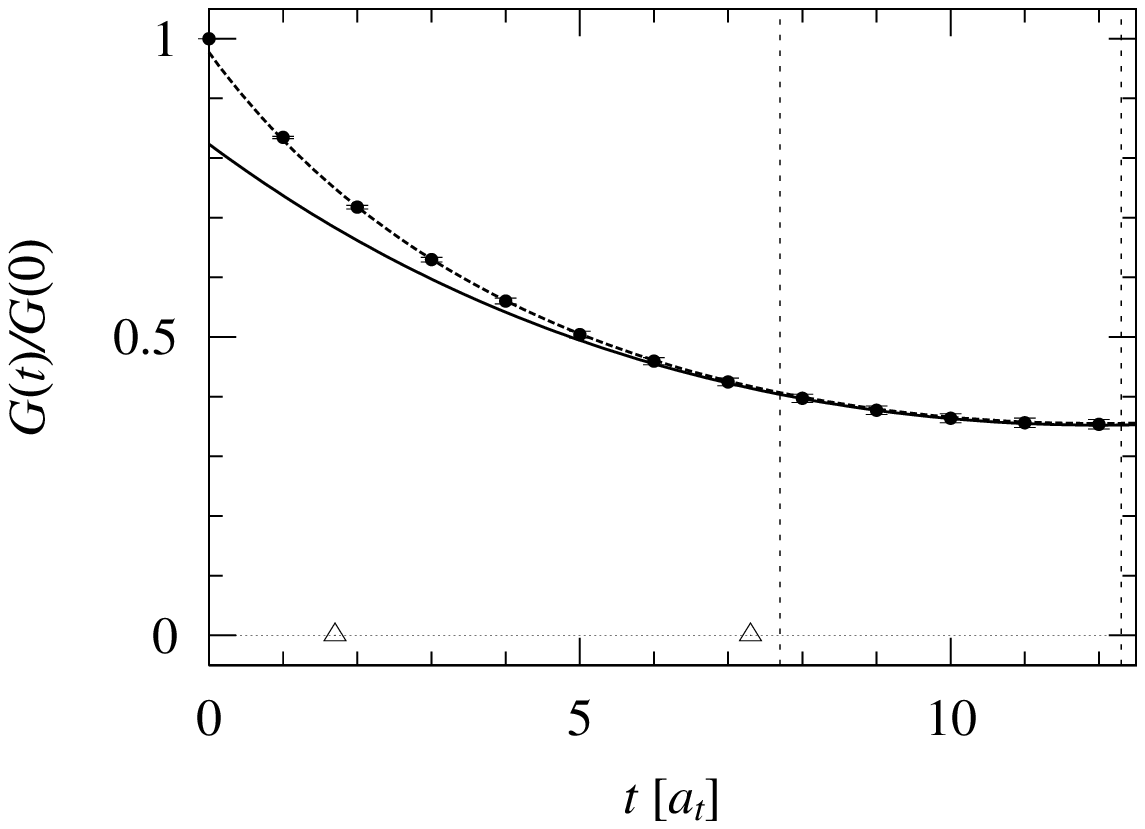}
\leftline{(b)}
\includegraphics[width=\figwidth]{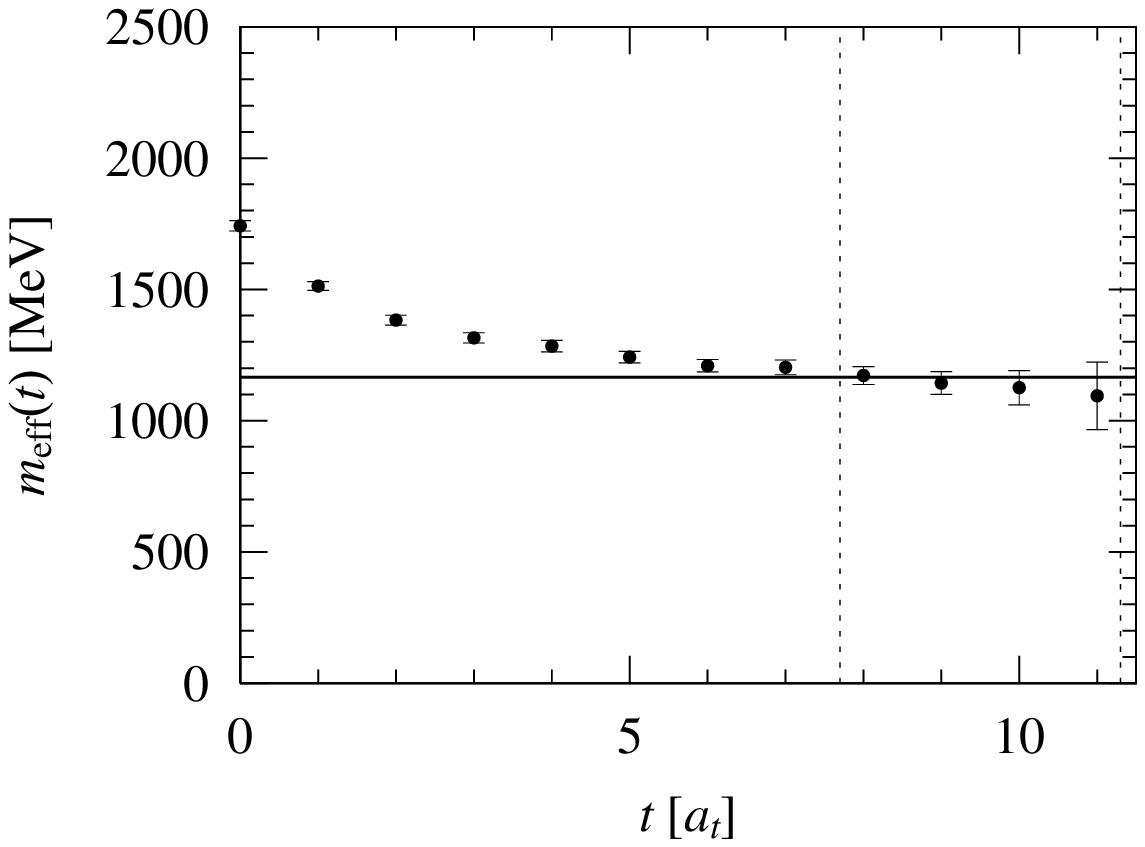}
\caption{ (a) The $0^{++}$  glueball correlator $G(t)/G(0)$ at $T=390$
MeV above $T_c$.  (b) The corresponding cosh-type effective-mass plot.
The meaning of the solid lines in both figures and the dashed curve in
(a) are the same as in \Fig{green.72}.  }
\label{green.24}
\setcounter{subfigure}{0}
\refstepcounter{subfigure}
\label{correlator.24}
\refstepcounter{subfigure}
\label{effmass.24}
\end{figure}

We   show,   in   Figs.~\ref{green.30},~\ref{green.24},   the   temporal
correlators  $G(t)/G(0)$ together with  the associated  effective masses
$m_{\rm  eff}(t)$ at $T=312,  390$ MeV,  respectively, for  the $0^{++}$
glueball-like  color-singlet  mode.   In  each  effective-mass  plot,  a
plateau  is observed  in the  region  indicated by  the vertical  dotted
lines.
By simply  applying the single-cosh fit within  the fit-range determined
from  the  plateau,  the  ``pole-mass''  of  the  $0^{++}$  glueball  is
extracted above  $T_c$, which is listed in  \Table{table-S}.  We observe
that  the ``pole-mass''  is of  about $1200$  MeV, and  that  it changes
rather continuously across the critical temperature $T_c$.
\begin{figure}
\leftline{(a)}
\includegraphics[width=\figwidth]{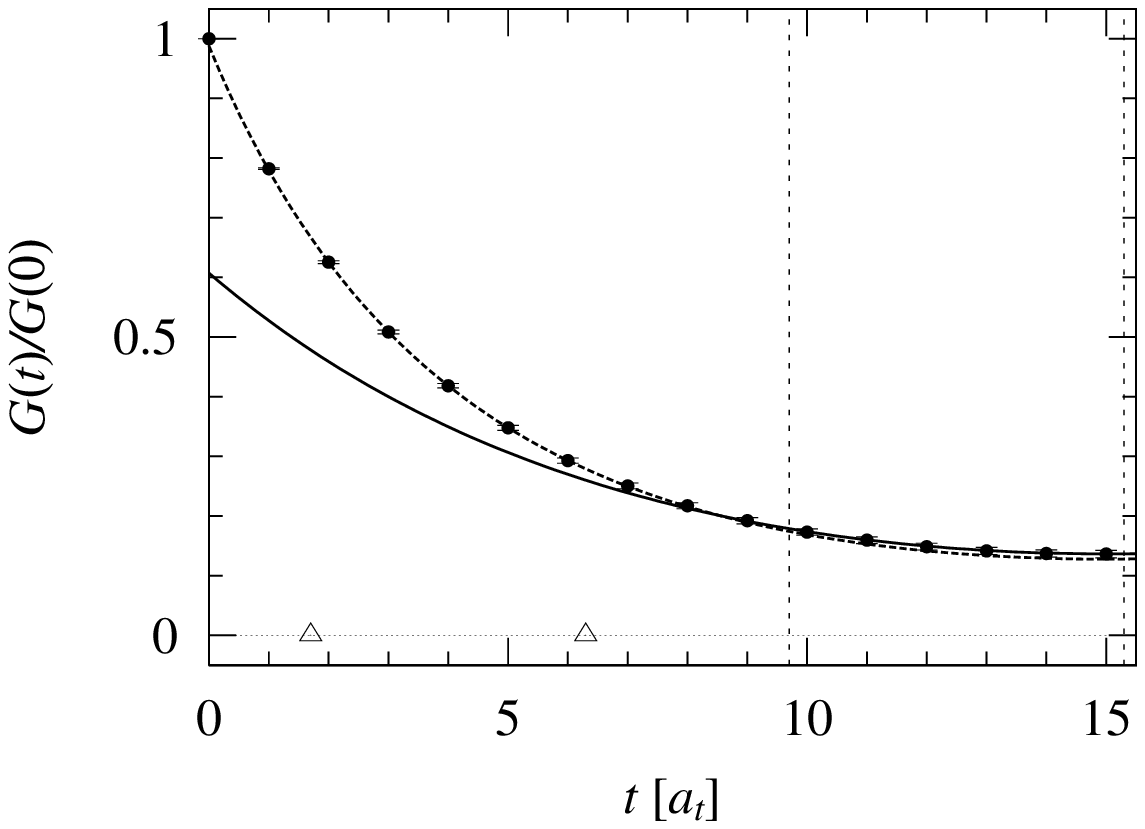}
\leftline{(b)}
\includegraphics[width=\figwidth]{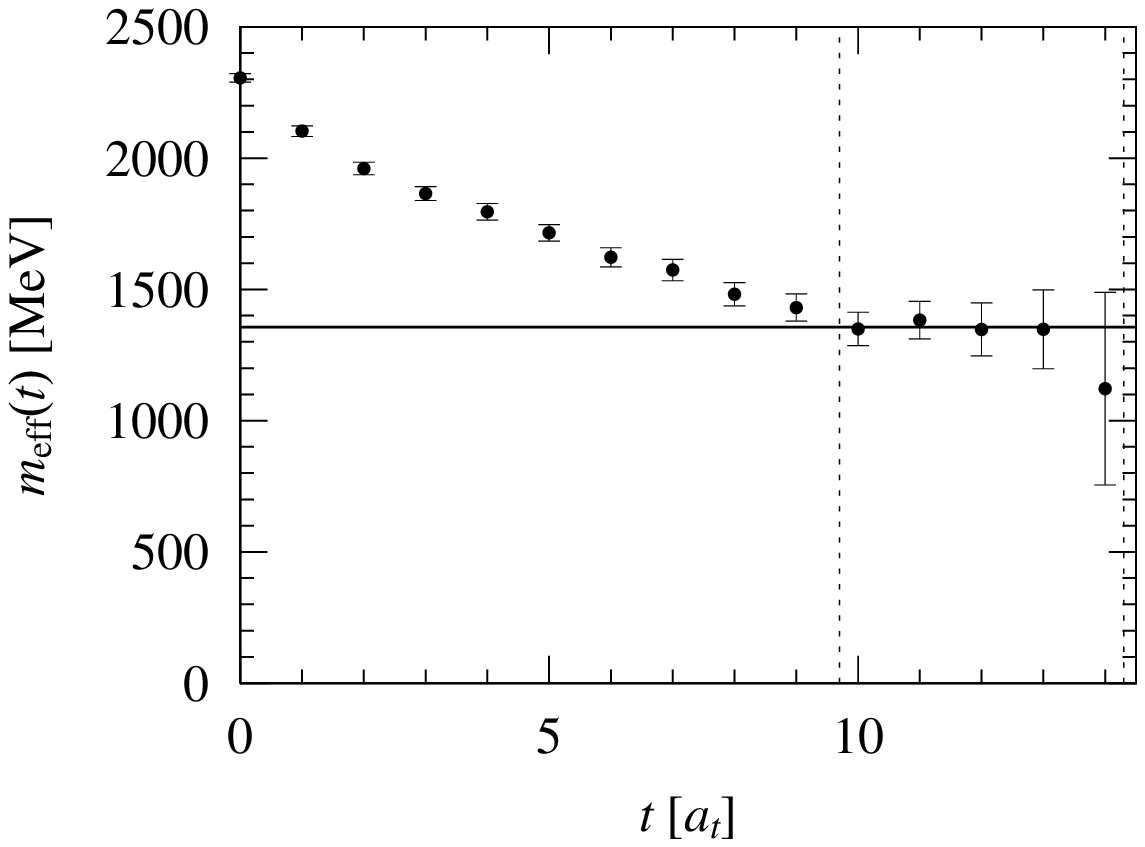}
\caption{ (a) The $2^{++}$  glueball correlator $G(t)/G(0)$ at $T=312$
MeV above $T_c$.  (b) The corresponding cosh-type effective-mass plot.
The meaning of the solid lines in both figures and the dashed curve in
(a) are the same as in \Fig{green.72}.  }
\label{green.30-T}
\setcounter{subfigure}{0}
\refstepcounter{subfigure}
\label{correlator.30-T}
\refstepcounter{subfigure}
\label{effmass.30-T}
\end{figure}
\begin{figure}
\leftline{(a)}
\includegraphics[width=\figwidth]{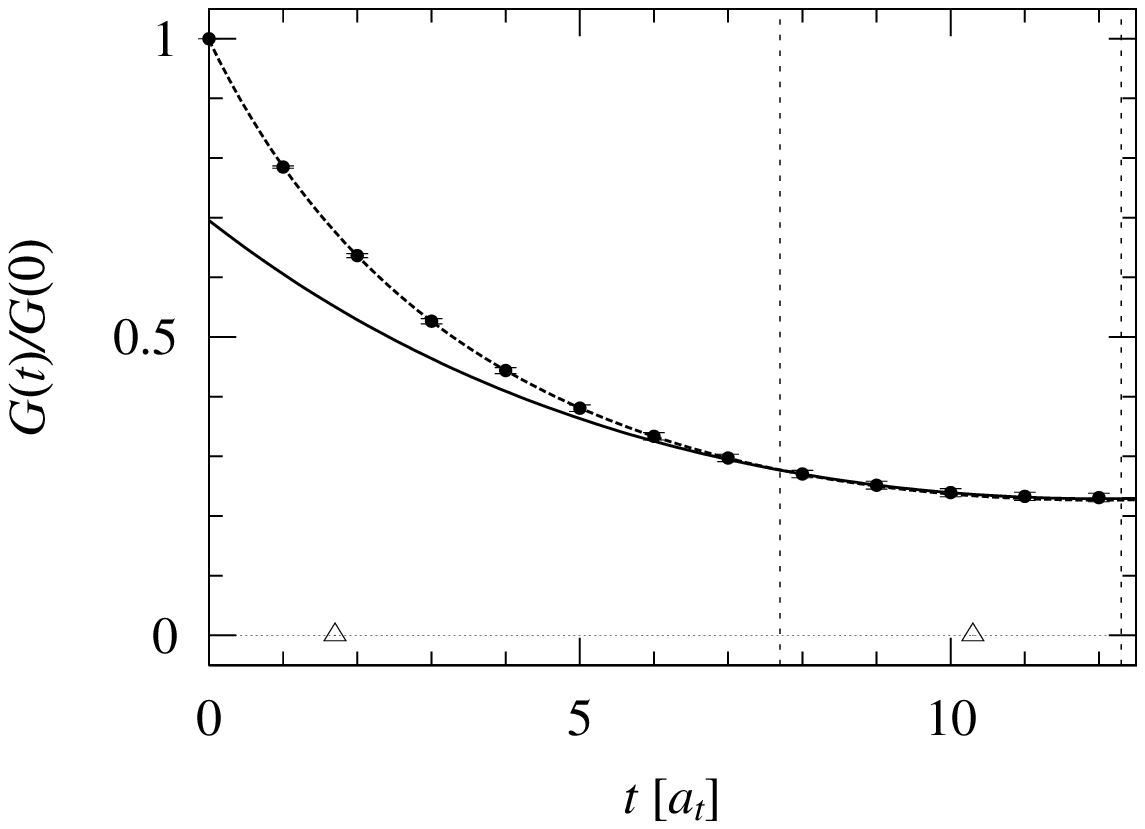}
\leftline{(b)}
\includegraphics[width=\figwidth]{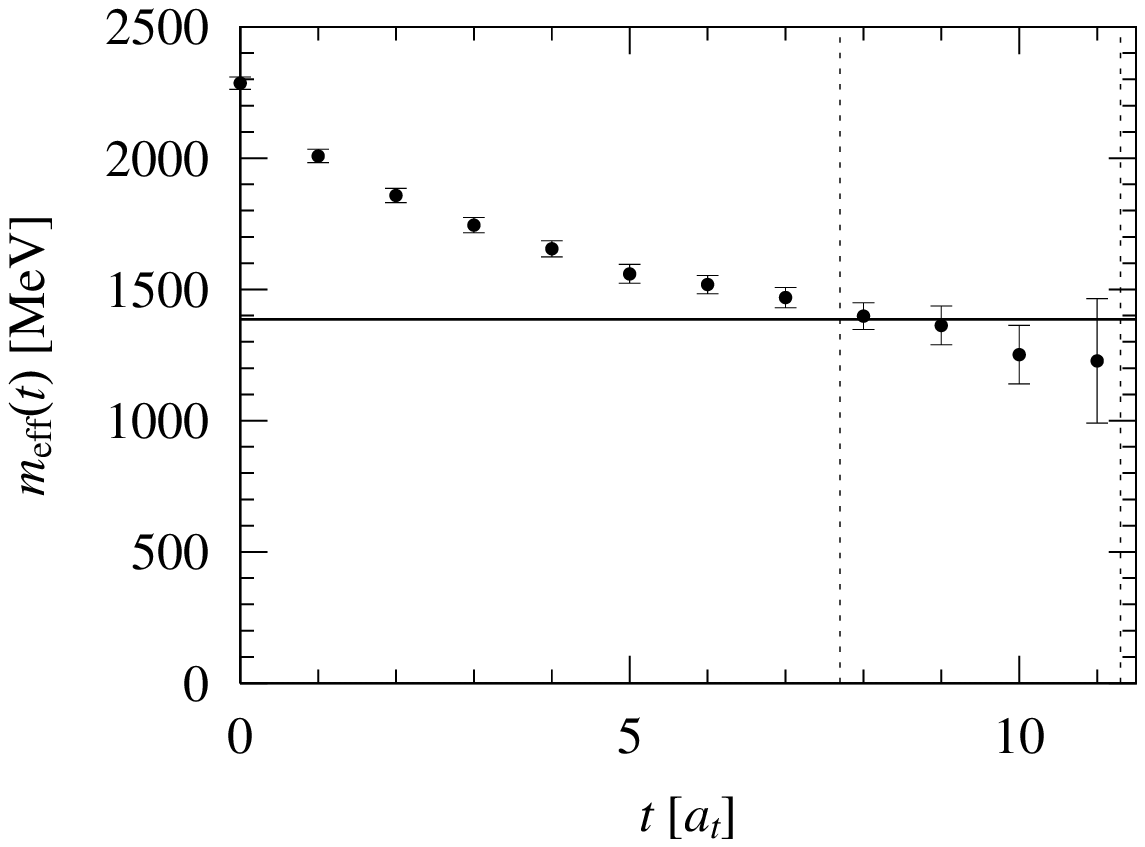}
\caption{ (a) The $2^{++}$  glueball correlator $G(t)/G(0)$ at $T=390$
MeV above $T_c$.  (b) The corresponding cosh-type effective-mass plot.
The meaning of the solid lines in both figures and the dashed curve in
(a) are the same as in \Fig{green.72}.  }
\label{green.24-T}
\setcounter{subfigure}{0}
\refstepcounter{subfigure}
\label{correlator.24-T}
\refstepcounter{subfigure}
\label{effmass.24-T}
\end{figure}

We perform the similar analysis  for the $2^{++}$ glueball-like mode. In
Figs.~\ref{green.30-T}  and \ref{green.24-T},  the  temporal correlators
$G(t)/G(0)$  are plotted  together  with the  associated effective  mass
plots at $T=312, 390$ MeV, respectively. In each effective mass plot, we
observe  a plateau in  the region,  which is  indicated by  the vertical
dotted  lines.   Similar to  the  $0^{++}$  case,  we simply  apply  the
single-cosh  fit within  the fit-range  determined from  the  plateau to
extract the ``pole-mass''. The results are listed in \Table{table-T}. We
observe that the ``pole-mass'' is  about $1400$ MeV, and that its change
is again rather continuous across the critical temperature $T_c$.

In both  of these $0^{++}$ and  $2^{++}$ cases, the size  of the plateau
tends to shrink at high temperatures  above $T_c$. We note that, at much
higher  temperatures, the  plateau finally  disappears,  which, however,
seems to be due to the  failure of the ground-state saturation in such a
limited temporal  size.  At any  rate, these phenomena suggest  that the
single-cosh  ansatz becomes  inappropriate for  the fit-function  of the
temporal correlator $G(t)$ above $T_c$ at such a high temperature.
However, because of  the existence of the plateau  in the effective-mass
plot, we  do not exclude so  far the possibility  that the color-singlet
modes exist as meta-stable modes just above $T_c$.
According to  the lattice data, the color-singlet  $0^{++}$ and $2^{++}$
modes  have  the   pole-mass  of  about  $1200$  MeV   and  $1400$  MeV,
respectively, provided that it can be regarded as a bound state.
However,  even  in this  case,  it  should be  kept  in  mind that  such
glueball-like  modes  can   decay  into  two  or  more   gluons  in  the
quark-gluon-plasma phase.  Since they cannot be stable above $T_c$, they
will acquire finite decay widths.  Hence, for more reliable analysis, it
is necessary to take into account the effect of such a width, which will
be attempted in \Sect{section.spectral-function}.

%
\subsection{Discussions and remarks}
In this subsection, we compare  our results of the pole-mass reduction
of the  thermal glueball with the  related lattice QCD  results on the
screening  masses as  well as  the pole-masses  of various  hadrons at
finite temperature.

We  begin  with  the  discussion  of the  pole-masses  of  mesons.   The
pole-masses of  various mesons have been measured  at finite temperature
by  using  the anisotropic  SU(3)  lattice  QCD  at the  quenched  level
in Refs.~\cite{taro,umeda}.
From the same pole-mass analysis as ours, they have found that, both for
light mesons  and for heavy mesons  such as the  charmonium, the thermal
pole-masses  are  almost unchanged  from  their zero-temperature  values
within the  error bar  below $T_c$. This  tendency persists even  in the
vicinity of $T_c$ as
\begin{equation}
	m_{\rm meson}(T) \simeq m_{\rm meson}(T=0),
\Hs
	{\rm for}
\Hs
	T\le T_c.
\end{equation}
In  contrast, with  the  same pole-mass  analysis,  the lowest  $0^{++}$
glueball shows  the significant pole-mass  reduction of about  300MeV in
the vicinity  of $T_c$.
Hence, the  pole-mass reduction of  the thermal $0^{++}$  glueball may
serve as a better pre-critical  phenomenon of the QCD phase transition
at  finite  temperature  than  the  pole-mass shifts  of  the  thermal
mesons. Thus  the thermal glueball may become  an interesting particle
in the RHIC project in the future.

We  next  discuss  the  related  results on  the  screening  mass.   The
screening mass  is defined as  the reciprocal of the  correlation length
along the spatial direction.  It  can also reflect some of the important
nonperturbative  features of  the  QCD  vacuum.  Hence,  it  is also  an
interesting  quantity, which  may  provide  a signal  of  the QCD  phase
transition.  However,  unlike the pole-mass,  which is considered  to be
directly  measurable as  a mass  of an  elementary excitation  at finite
temperature,  the  screening  mass  is  not thought  of  as  a  directly
measurable quantity in high energy experiments.
In  the  $0^{++}$  glueball  sector,  a  significant  reduction  of  the
screening mass is reported as
\begin{equation}
	m_{\rm G}^{\rm scr}(T = 0.75 T_c) / m_{\rm G}^{\rm scr}(T \sim 0)
=
	0.6 \pm 0.1 
\end{equation} 
in  SU(3) isotropic  lattice QCD  with $\beta_{\rm  lat}=5.93$  over the
lattice of the size $16^3\times N_t$  with $N_t = 4,6,8$ at the quenched
level \cite{gupta,gupta2}.
This significant reduction  of the screening mass of  the glueball would
be an  indication of the decrease  of the nonperturbative  nature of the
QCD vacuum.
In  this sense,  it  seems to  be  consistent with  our  results on  the
significant reduction of the  pole-mass of the thermal $0^{++}$ glueball
near the critical temperature $T_c$.
In contrast to  the $0^{++}$ glueball, in the  meson sector, the changes
of the screening  mass has been found again rather  small below $T_c$ at
the quenched level \cite{laermann}.

\section{New Analysis of Temporal Correlations ---Spectral Function}
\label{section.spectral-function}
\begin{figure}[h]
\leftline{(a) $T=130$ MeV}
\includegraphics[width=0.48\figwidth]{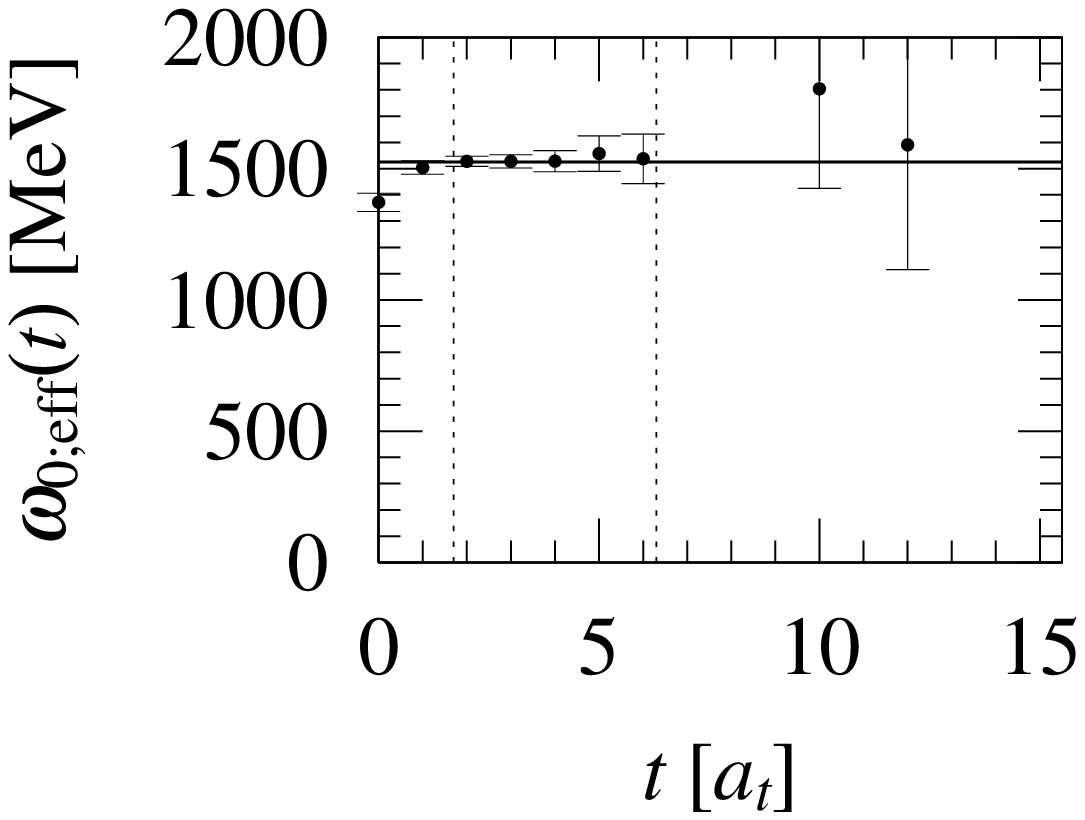}
\includegraphics[width=0.48\figwidth]{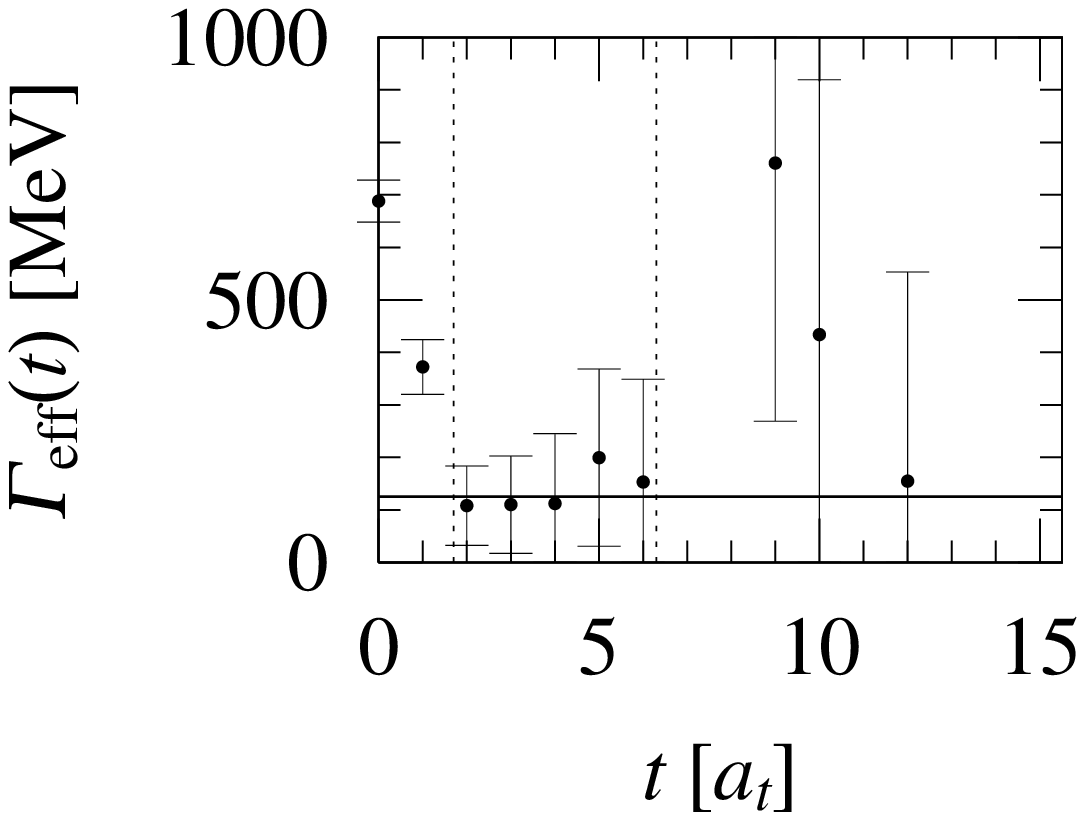}
\leftline{(b) $T=253$ MeV}
\includegraphics[width=0.48\figwidth]{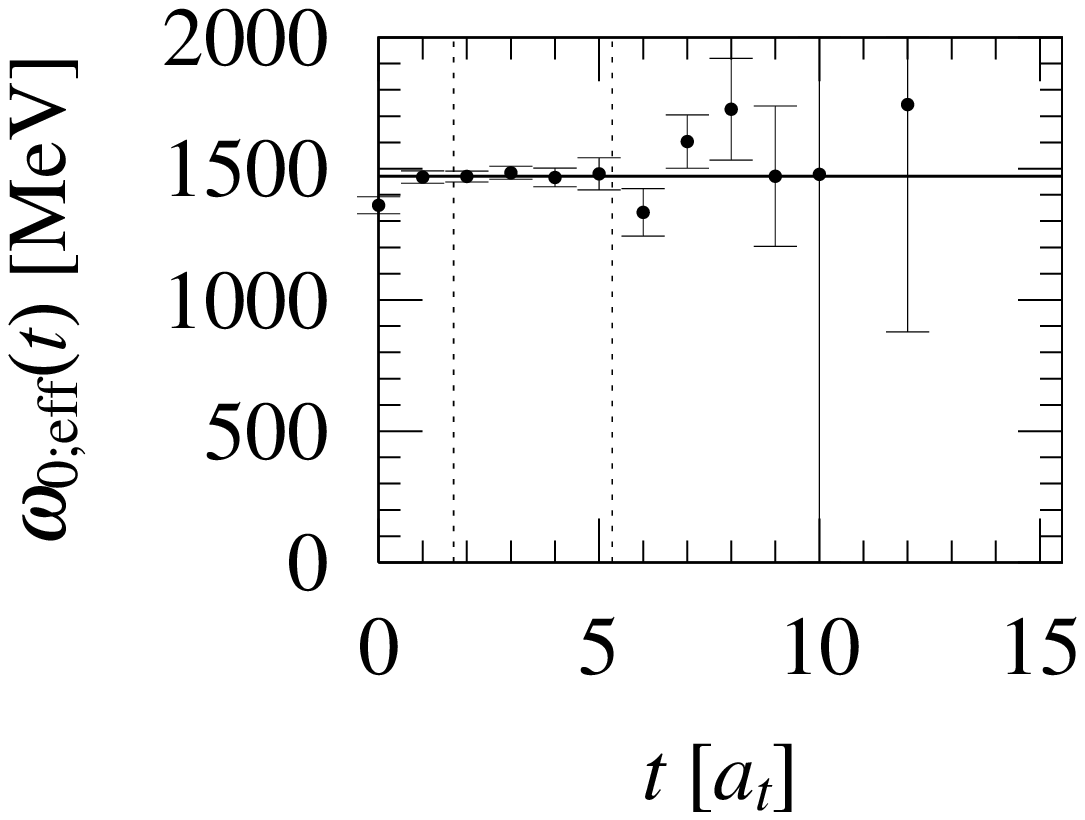}
\includegraphics[width=0.48\figwidth]{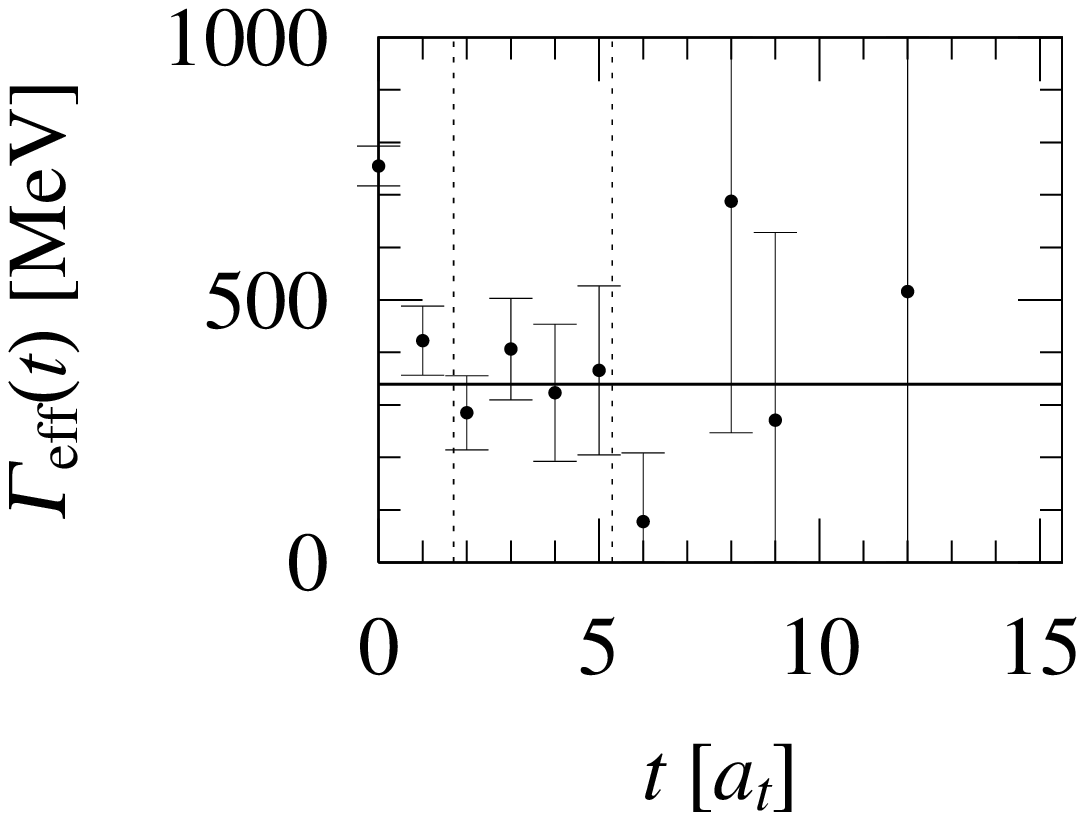}
\leftline{(c) $T=390$ MeV}
\includegraphics[width=0.48\figwidth]{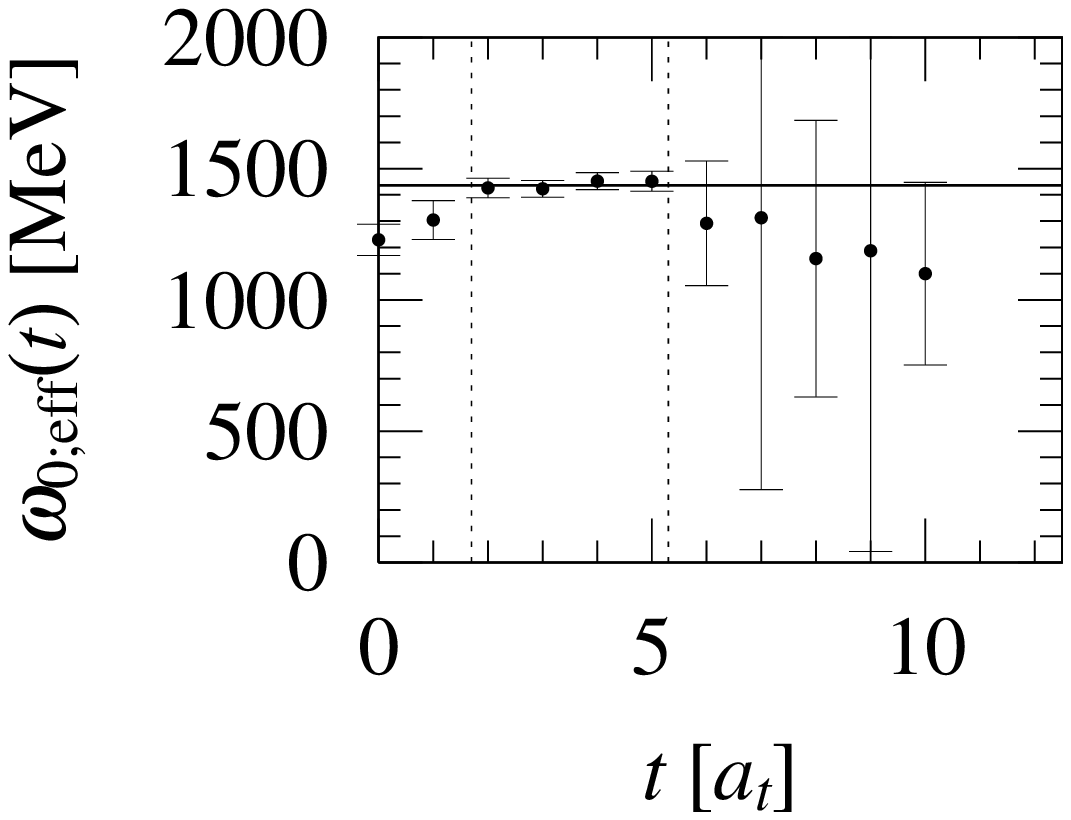}
\includegraphics[width=0.48\figwidth]{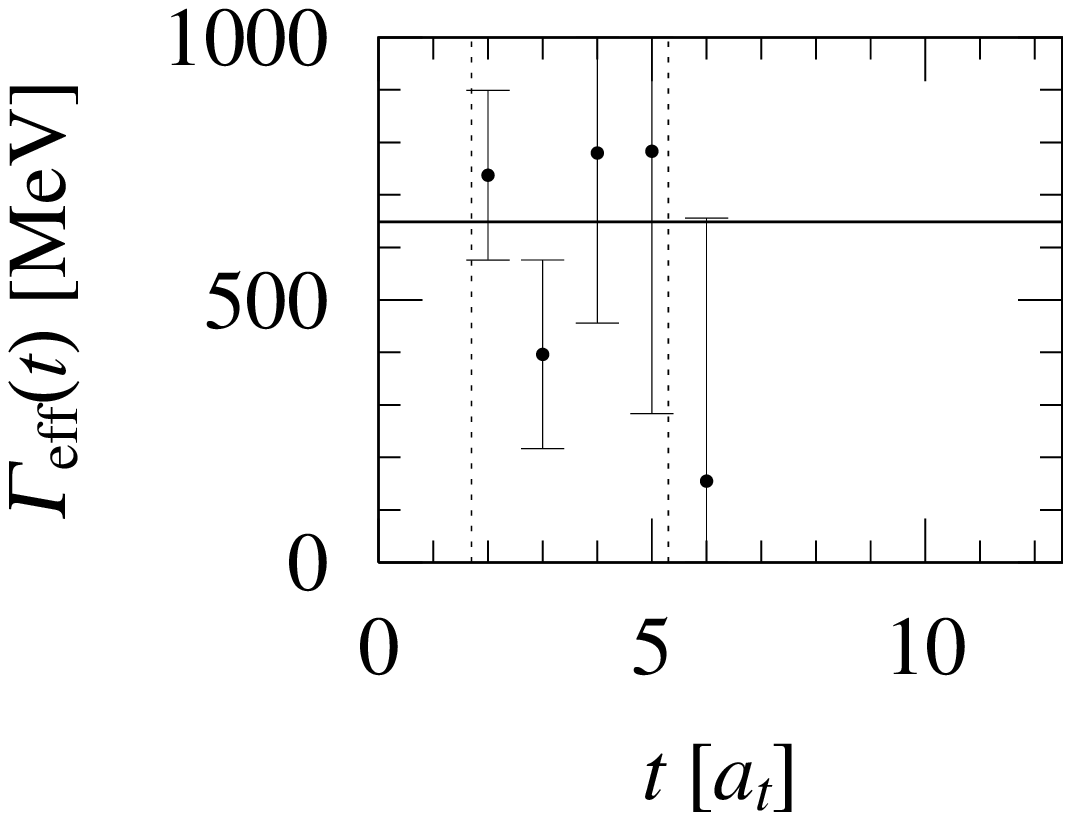}
\caption{  The  effective   center  $\omega_{0;\rm{eff}}(t)$  and  the
effective  width  $\Gamma_{\rm{eff}}(t)$   of  the  $0^{++}$  glueball
correlator at  various temperatures (a)  $T=130$ MeV, (b)  $T=253$ MeV
and (c)  $T=390$ MeV.  The solid  lines represents the  results of the
best-fit  analysis of  Breit-Wigner  type.  Here,  the fit-ranges  are
determined from  the simultaneous  plateaus, which are  indicated with
the vertical dotted lines.}
\label{local.width}
\end{figure}
\begin{figure}[h]
\includegraphics[width=0.9\figwidth]{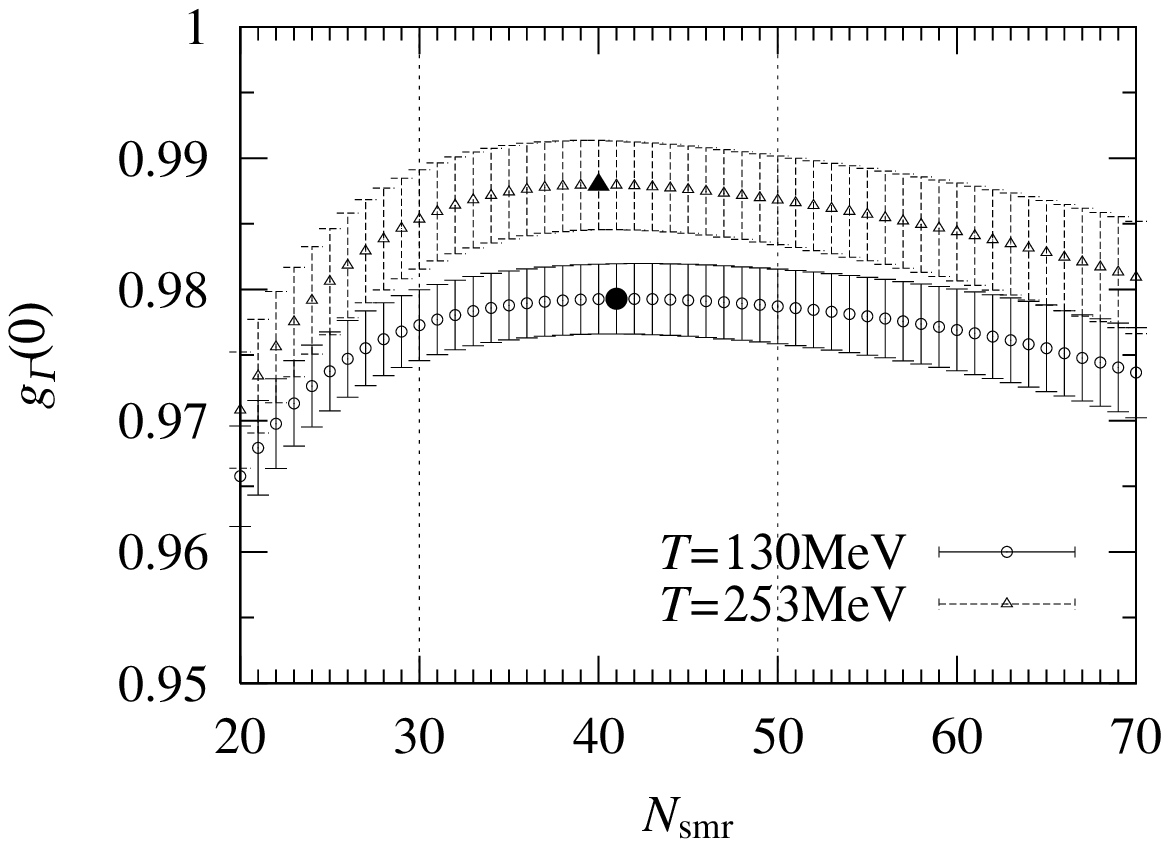}
\includegraphics[width=0.9\figwidth]{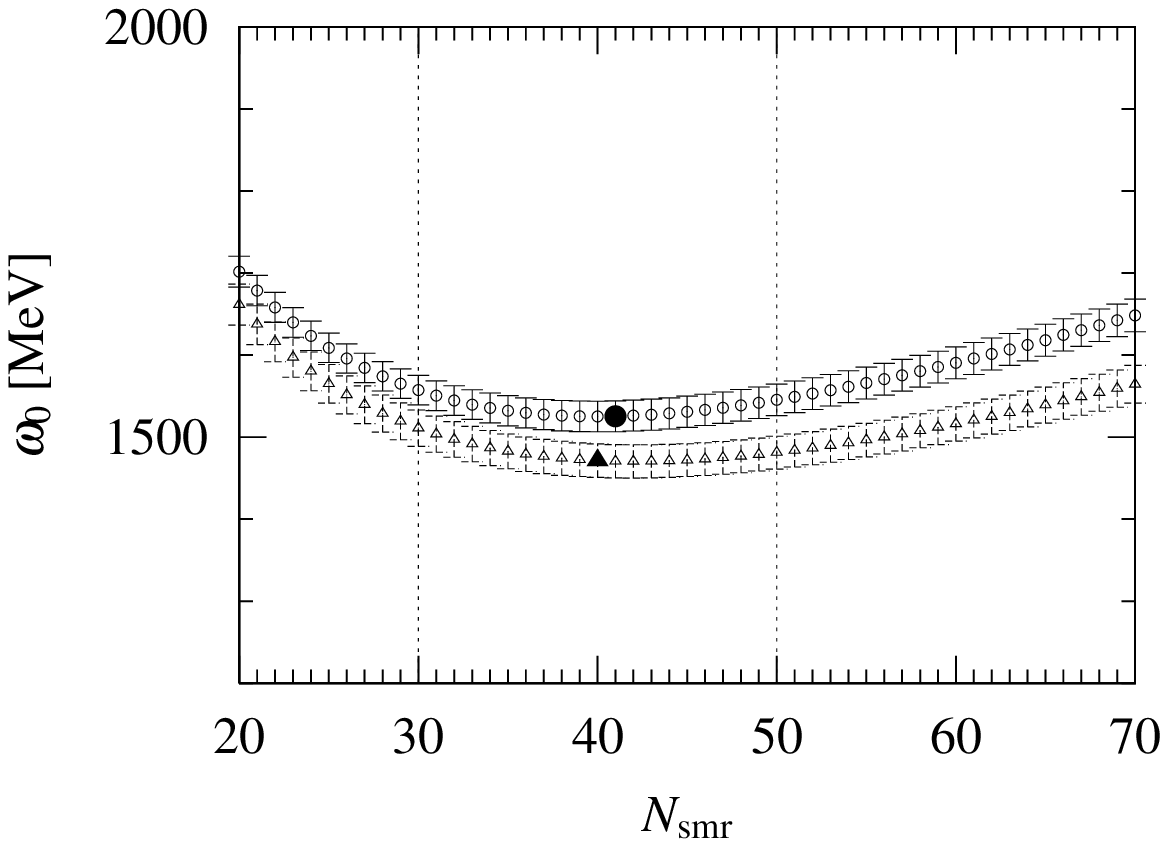}
\includegraphics[width=0.9\figwidth]{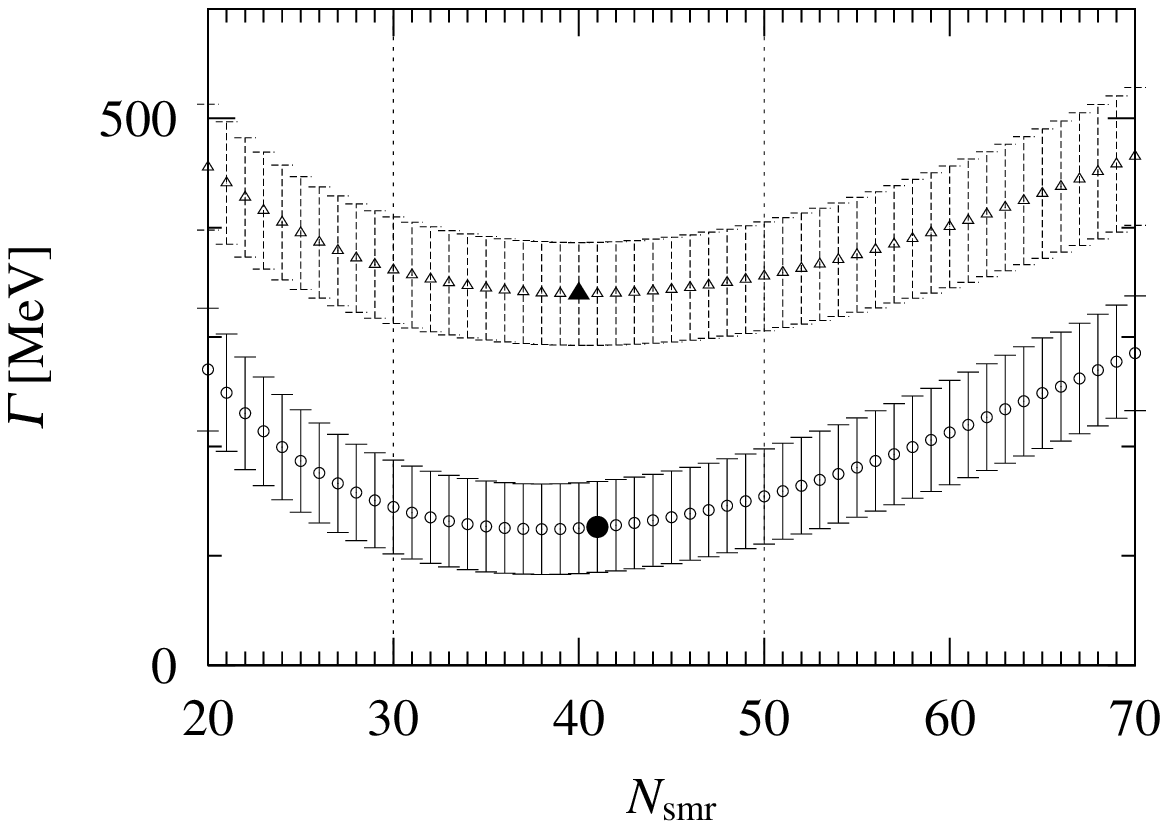}
\caption{The normalized  overlap $g_{\Gamma}(0)$, the  center $\omega_0$
and  the thermal  width  $\Gamma$ of  the  lowest peak  of the  $0^{++}$
glueball against the smearing  number $\Nsmear$ in the confinement phase
at $T=130$  MeV and $250$ MeV, which  are denoted by the  circle and the
triangle,  respectively.   The  solid  symbols  denote  the  data  where
$g_{\Gamma}(0)$ becomes  maximum. Both for $\omega_0$  and $\Gamma$, the
$\Nsmear$-dependence is small in the region of $30 \le \Nsmear \le 50$.}
\label{fig.width-nsmear}
\end{figure}
\begin{figure}[h]
\includegraphics[width=0.9\figwidth]{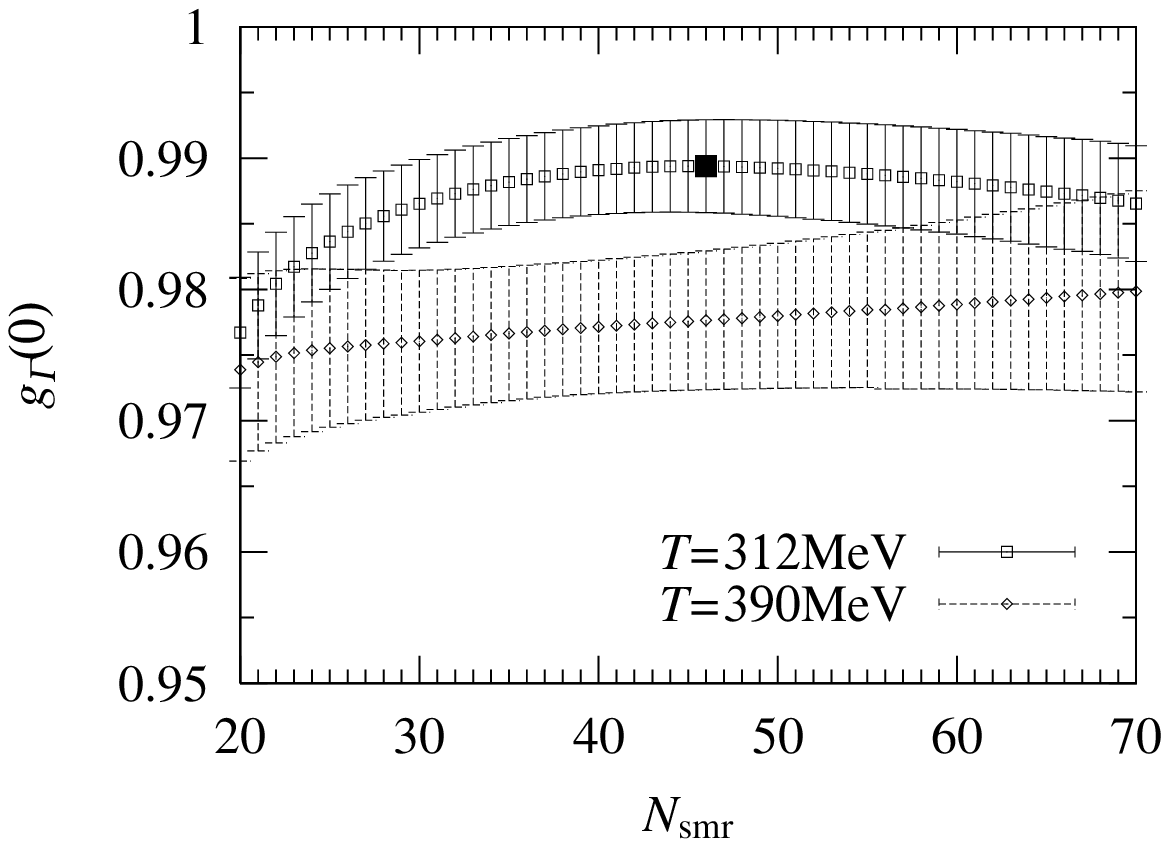}
\includegraphics[width=0.9\figwidth]{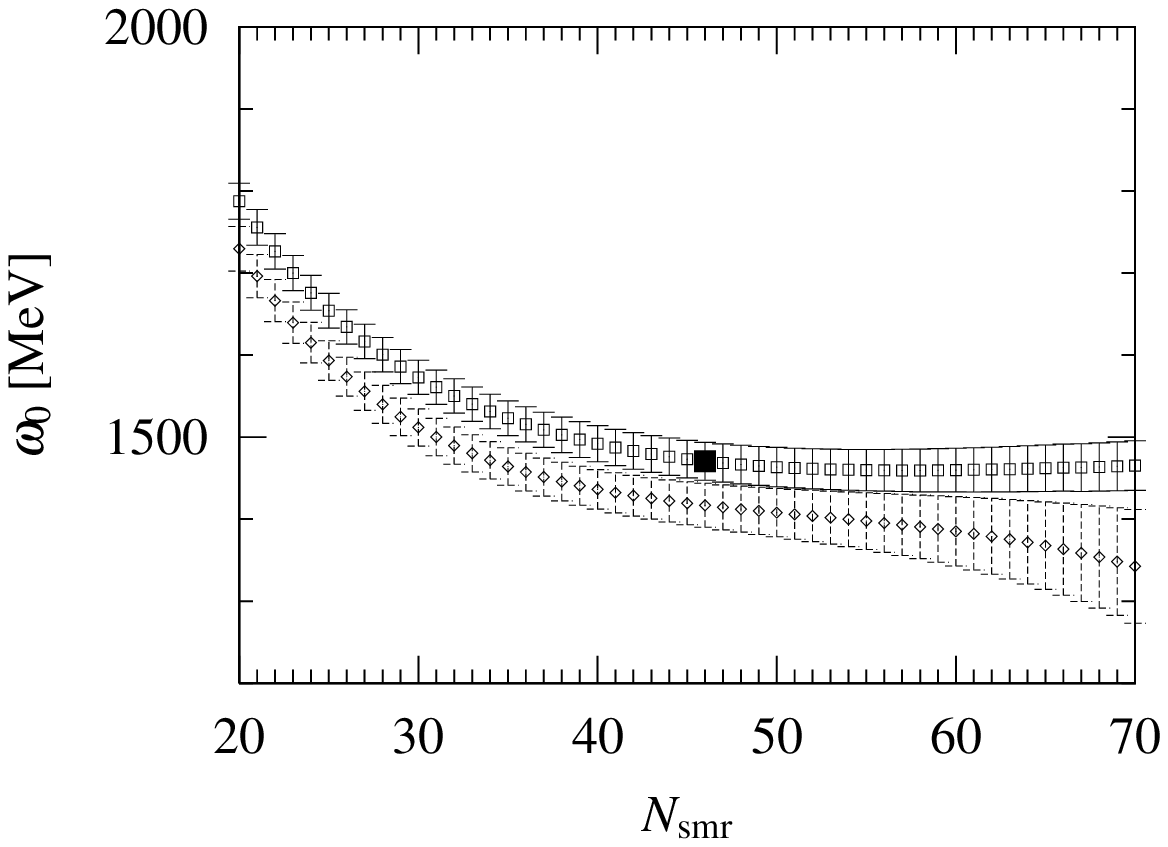}
\includegraphics[width=0.9\figwidth]{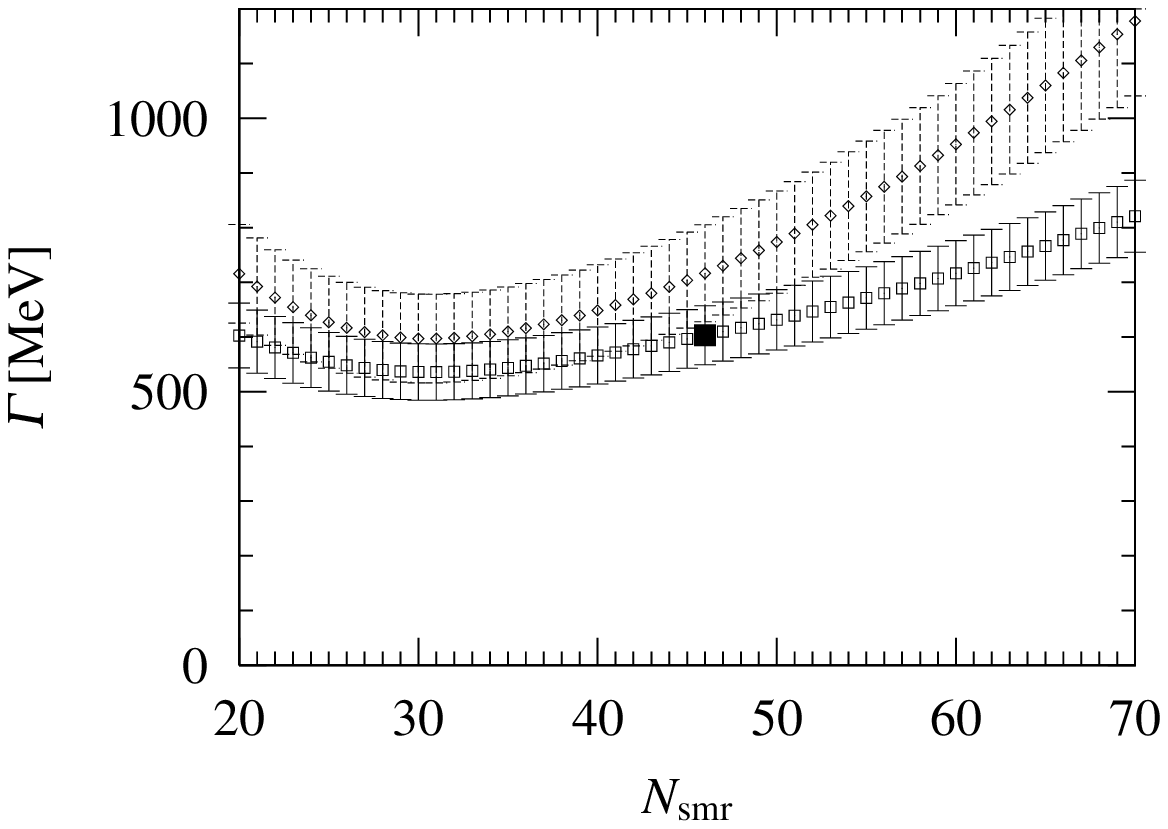}
\caption{The   similar   figure   as   \Fig{fig.width-nsmear}   in   the
deconfinement phase at $T=312$ MeV  and $390$ MeV, denoted by the square
and diamond,  respectively. The absence of  the solid diamond  is due to
the  missing maximum of  the normalized  overlap $g_{\Gamma}(0)$  in the
region   $\Nsmear  <   70$.
}
\label{fig.width-nsmear2}
\end{figure}
\begin{figure}[h]
\includegraphics[width=\figwidth]{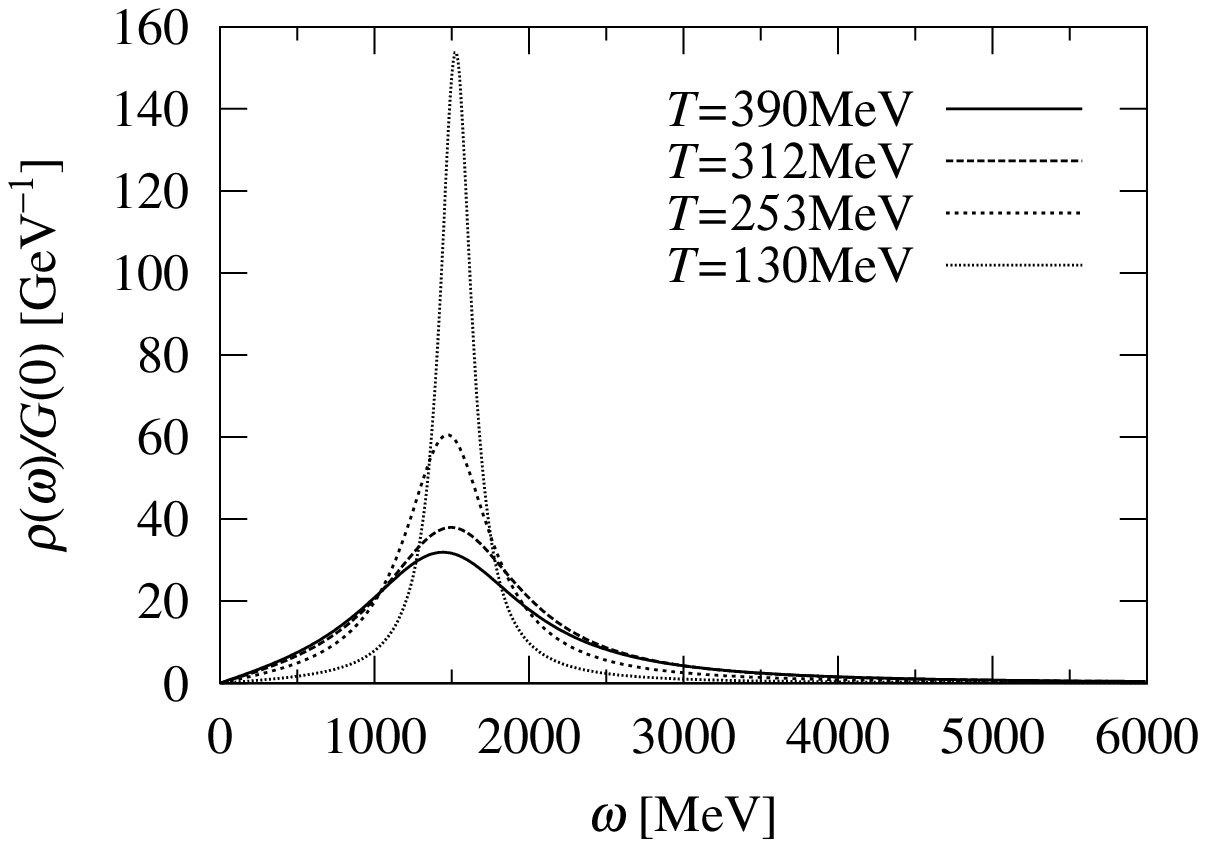}
\caption{ The spectral function  $\rho(\omega)$ of the lowest $0^{++}$
glueball at $T=130, 253, 312, 390$ MeV, obtained from the Breit-Wigner
fit   analysis   of    the   $0^{++}$-glueball   temporal   correlator
$G(t)/G(t)$. }
\label{fig.spec-S}
\end{figure}
\begin{figure}[h]
\includegraphics[width=\figwidth]{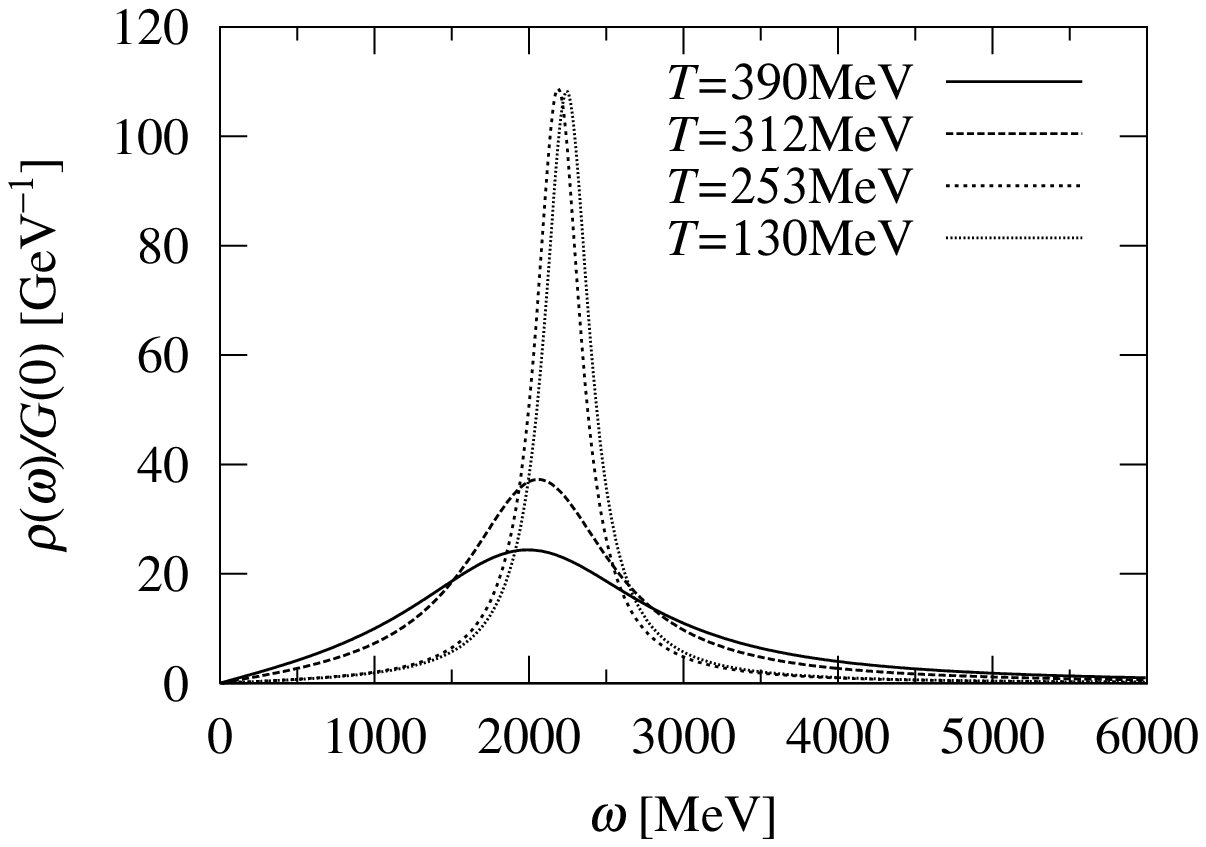}
\caption{ The spectral function  $\rho(\omega)$ of the lowest $2^{++}$
glueball at $T=130, 253, 312, 390$ MeV, obtained from the Breit-Wigner
fit   analysis   of    the   $2^{++}$-glueball   temporal   correlator
$G(t)/G(t)$. }
\label{fig.spec-T}
\end{figure}
\begin{figure}[h]
\includegraphics[width=\figwidth]{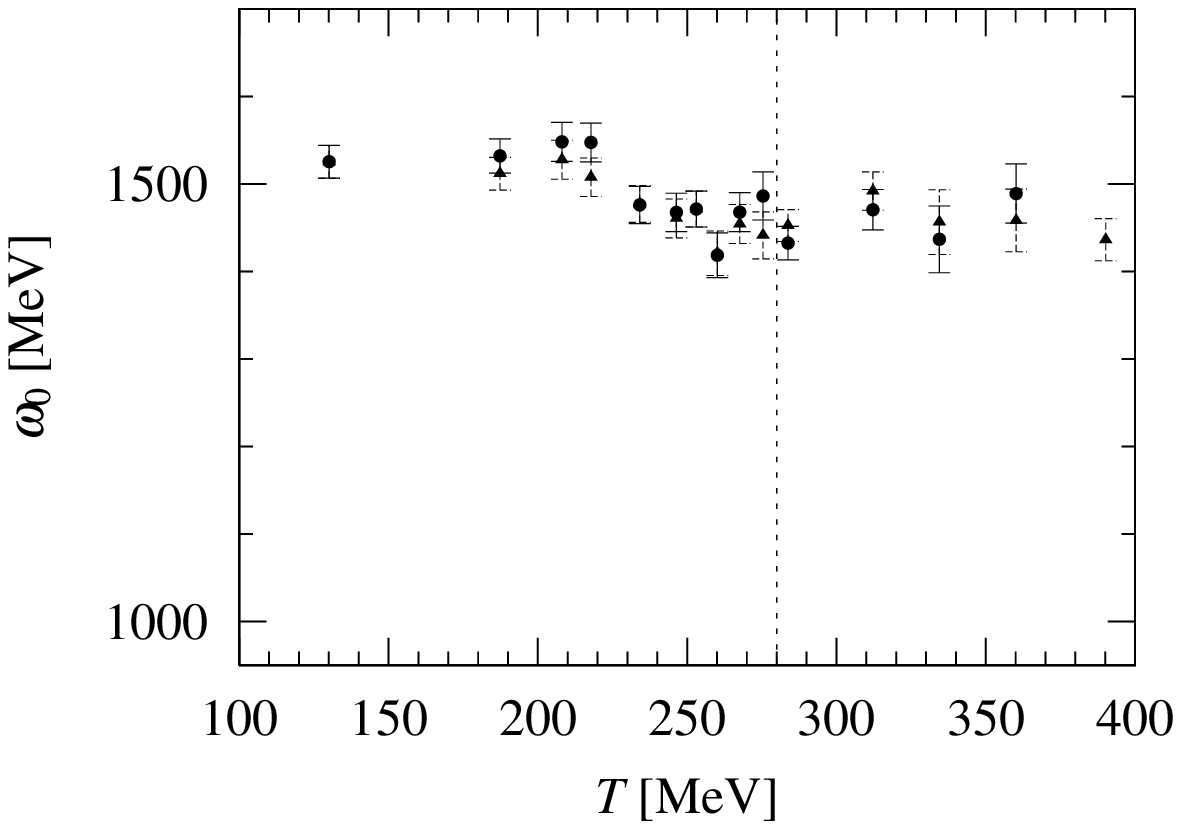}
\includegraphics[width=\figwidth]{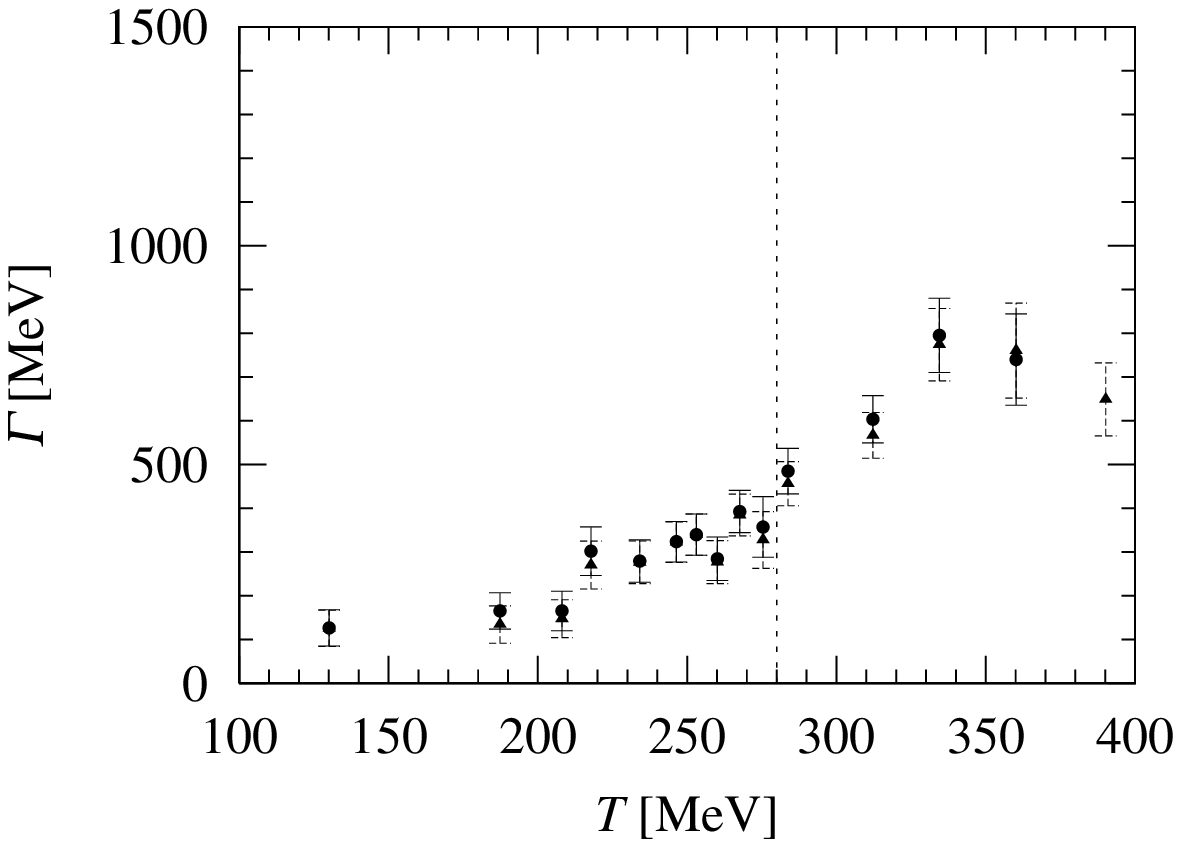}
\caption{ The  center $\omega_0(T)$ and  the width $\Gamma(T)$  of the
lowest-state  peak in  the  spectral function  $\rho(\omega)$ for  the
$0^{++}$  glueball against  temperature  $T$ in  the Breit-Wigner  fit
analysis.   The  triangle  denotes  the results  associated  with  the
suitable  smearing  as  \Eq{suitable.parameter.set},  and  the  circle
denote  the results  with $N_{\rm  smr}$ listed  in \Table{L-table-S},
which maximizes the normalized overlap $g_\Gamma(0)$. }
\label{fig.lorentzian.s}
\setcounter{subfigure}{0}
\refstepcounter{subfigure}
\label{fig.lorenzian.s.a}
\refstepcounter{subfigure}
\label{fig.lorenzian.s.b}
\end{figure}
\begin{figure}[h]
\includegraphics[width=\figwidth]{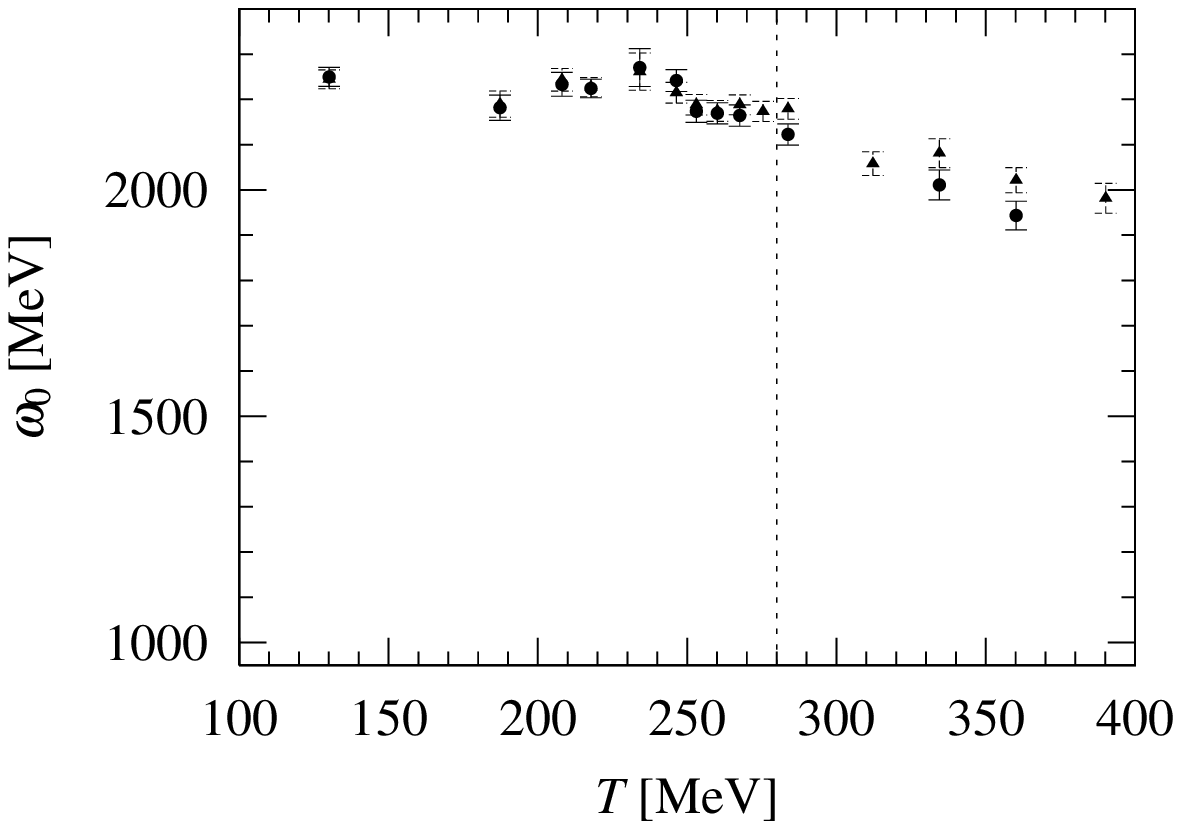}
\includegraphics[width=\figwidth]{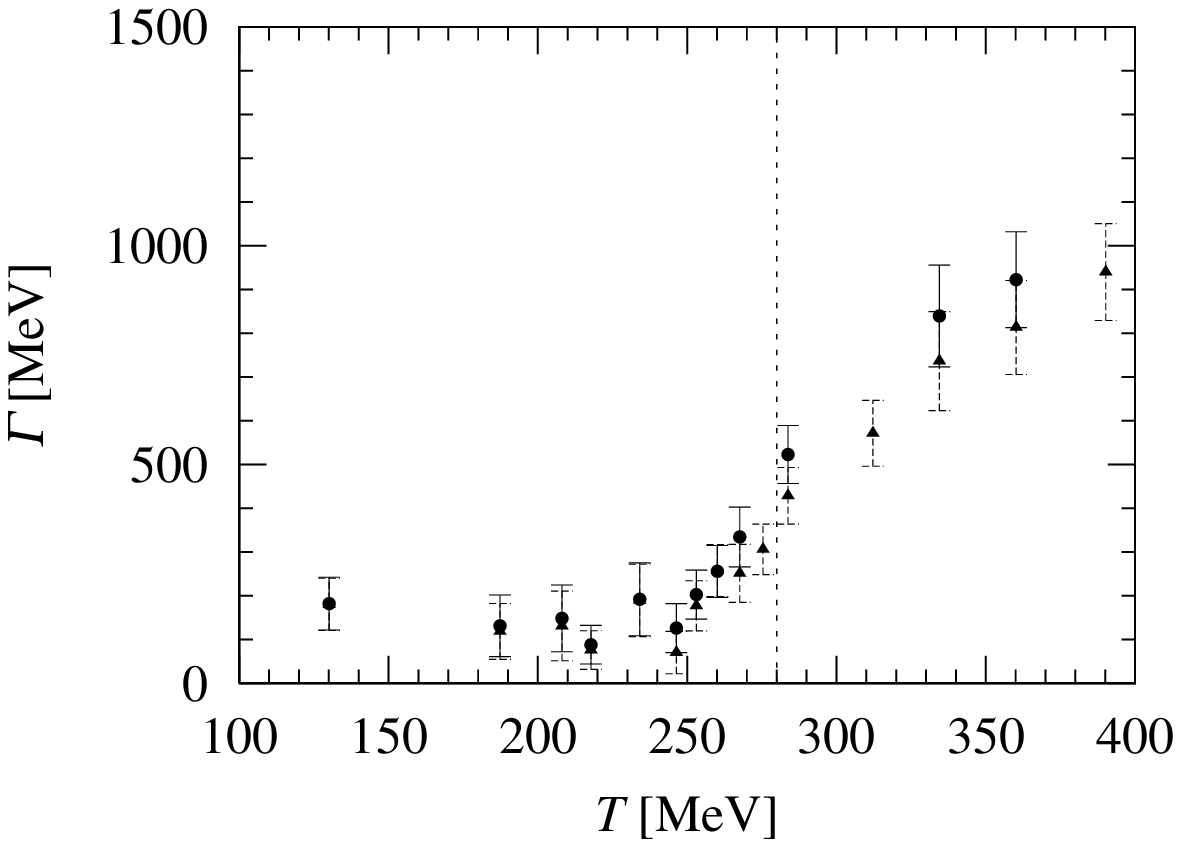}
\caption{ The  center $\omega_0(T)$ and  the width $\Gamma(T)$  of the
lowest-state  peak in  the  spectral function  $\rho(\omega)$ for  the
$2^{++}$  glueball against  temperature  $T$ in  the Breit-Wigner  fit
analysis.  The  triangle denotes the results in  the suitable smearing
as \Eq{suitable.parameter.set}, and the circle denote the results with
$N_{\rm  smr}$  listed   in  \Table{L-table-T},  which  maximizes  the
normalized overlap $g_\Gamma(0)$.  }
\label{fig.lorentzian.t}
\setcounter{subfigure}{0}
\refstepcounter{subfigure}
\label{fig.lorenzian.t.a}
\refstepcounter{subfigure}
\label{fig.lorenzian.t.b}
\end{figure}
In this section,  we attempt to generalize the  pole-mass analysis given
in  \Sect{section.pole-mass-measurement}.   Our aim  is  to analyze  our
lattice QCD data of the temporal correlator $G(t)/G(0)$ without assuming
the  narrowness of  the thermal  width of  the bound-state  peak  in the
spectral function.   We first  consider the functional  form of  the new
fit-function, which takes into account  the effect of the thermal width.
We  also  provide  its  theoretical  background.  We  then  present  the
numerical  results of  this new  analysis.  The  section is  closed with
several comments and cautions.

\subsection{Theoretical consideration on the temporal correlator 
and the Breit-Wigner fit-function}
\label{theory.breit-wigner}

In \Sect{section.pole-mass-measurement},  we have presented  the results
of the  pole-masses of  the thermal $0^{++}$  and $2^{++}$  glueballs at
finite  temperature obtained in  the best-fit  analysis of  the temporal
correlator $G(t)/G(0)$ with the fit function of the single-cosh type.
In  this  analysis,  we  have   regarded  each  thermal  glueball  as  a
quasi-particle  and   have  assumed  that  the  thermal   width  of  the
corresponding peak is sufficiently  narrow in the spectral function.  In
this case, the ground-state contribution to the spectral function can be
well  approximated  as  $\rho(\omega) \simeq  2\pi\{\delta(\omega-m_{\rm
G}(T))   -  \delta(\omega   +  m_{\rm   G}(T))\}$  by   introducing  the
temperature-dependent pole-mass $m_{\rm  G}(T)$ of the thermal glueball.
As a  consequence, in order to  measure the pole-mass, we  can adopt the
standard  single  hyperbolic  cosine  in  \Eq{single.cosh}  as  the  fit
function, and perform  the best-fit analysis in exactly  the same way as
the zero-temperature case.

However, to be  strict, at finite temperature, each  bound-state peak in
the spectral  function $\rho(\omega)$ acquires a  non-zero width through
the  thermal  fluctuations.
Since the thermal width would  grow with the temperature, it is expected
to play the more significant roles and become the less negligible at the
higher  temperature. Hence,  it is  desirable to  take into  account the
effect of the thermal width in the best-fit analysis.
We can directly see how such a thermal width $\Gamma(T)$ affects the
temporal correlator $G(t)$ in the spectral representation
\Eq{spectral.rep}.

There  may have already  appeared a  signal of  such a  non-zero thermal
width        in         the        glueball        correlators        in
\Figs~\ref{green.72},~\ref{green.37},~\ref{green.30},~\ref{green.24},
where  the ground-state  enhancement works  the less  effective  for the
higher  temperature.  In \Table{table-S},  we  observe  a tendency  that
$g(0)$ declines  with the  increasing temperature.  For  instance, $g(0)
\simeq  0.95$ for  the low  temperature $T=130$  MeV becomes  $g(0) \alt
0.90$  for the  higher temperature  $T \agt  250$ MeV  for  the $0^{++}$
glueball.        In        the       effective-mass       plots       in
\Figs~\ref{green.72},~\ref{green.37},~\ref{green.30},
we observe  that the plateau tends to  start with the larger  $t$ at the
higher temperature.
These phenomena  may indicate that  the existence of the  other spectral
components become the less negligible at the higher temperature.

In fact, at finite temperature, each bound state peak acquires a thermal
width.  The thermal width is one of the most feasible candidates for the
source of these other  spectral components, since the contributions from
such a  bunch of  spectral components cannot  be well separated  even by
means of the smearing procedure.
In this section, we perform a  more general new analysis of the temporal
correlator $G(t)/G(0)$ as an attempt to take into account the effects of
the thermal width  $\Gamma(T)$ of the ground-state peak  in the spectral
function $\rho(\omega)$.

At  zero  temperature, the  bound-state  poles  of  the Green  functions
$G_{\rm R}(\omega)$ and $G_{\rm A}(\omega)$ in \Eq{real.time} lie on the
real  axis  in  the  complex $\omega$-plane.   Hence,  the  lowest-state
contribution to the spectral function $\rho(\omega)$ is expressed as
$\displaystyle \rho(\omega) = 2\pi  A\left\{\delta(\omega - m_{\rm G}) -
\delta(\omega + m_{\rm G}) \right\}$,
where  $A$  and  $m_{\rm G}$  are  the  strength  and  the mass  of  the
lowest-state  pole,  respectively.  By  inserting  this expression  into
\Eq{spectral.representation},  and  taking  the  limit $\beta  =  1/T\to
\infty$, we recover the familiar expression as
\begin{equation}
	G(t)
=
	\lim_{\beta \to \infty}
	A
	\frac{
		\cosh\left(\Tate m_{\rm G}(\beta/2 - t) \right)
	}{
		\sinh(\beta m_{\rm G}/2)
	}
=
	A e^{-m_{\rm G}t}.
\end{equation}

With the increasing temperature,  these bound-state poles are moving off
the  real  axis  into  the   second  Riemannian  sheet  of  the  complex
$\omega$-plane.   Suppose  that  the  ground-state pole  is  located  at
$\omega = \omega_0 -  i\Gamma$ ($\omega_0,\Gamma \in \mathbb{R}$) in the
second Riemannian sheet.  Since  the spectral function $\rho(\omega)$ is
the imaginary part of  the Green function, the ground-state contribution
in  the  spectral  function  $\rho(\omega)$  can  be  expressed  in  the
following way:
\begin{eqnarray}
	\rho(\omega)
&=&
	- 2 \mbox{Im}\left( G_{\rm R}(\omega) \right)
\label{width.width}
\\\nonumber
&=&
	2\pi A\left(\Tate
		\delta_{\Gamma}(\omega - \omega_0)
	-	\delta_{\Gamma}(\omega + \omega_0)
	\right)
	+ \cdots.
\end{eqnarray}
Here,  $A$ represents  the residue  of the  pole, $\delta_{\epsilon}(x)$
denotes the Lorentzian at $x=0$  with the width $\epsilon$ ($>0$), which
is defined as
\begin{equation}
	\delta_{\epsilon}(x)
\equiv
	\frac1{\pi}
	\mbox{Im}\left(
		\frac1{x - i\epsilon}
	\right)
=
	\frac1{\pi}
	{ \epsilon \over x^2 + \epsilon^2 }.
\end{equation}
The appearance of the second term in \Eq{width.width} is due to the fact
that  the  spectral  function  $\rho(\omega)$  is  an  odd  function  in
$\omega$, reflecting  the bosonic nature of  the glueball.  ``$\cdots$''
in  \Eq{width.width}  represents  the  contributions  from  the  excited
states,  which  are expected  to  be  suppressed  after the  appropriate
smearing procedure.  Corresponding to \Eq{width.width}, the ground state
contribution in the temporal glueball correlator $G(t)$ is expressed as
\begin{eqnarray}
\renewcommand{\arraystretch}{1.5}
	G(t)
&=&
	\int_{-\infty}^{\infty}
	\Bs
	\begin{array}[t]{l}\displaystyle
		{d\omega\over 2\pi}
		\frac{ \cosh(\omega(\beta/2 - t)) }{\sinh(\beta\omega/2)}
	\\\displaystyle
		\times
		2\pi A
		\left(\Tate
			\delta_{\Gamma}(\omega - \omega_0)
		-	\delta_{\Gamma}(\omega + \omega_0)
		\right)
		+ \cdots.
	\end{array}
\label{breit.wigner.correlator}
\end{eqnarray}
Hence,  to  extract  the  center  $\omega_0(T)$ and  the  thermal  width
$\Gamma(T)$  of the  lowest-state peak  in $G(t)/G(0)$,  the appropriate
fit-function is given as
\begin{equation}
\renewcommand{\arraystretch}{1.5}
	g_{\Gamma}(t)
\equiv
	\int_{-\infty}^{\infty}
	\Bs
	\begin{array}[t]{l}\displaystyle
		{d\omega\over 2\pi}
		\frac{ \cosh(\omega(\beta/2 - t)) }{\sinh(\beta\omega/2)}
	\\\displaystyle
		\times
		2\pi \tilde A
		\left(\Tate
			\delta_{\Gamma}(\omega - \omega_0)
		-	\delta_{\Gamma}(\omega + \omega_0)
		\right),
	\end{array}
\label{single.lorentz.fit}
\end{equation}
where  $\tilde A$,  $\omega_0$  and the  $\Gamma$  are used  as the  fit
parameters.    Here,    $\tilde   A$   corresponds    to   $A/G(0)$   in
\Eq{breit.wigner.correlator}, and  will be  referred to as  the strength
parameter.  We will refer to \Eq{single.lorentz.fit} as the fit-function
of ``Breit-Wigner'' type.  Note that $g_{\Gamma}(t)$ is a generalization
of the  fit-function $g(t)$  of the single-cosh  type in the  sense that
$g_\Gamma(t)$ reduces to $g(t)$ in a special limit as
\begin{equation}
	\lim_{\Gamma \to +0} g_{\Gamma}(t)=g(t).
\end{equation}
In this  sense, the best-fit analysis  of Breit-Wigner type  serves as a
generalization of  the ordinary pole-mass analysis  of single-cosh type.
Note that the analysis of  Breit-Wigner type is rather general, which is
also  applicable to  the  analysis of  temporal  correlators of  various
thermal hadrons.

We note that, after  some calculations, the fit-function $g_{\Gamma}(t)$
in \Eq{single.lorentz.fit} can be expressed by the infinite series as
\begin{widetext}
\begin{equation}
	g_{\Gamma}(t)
=
	{\tilde A}
	\left[
		\mbox{Re}\left(
			\frac{
				\cosh\left\{(\omega_0 + i\Gamma)t\right\}
			}{
				\sinh\left\{
					{\beta\over 2}
					(\omega_0 + i\Gamma)
				\right\}
			}
		\right)
	+
		2\beta\omega_0
		\sum_{n=1}^{\infty}
		\cos\left({2\pi n \over \beta}t\right)
		\left\{
			\frac1{ (2\pi n + \beta\Gamma)^2 + \beta^2 \omega_0^2 }
		-	\mbox{($n \to -n$)}
		\right\}
	\right]
\end{equation}
\end{widetext}

We make a comment on the role  of the smearing method in the presence of
the non-zero width  of the lowest state (or the  ground state) at finite
temperature.  As was  explained in \Sect{section.smearing}, the smearing
method aims at a suitable choice of the glueball operators, by providing
a series of operators with  different sizes, all of which, however, hold
the identical quantum numbers in common.
Whereas  the residues  of  the  bound-state poles  are  affected by  the
different  choices of  the glueball  operators, their  positions  in the
complex $\omega$-plane  are not affected.   One can thereby  enhance the
residue of  the particular  bound-state pole by  a clever choice  of the
operator, while keeping its position unaffected.  Note that the position
of the pole determines the characteristics of the peak, i.e., the center
and the width.   As this clever choice of the operator,  we will use the
smearing method,  which is expected to  pick up and then  to enhance the
contribution of the ground-state peak.
\subsection{Setup for the Breit-Wigner fit analysis}

In this  subsection, we consider  the fit-range of the  Breit-Wigner fit
analysis for  the temporal glueball correlator $G(t)$.   We examine also
the $\Nsmear$-dependence of the best-fit parameters in order to estimate
the  systematic error originating  from a  particular choice  of $N_{\rm
smr}$.

\subsubsection{The fit range}
We consider the  determination of the fit-range of  the Breit-Wigner fit
analysis.   As a consequence  of the  smearing method,  the contribution
from the  ground-state peak is expected  to be enhanced  in the spectral
function $\rho(\omega)$.   However, even  with the smearing  method, the
complete   elimination  of  all   the  excited-state   contributions  is
practically  impossible.   Therefore,  we  have to  reduce  further  the
remaining contributions  from the higher spectral components  as much as
possible.  To  this end,  we seek for  the appropriate  fit-range, where
$G(t)$  consists of  nearly a  single-peak contribution.   We  adopt the
strategy, which is a straightforward extension to the one adopted in the
single-cosh  analysis, i.e.,  the analysis  based on  the effective-mass
plot  in  the  following  way.   We consider  the  ``effective  center''
$\omega_{0;\rm{eff}}(t)$      and      the      ``effective      width''
$\Gamma_{\rm{eff}}(t)$,  which  are  defined  as the  solutions  to  the
following equations:
\begin{eqnarray}
	{G(t) / G(t+1)}
&=&
	{g_{\Gamma}(t) / g_{\Gamma}(t+1)}
\\\nonumber
	{G(t+1) / G(t+2)}
&=&
	{g_{\Gamma}(t+1) / g_{\Gamma}(t+2)},
\end{eqnarray}
for  given $G(t)/G(t+1)$  and  $G(t+1)/G(t+2)$ at  each  fixed $t$.   In
\Fig{local.width},  we   show  the   plots  of  the   effective  centers
$\omega_{0;{\rm{eff}}}(t)$       and      the       effective      width
$\Gamma_{\rm{eff}}(t)$ of  the $0^{++}$ glueball  correlators at various
temperatures  $T=130,  253,  390$   MeV.   The  statistical  errors  for
$\omega_{0;{\rm{eff}}}(t)$ and $\Gamma_{\rm{eff}}(t)$ are estimated with
the jackknife analysis.  In  each pair of $\omega_{0;{\rm{eff}}}(t)$ and
$\Gamma_{\rm{eff}}(t)$  plots, there appears  a simultaneous  plateau, a
region  where $\omega_{0;{\rm{eff}}}(t)$ and  $\Gamma_{\rm{eff}}(t)$ are
almost constant  simultaneously.  In this region, $G(t)$  is expected to
consist of a single  peak contribution, and $g_{\Gamma}(t)$ can properly
represent    the   contribution    of   the    ground-state    peak   in
\Eq{breit.wigner.correlator}.   Note  that  a nontrivial  $t$-dependence
indicates  the existence of  the contributions  from the  other spectral
components.  In this sense, the existence of the simultaneous plateau in
the pair of  $\omega_{0;{\rm{eff}}}(t)$ and $\Gamma_{\rm{eff}}(t)$ plots
works as  a necessary  condition to determine  whether the  fit function
$g_{\Gamma}(t)$ of Breit-Wigner type is appropriate or not.

\begin{table*}
\caption{The  analysis of  the  temporal correlator  $G(t)/G(0)$ of  the
thermal $0^{++}$  glueball based on Breit-Wigner ansatz  of the spectral
function $\rho(\omega)$  as \Eq{width.width}.  The  temperature $T$, the
temporal lattice  size $N_t$, the center $\omega_0$,  the width $\Gamma$
and  the strength  parameter  $\tilde  A$ of  the  peak, the  correlated
$\chi^2/\Ndf$, the  fit-range $(t_1,t_2)$, the  overlap $g_{\Gamma}(0)$,
the  smearing number  $\Nsmear$ which  achieves  the $g_{\Gamma}(0)$-max
condition, and the gauge configuration number $\Nconfig$ are listed. The
asterisk ``$*$'' in the column  of $\Nsmear$ indicates that the original
value  of $\Nsmear$ is  $70$, which  had been  replaced by  the suitable
smearing number $40$. We had  done this, because $\Nsmear=70$ is the end
point   and   because   the    situation   similar   to   the   one   in
\Fig{fig.width-nsmear2} at $T=390$ is observed. }
\label{L-table-S}
\begin{ruledtabular}
\begin{tabular}{cccccccccc}
$T$[MeV] & 
$N_t$ & 
$\omega_0$[MeV] &  
$\Gamma$[MeV] & 
$\tilde A$ & 
$\chi^2/\Ndf$ & 
$(t_1,t_2)$ & 
$g_{\Gamma}(0)$ & 
$\Nsmear$ & 
$\Nconfig$ \\
\hline
130 & 72 & 1525(19) &  126( 42) & 1.03( 2) & 0.14 & (2, 8) & 0.979(3) &$41$ & 5500 \\
187 & 50 & 1532(20) &  165( 42) & 1.04( 2) & 0.65 & (2,16) & 0.977(3) &$34$ & 5700 \\
208 & 45 & 1548(22) &  165( 45) & 1.05( 2) & 1.48 & (2, 9) & 0.985(3) &$33$ & 6400 \\
218 & 43 & 1547(22) &  302( 55) & 1.11( 3) & 0.94 & (3, 8) & 0.992(6) &$31$ & 9200 \\
234 & 40 & 1476(21) &  279( 48) & 1.09( 2) & 1.23 & (2, 6) & 0.988(3) &$42$ & 8600 \\
246 & 38 & 1467(22) &  323( 46) & 1.11( 2) & 0.37 & (2, 9) & 0.990(3) &$37$ & 8900 \\
253 & 37 & 1471(20) &  339( 47) & 1.11( 2) & 0.53 & (2, 7) & 0.988(3) &$40$ & 8900 \\
260 & 36 & 1418(26) &  284( 50) & 1.07( 3) & 0.18 & (3,10) & 0.976(5) &$42$ & 9900 \\
268 & 35 & 1468(22) &  392( 48) & 1.13( 3) & 0.36 & (2, 8) & 0.989(3) &$35$ & 9900 \\
275 & 34 & 1486(27) &  357( 69) & 1.10( 4) & 2.04 & (3, 8) & 0.980(8) &$32$ & 9900 \\
\hline
284 & 33 & 1432(19) &  484( 52) & 1.17( 3) & 1.89 & (2,10) & 0.988(3) &$45$ & 9900 \\
312 & 30 & 1471(23) &  603( 54) & 1.21( 3) & 1.40 & (2,13) & 0.989(4) &$46$ & 9900 \\
334 & 28 & 1437(38) &  795( 85) & 1.32( 6) & 1.00 & (2,12) & 0.997(5) &$43$ & 6200 \\
360 & 26 & 1489(34) &  740(104) & 1.25( 7) & 0.83 & (2, 6) & 0.992(6) &$36$ & 6800 \\
390 & 24 & 1436(24) &  648( 84) & 1.16( 5) & 1.00 & (2, 7) & 0.977(5) &$40^*$ & 7700 \\
\end{tabular}
\end{ruledtabular}
\end{table*}

\begin{table*}
\caption{The  analysis  of  the  thermal  $2^{++}$  glueball  correlator
$G(t)/G(0)$   with  Breit-Wigner   ansatz  of   the   spectral  function
$\rho(\omega)$.   The  meaning  of  $T$,  $N_t$,  $\omega_0$,  $\Gamma$,
$\tilde{A}$, $\chi^2/\Ndf$,  $(t_1,t_2)$, $g_{\Gamma}(0)$ and $\Nconfig$
are the same as those  in \Table{L-table-S}.}
\label{L-table-T}
\begin{ruledtabular}
\begin{tabular}{cccccccccc}
$T$[MeV] & 
$N_t$ & 
$\omega_0$[MeV] &  
$\Gamma$[MeV] & 
$\tilde A$ & 
$\chi^2/\Ndf$ & 
$(t_1,t_2)$ & 
$g_{\Gamma}(0)$ & 
$\Nsmear$ & 
$\Nconfig$ \\
\hline
130 & 72 & 2250(21) &  182( 60) & 1.04( 2) & 1.23 & (2, 7) & 0.989(5) &$39$ & 5500 \\
187 & 50 & 2182(28) &  131( 70) & 1.02( 3) & 0.59 & (2,10) & 0.980(7) &$48$ & 5700 \\
208 & 45 & 2233(26) &  148( 76) & 1.02( 3) & 1.21 & (2, 6) & 0.983(7) &$47$ & 6400 \\
218 & 43 & 2224(21) &   88( 44) & 1.00( 2) & 1.42 & (2,10) & 0.978(4) &$49$ & 9200 \\
234 & 40 & 2270(42) &  192( 83) & 1.05( 4) & 0.20 & (3, 7) & 0.998(12) &$38$ & 8600 \\
246 & 38 & 2242(24) &  126( 56) & 1.02( 2) & 1.40 & (2,18) & 0.983(5) &$58$ & 8900 \\
253 & 37 & 2174(25) &  202( 56) & 1.04( 2) & 0.10 & (2, 7) & 0.981(5) &$54$ & 8900 \\
260 & 36 & 2169(23) &  255( 59) & 1.05( 2) & 0.46 & (2, 8) & 0.983(6) &$41$ & 9900 \\
268 & 35 & 2164(24) &  334( 68) & 1.08( 3) & 1.24 & (2, 7) & 0.987(6) &$55$ & 9900 \\
275 & 34 & 2173(22) &  306( 58) & 1.06( 2) & 1.54 & (2, 8) & 0.980(5) &$40^*$ & 9900 \\
\hline
284 & 33 & 2123(24) &  523( 66) & 1.15( 3) & 0.16 & (2, 7) & 0.997(5) &$55$ & 9900 \\
312 & 30 & 2058(26) &  571( 75) & 1.16( 3) & 0.48 & (2, 6) & 0.988(6) &$40^*$ & 9900 \\
334 & 28 & 2011(33) &  839(116) & 1.28( 6) & 0.26 & (2, 6) & 1.002(8) &$51$ & 6200 \\
360 & 26 & 1943(32) &  922(109) & 1.31( 6) & 0.03 & (2, 7) & 1.001(8) &$52$ & 6800 \\
390 & 24 & 1981(33) &  940(111) & 1.29( 6) & 0.41 & (2,10) & 0.998(8) &$40^*$ & 7700 \\
\end{tabular}
\end{ruledtabular}
\end{table*}

\subsubsection{The \boldmath{$\Nsmear$} dependence}
We consider  the $\Nsmear$-dependences  of the best-fit  parameters, the
center  $\omega_0(T)$ and the  thermal width  $\Gamma(T)$ of  the lowest
peak.
Since  the physical  meaning of  $\tilde  A$ is  not so  obvious in  the
presence of  the thermal width,  we consider $g_{\Gamma}(0)$  instead of
$\tilde A$.  Note that $g_{\Gamma}(0)$  is a generalization of $g(0)$ in
the  single-cosh analysis.   We  will refer  to  $g_{\Gamma}(0)$ as  the
``(normalized)  overlap'' with  the  state corresponding  to the  lowest
peak.

In \Fig{fig.width-nsmear},  the normalized overlap  $g_{\Gamma}(0)$, the
center  $\omega_0(T)$  and the  thermal  width  $\Gamma(T)$ are  plotted
against  $\Nsmear$ for fixed  $\alpha =  2.1$ at  $T=130$ and  $253$ MeV
below $T_c$ with circles and  triangles, respectively.  In the region of
$30 \le \Nsmear \le 50$, $g_{\Gamma}(0)$ is maximized, and $\omega_0(T)$
and $\Gamma(T)$  are minimized.  These behaviors  of $g_{\Gamma}(0)$ and
$\omega_0(T)$ are  analogous to those  of the normalized  overlap $g(0)$
(or the overlap $C$) and the pole-mass $m_{\rm G}(T)$ in the single-cosh
analysis.  As for the thermal width $\Gamma(T)$, since the contamination
of  the higher spectral  components is  expected to  make it  wider, the
minimal $\Gamma(T)$ would characterize the lowest-peak saturation.

According  to  these  considerations,  we expect  that  the  lowest-peak
contribution is maximally enhanced in the region of $30 \le  \Nsmear \le
50$, and that the normalized overlap $g_{\Gamma}(0) \simeq 1$ provides a
saturation rate of the lowest-peak contribution, in exactly the same way
as $g(0) \simeq 1$ does in the single-cosh analysis.
Note that both for the center $\omega_0$ and the thermal width $\Gamma$,
the  $\Nsmear$-dependence is  rather  small  in the  region  of $30  \le
\Nsmear \le 50$, resulting in the small systematic error on the specific
choice of $\Nsmear$.
%
The behaviors are qualitatively the same at the other temperatures below
$T_c$. This is also the case for the $2^{++}$ glueball.

In contrast, qualitatively different behaviors appear above $T_c$.
In \Fig{fig.width-nsmear2}, the  normalized overlap $g_{\Gamma}(0)$, the
center  $\omega_0$ and the  thermal width  $\Gamma$ are  plotted against
$\Nsmear$ for $\alpha=2.1$ at $T=312$ and $390$ MeV above $T_c$ with the
square and  the diamond,  respectively.
We  see that  the  overlap $g_{\Gamma}(0)$  becomes  too insensitive  to
$\Nsmear$ to determine its maximum.  In some cases, it does not take the
maximum in  the region $\Nsmear <  70$.  In addition,  there are sizable
$\Nsmear$-dependences  in  $\omega_0$   and  $\Gamma$  in  this  region,
although $\Gamma$ seem  to take its minimum around  the typical smearing
number  as $N_{\rm  smr}\sim  40$.   As for  the  center $\omega_0$,  it
continues  to  decline and  does  not take  the  minimum  in the  region
$\Nsmear < 70$.
We  note that  qualitatively  the  same behaviors  are  observed in  the
$2^{++}$ channel.
Considering  these  behaviors, it  would  be  better  to consider  the
Breit-Wigner analysis  both in the  case with $\Nsmear=40$ and  in the
case with $\Nsmear$ which maximizes $g_{\Gamma}(0)$ above $T_c$.

\subsubsection{The smearing and the single-pole saturation}
In this section,  we use the smearing method to  enhance the residues of
the pair of the complex poles associated with the low-lying peak.
However, in the  presence of the non-zero thermal  width, the meaning of
the  ``low-lying'' contribution  may become  less definite  in  a strict
sense, since these  peaks can overlap with each other.
With  the  increasing  temperature,  the thermal  width  becomes  wider.
Hence,  each  bound-state peak  becomes  less  distinguishable from  one
another due to the possibly strong overlap with the neighboring peaks.
In some  cases, one may wonder  if the smearing really  extract only the
contribution from the lowest-state peak among the overlapping peaks.
Therefore,  we  have  to   check  the  smallness  of  the  excited-state
contamination  by examining the  quality of  the fitting  of $G(t)/G(0)$
with $g_\Gamma(t)$.
If $G(t)/G(0)$  can be well  fitted with $g_\Gamma(t)$  corresponding to
the  single  Lorentzian  peak,
we may expect  that the lowest-state dominates in  the spectral function
$\rho(\omega)$ as a result of the smearing.


\subsection{Numerical results of the Breit-Wigner fit}

In this  subsection, we  perform the Breit-Wigner  fit analysis  for the
lattice QCD  data of the  temporal glueball correlator. We  also present
the  numerical results  on the  spectral  function of  the glueballs  at
various temperatures.

\subsubsection{The glueball correlators and the spectral functions}
In    \Figs~\ref{correlator.72},~\ref{correlator.37},%
~\ref{correlator.30},~\ref{correlator.24}, we show the best-fit function
$g_{\Gamma}(t)$ for the $0^{++}$ glueball correlator $G(t)/G(0)$ 
with dashed curves at
various temperatures as $T=130, 253, 312, 390$ MeV, respectively.
Similarly        for       the        $2^{++}$        glueball,       in
\Figs~\ref{correlator.72-T},~\ref{correlator.37-T},~\ref{correlator.30-T},%
~\ref{correlator.24-T}, the best-fit functions $g_{\Gamma}(t)$ are added
with dashed  curves at  various temperatures as  $T=130, 253,  312, 390$
MeV, respectively.  It is remarkable that, at all of these temperatures,
the lattice  QCD data of the temporal glueball correlators $G(t)/G(0)$  are in a
good agreement with $g_{\Gamma}(t)$ in the whole region of $t$.

In  \Figs~\ref{fig.spec-S} and  \ref{fig.spec-T}, we  plot  the spectral
functions $\rho(\omega)$ in the  $0^{++}$ and $2^{++}$ glueball channels
at various temperatures.  In both  the channels, we observe the tendency
that the thermal width grows  with the increasing temperature (up to the
statistical errors).
To be strict, each of these curves corresponds to actually a part of the
spectral   function  $\rho(\omega)$,  i.e.,   it  is   the  ground-state
contribution in the spectral  function rather than the original spectral
function itself.  This can be  understood in the following  way.  Recall
that the best-fit analysis is performed in the fit-range determined from
the  simultaneous   plateau  in  the  plots  of   the  effective  center
$\omega_{0;\rm{eff}}(t)$ and the effective width $\Gamma_{\rm{eff}}(t)$.
In  this fit-range,  $G(t)$  is expected  to  be less  sensitive to  the
existence  of the  higher  spectral components,  and  hence, it  becomes
possible to  extract only the  contribution from the  ground-state peak.
However,  in  order  to   express  $G(t)$  with  the  spectral  function
$\rho(\omega)$ (cf.   \Eq{spectral.representation}) in the  whole region
of $t$,  the higher spectral  components should be necessary  even after
the smearing  procedure is applied.   Nevertheless, the goodness  of the
fitting                                                                in
Figs.~\ref{green.72},~\ref{green.37},~\ref{green.30},~\ref{green.24},%
~\ref{green.72-T},~\ref{green.37-T},~\ref{green.30-T}                 and
\ref{green.24-T}  indicates  that  the  contributions  from  the  higher
spectral components may be actually rather small.
\subsubsection{The main result of the Breit-Wigner fit}
In  \Fig{fig.lorentzian.s},  the center  $\omega_0(T)$  and the  thermal
width  $\Gamma(T)$ of  the lowest-state  peak of  the  spectral function
$\rho(\omega)$  in the  $0^{++}$  glueball channel  are plotted  against
temperature  $T$.   Here,  the  triangles  denote the  results  for  the
suitable  smearing with  $N_{\rm smr}=40$,  and the  circles  denote the
results  with  $N_{\rm smr}$,  which  maximizes  the normalized  overlap
$g_\Gamma(0)$.   The  agreement  of  both  results  indicates  that  the
systematic error originating from the particular choice of $N_{\rm smr}$
is small.

We observe that  the thermal width $\Gamma(T)$ grows  gradually with the
increasing temperature,  resulting in the significant  increase near the
critical temperature $T_c$ as
\begin{eqnarray}
	\Gamma(T\sim T_c) \sim 300 {\rm MeV}.
\end{eqnarray}
The center $\omega_0(T)$ is observed  to decline modestly by about $100$
MeV near  $T_c$ as $\omega_0(T\sim  T_c) \simeq 1440$ MeV  in comparison
with $\omega_0(T\sim 0) \simeq 1530$ MeV.

In  \Fig{fig.lorentzian.t},  the center  $\omega_0(T)$  and the  thermal
width  $\Gamma(T)$ of  the lowest-state  peak of  the  spectral function
$\rho(\omega)$  in the  $2^{++}$  glueball channel  are plotted  against
temperature $T$.  In comparison with the $0^{++}$ glueball case, changes
are  observed less significant  in the  $2^{++}$ glueball  channel below
$T_c$, which, however, is followed  by a sudden expansion of the thermal
width $\Gamma(T)$ around $T \sim T_c$.
This sudden expansion of $\Gamma(T)$ may indicate the instability of the
thermal $2^{++}$ glueball around $T_c$.

In  Tables~\ref{L-table-S}  and   \ref{L-table-T},  we  summarize  the
results  of the  Breit-Wigner fit  analysis for  the $0^{++}$  and the
$2^{++}$  glueballs, respectively, such  as the  center $\omega_0(T)$,
the thermal width $\Gamma(T)$ and the strength parameter $\tilde A$ of
the lowest-state peak, which play  the role of the fit parameters, and
so on.

As a remarkable  result of the Breit-Wigner analysis,  we emphasize that
the  gradual growth  of the  thermal width  $\Gamma(T)$ of  the $0^{++}$
glueball begins already  far below $T_c$ in the  confinement phase. This
may be  an attractive  feature for the  experimental observation  of the
thermal effect.

\subsubsection{Comparison of the Breit-Wigner analysis with  the pole-mass
analysis}
Finally, we  consider the relations  between the pole-mass  analysis and
the Breit-Wigner analysis.
As is  mentioned before, the Breit-Wigner analysis  is a straightforward
extension  to the  pole-mass analysis.   In fact,  they coincide  in the
vanishing  width limit  $\Gamma \to  +0$.  In  this limit,  we  have the
equality, 
\begin{equation}
	m_{\rm G}(T) = \omega_0(T).
\end{equation}
However, in  the presence  of the width  $\Gamma$, the  relation between
$m_{\rm  G}(T)$ and $\omega_0(T)$  becomes nontrivial,  and we  are left
with the inequality,
\begin{equation}
	m_{\rm G}(T) \le \omega_0(T).
\label{massinequality}
\end{equation}
To see this, we  consider the spectral representation \Eq{spectral.rep},
where the temporal  correlator $G(t)$ is expressed as  an average of the
hyperbolic cosine of frequency $\omega$ with the weight as
\begin{equation}
	\frac{\rho(\omega)}{2\sinh(\beta\omega/2)}.
\label{average.1}
\end{equation}
Suppose the idealized case where the spectral function $\rho(\omega)$ is
completely  dominated by a  single peak.   In this  case, \Eq{average.1}
reads
\begin{equation}
	W(\omega) \equiv \frac{
		2\pi A\left(\Tate
			\delta_{\Gamma}(\omega - \omega_0)
		-	\delta_{\Gamma}(\omega + \omega_0)
		\right)
	}{2\sinh(\beta \omega/2)}.
\label{average.1a}
\end{equation}
To define the pole-mass, we  approximate the weight $W(\omega)$ with the
$\delta$-functions as
\begin{equation}
	\frac{
		2\pi A\left(\Tate
			\delta(\omega - m_{\rm G}) + \delta(\omega + m_{\rm G})
		\right)
	}{2\sinh(\beta m_{\rm G}/2)},
\label{average.2}
\end{equation}
where the  pole-mass $m_{\rm  G}$ is determined so as to reproduce the
shape of $G(t)/G(0)$ in a fit range as faithful as possible.
Note  that $\sinh(\beta\omega/2)$ 
in the denominator  of \Eq{average.1a} works as a
biased  factor,  which  enhances   the  smaller  $\omega$  region  while
suppressing the larger $\omega$ region.  As a consequence of this biased
factor, the peak position of $W(\omega)$ is shifted from $\omega_0$ to a
smaller value, which leads to the inequality \Eq{massinequality}.

To make a  rough estimate of the pole-mass $m_{\rm  G}$, we consider the
peak position $\omega=\omega_{\rm max}$ ($>0$) of the weight $W(\omega)$
in \Eq{average.1a}, which is defined as
\begin{equation}
	\left.
		{d\over d\omega} W(\omega)
	\right|_{\omega=\omega_{\rm max}}
=
	0.
\label{peaklocation}
\end{equation}
We adopt the simple identification as
\begin{equation}
	m_{\rm G} \simeq \omega_{\rm max}.
\label{simple.identification}
\end{equation}
In the confinement phase $T <  T_c$, the results of the Breit-Wigner fit
indicate the inequality as
\begin{equation}
	\omega_0 \gg \Gamma, T.
\end{equation}
In this case, \Eq{average.1a} can be approximated as 
\begin{equation}
	W(\omega)
\simeq
	2\pi A e^{-\beta\omega/2} \delta_{\Gamma}(\omega - \omega_0)
\label{approx.weight}
\end{equation}
in  the  region  of  $\omega \sim  \omega_0$.  Hence,  \Eq{peaklocation}
reduces to
\begin{equation}
	(\omega_{\rm        max}-\omega_0)
+
	\frac{\beta}{4}
	\left\{
		(\omega_{\rm max}-\omega_0)^2
	+	\Gamma^2
	\right\} \simeq 0.
\end{equation}
As a consequence of the simple identification in \Eq{simple.identification}, 
we find the relation among $m_{\rm G}(T)$, $\omega_0(T)$ and $\Gamma(T)$ as
\begin{equation}
	m_{\rm G}(T)
\simeq 
	\omega_0(T)
-
	\left\{
		2T -\sqrt{4T^2-\Gamma(T)^2}
	\right\}
\le
	\omega_0(T),  
\end{equation}
which seems consistent with our numerical results below $T_c$.

We make a comment on the experimentally observed particle mass at finite
temperature in the  presence of the width $\Gamma$.   In the high-energy
experiments, the spectral  function $\rho(\omega)$ is actually observed,
providing  the relevant information  of the  mass and  the width  of the
particle at  finite temperature.   In fact, the  experimentally observed
mass is expected to be distributed around $m=\omega_0(T)$ with the width
$\Gamma(T)$.
As long as the width  $\Gamma$ is enough narrow, the pole-mass $m_{\rm
G}(T)$  provides  a  good   approximation  of  the  peak  position  of
$\omega_0(T)$ in $\rho(\omega)$.
In  this case, the  pole-mass $m_{\rm  G}(T)$ can  be regarded  as the
thermal particle mass.
In  fact, in  all of  the previous  studies of  the  hadronic temporal
correlations  \cite{taro,umeda,ishii},  the  single-cosh analysis  has
been used assuming that the thermal width is enough narrow.
However, when  the width $\Gamma$ becomes wide,  the pole-mass $m_{\rm
G}(T)$ obtained from the  single-cosh analysis suffers from the effect
of the biased factor  ``$\sinh(\beta\omega/2)$'' in the denominator in
\Eq{average.1}.  Therefore, $m_{\rm  G}(T)$ is shifted from $\omega_0$
to  a smaller  value.   In  this case,  the  Breit-Wigner analysis  is
preferable.
As is  demonstrated before,  the Breit-Wigner analysis  physically means
the direct measurement of the pole position of the low-lying particle in
the complex $\omega$-plane, i.e., the shape of the low-lying peak in the
spectral function $\rho(\omega)$. It provides  the mass and the width of
the particle at  finite temperature as the center  $\omega_0(T)$ and the
width $\Gamma(T)$, respectively.

\section{Summary and Concluding Remarks}
\label{section.summary}
We have  studied the thermal  properties of the $0^{++}$  and $2^{++}$
glueballs    using     SU(3)    anisotropic    lattice     QCD    with
$\beta_{\rm{lat}}=6.25$,  the   renormalized  anisotropy  $\xi  \equiv
a_s/a_t=4$ ($a_s \simeq 0.084$ fm  and $a_t \simeq 0.021$ fm) over the
lattice of the size $20^3 \times N_t$  with $N_t = 24, 26, 28, 30, 33,
34, 35, 36, 37, 38, 40, 43, 45, 50, 72$ at the quenched level.

To begin with, we have  measured the temporal correlators $G(t)$ for the
lowest  $0^{++}$ and  $2^{++}$  glueballs using  more  than 5,500  gauge
configurations  at  each  temperature.   In this  calculation,  we  have
adopted the smearing  method to construct the suitable  operator for the
lowest-lying $0^{++}$  and $2^{++}$ glueballs  on the lattice.   We have
also provided an analytical consideration on the physical meaning of the
smearing procedure based on the  spatial distribution of the gluon field
in the smeared operator.

Next, we  have performed  the  pole-mass  measurements  of the  thermal
glueballs from  $G(t)$ by  adopting the procedure  used in  the standard
hadron mass measurements.
For  the  lowest  $0^{++}$  glueball,  we have  observed  a  significant
pole-mass reduction of about $300$  MeV near $T_c$ or $m_G(T \simeq T_c)
\simeq 0.8  m_G(T \sim 0)$, while  its size remains  almost unchanged as
$\rho(T) \simeq 0.4$  fm.  This pole-mass shift is  actually much larger
than any  other pole-mass  shifts which have  been ever observed  in the
meson sector \cite{taro,umeda} with the similar analysis in lattice QCD.


Finally,  for  completeness,  we  have  performed  a  more  general  new
fit-analysis of the temporal glueball correlator $G(t)$ as an attempt to
take into account the effects  of the possible appearance of the thermal
width $\Gamma(T)$ of the bound-state peak.
We  have proposed the  Breit-Wigner form  as a  generalized fit-function
$g_{\Gamma}(t)$   for   the  lowest-peak   in   the  spectral   function
$\rho(\omega)$ of the temporal correlator $G(t)$ at finite temperature.
This ansatz  is also applicable  to the temporal correlators  of various
thermal hadrons.  In this advanced analysis, we have found a significant
broadening of the lowest-glueball peak  as $\Gamma(T) \sim 300$ MeV near
$T_c$ as well as a rather  modest reduction in its center of about $100$
MeV.

We have investigated also  the temporal correlators of the color-singlet
modes  corresponding to the  glueballs in  the deconfinement  phase.  We
have found  that the  thermal width $\Gamma(T)$  increases monotonically
with  increasing  temperature both  in  the  $0^{++}$  and the  $2^{++}$
glueball channels.
However,  from only  our  lattice  data, it  is  difficult to  determine
whether such color singlet modes really  survive above $T_c$ or not as a
meta stable modes, which is left for the future studies.

%
%

In this way, these two analyses have indicated the significant thermal
effects to the lowlying thermal  $0^{++}$ glueball near $T_c$, such as
the considerable  pole-mass reduction  or the width  broadening.
Note  that  the  actual   decay  width  of  the  glueball  candidates,
$f_0(1500)$ and $f_0(1710)$, are known to be about $100$ MeV, which is
rather small.   Hence, either the  300 MeV pole-mass reduction  or the
thermal width broadening  of 300 MeV, if happens, will  be seen in the
change of the spectral function in the high-energy experiment.
Therefore,  the   thermal  properties  of  glueball   may  provide  an
interesting signal  of the precritical phenomenon of  the QGP creation
in the future experiment in RHIC.  For more direct comparison with the
experiment, it is desirable to include the dynamical-quark effect.

\begin{center}{\bf Acknowledgement}\end{center}
We would  like to thank  T.~Doi for his useful comments.
H.~S. is supported by Grant for Scientific Research (No.12640274) from
Ministry    of   Education,    Culture,   Science    and   Technology,
Japan.  H.~M. is  supported  by  Japan Society  for  the Promotion  of
Science for Young Scientists.
The lattice QCD Monte Carlo calculations have been performed partly on
NEC-SX5 at Osaka University and partly on SR8000 at KEK.

\end{document}